\documentclass{aa}  
\usepackage[colorlinks=true,allcolors=blue]{hyperref}
\usepackage{graphicx}
\usepackage{txfonts}
\usepackage{color}
\usepackage{multirow}
\usepackage{xcolor}
\usepackage[normalem]{ulem}

\begin{document} 

\title{Modelling the selection of galaxy groups\\ with end to end simulations
}
\author{
R. Seppi\inst{1}\thanks{E-mail: riccardo.seppi@unige.ch} \and
D. Eckert\inst{1} \and
A. Finoguenov\inst{2} \and
S. Shreeram\inst{3} \and
E. Tempel\inst{5,6} \and
G. Gozaliasl\inst{3,4} \and
M. Lorenz\inst{7} \and \\
J. Wilms\inst{7} \and
G. A. Mamon\inst{8} \and
F. Gastaldello\inst{9} \and
L. Lovisari\inst{10} \and
E. O'Sullivan\inst{11} \and
K. Kolokythas\inst{12,13} \and \\
M. A. Bourne\inst{14,15} \and
M. Sun\inst{16} \and
A. Pillepich\inst{17}
}
\institute{
Department of Astronomy, University of Geneva, Ch. d’Ecogia 16, CH-1290 Versoix, Switzerland \and
Department of Physics, University of Helsinki, Gustaf H\"{a}llstr\"{o}min katu 2, 00560 Helsinki, Finland \and
Max-Planck-Institut f\"{u}r extraterrestrische Physik (MPE), Giessenbachstraße 1, D-85748 Garching bei M\"unchen, Germany \and
Department of Computer Science, Aalto University, PO Box 15400, Espoo, FI-00076, Finland \and
Tartu Observatory, University of Tartu, Observatooriumi 1, 61602 T\~{o}ravere, Estonia \and
Estonian Academy of Sciences, Kohtu 6, 10130 Tallinn, Estonia \and
Dr. Karl Remeis-Observatory and Erlangen Centre for Astroparticle Physics, Friedrich-Alexander Universität Erlangen-Nürnberg, Sternwartstr. 7, 96049 Bamberg, Germany \and
Institut d’Astrophysique de Paris (UMR 7095: CNRS \& Sorbonne Universit\'{e}), 98 bis Bd Arago, F-75014 Paris, France \and
INAF/IASF-Milano, Via A. Corti 12, 20133 Milano, Italy \and 
INAF, Osservatorio di Astrofisica e Scienza dello Spazio, via Piero Gobetti 93/3, 40129 Bologna, Italy \and
Center for Astrophysics | Harvard \& Smithsonian, 60 Garden Street, Cambridge, MA 02138, USA \and
Centre for Radio Astronomy Techniques and Technologies, Department of Physics and Electronics, Rhodes University,
PO Box 94, Makhanda 6140, South Africa \and
South African Radio Astronomy Observatory, Black River Park North, 2 Fir St, Cape Town 7925, South Africa \and
Centre for Astrophysics Research, Department of Physics, Astronomy and Mathematics, University of Hertfordshire, College Lane, Hatfield AL10 9AB, UK \and 
Kavli Institute for Cosmology, University of Cambridge, Madingley Road, Cambridge, CB3 0HA, UK \and
Department of Physics and Astronomy, The University of Alabama in Huntsville, Huntsville, AL 35899, USA \and
Max-Planck-Institut f\"{u}r Astronomie, K\"{o}nigstuhl 17, 69117 Heidelberg, Germany
}

% These dates will be filled out by the publisher
\date{Accepted XXX. Received YYY; in original form ZZZ}

\titlerunning{X-GAP: mocks and selection function}

\abstract{
% Context.  
Feedback from supernovae and active galactic nuclei (AGN) shapes galaxy formation and evolution, yet its impact remains unclear. Galaxy groups offer a crucial probe, as their gravitational binding energy is comparable to that available from their central AGN. The XMM-Newton Group AGN Project (X-GAP) is a sample of 49 groups selected in X-ray (ROSAT) and optical (SDSS) bands, providing a benchmark for hydrodynamical simulations. }
{
%
% Aims.
In sight of such a comparison, understanding selection effects is essential. We aim to model the selection function of X-GAP by forward modelling the detection process in the X-ray and optical bands.}
{
% 
% Methods 
Using the Uchuu N-body simulation, we build a dark matter halo light cone, predict X-ray group properties with a neural network trained on hydrodynamical simulations, and assign galaxies matching observed properties. We compare the selected sample to the parent population in the light cone.}
{% 
% Results
Our method provides a sample that matches the observed distribution of X-ray luminosity and velocity dispersion.
The 50$\%$ completeness is reached at a velocity dispersion of 450 km/s in the X-GAP redshift range.
The selection is driven by X-ray flux, with secondary dependence on velocity dispersion and redshift. We estimate a 93$\%$ purity level in the X-GAP parent sample. We calibrate the velocity dispersion–halo mass relation. We find a normalisation and slope in agreement with the literature, and an intrinsic scatter of about 0.06 dex. The measured velocity dispersion is accurate within 10$\%$ only for rich systems with more than about 20 members, while the velocity dispersion for groups with less than 10 members is biased at more than 20$\%$.
}
{
%Conclusions
The X-ray follow-up refines the optical selection, enhancing purity but reducing completeness. In an SDSS-like setup, velocity dispersion measurement errors dominate over intrinsic scatter. Our selection model will enable unbiased comparisons of thermodynamic properties and gas fractions between X-GAP groups and hydrodynamical simulations.}

\maketitle

\section{Introduction}
The large scale structure (LSS) of the Universe evolves under the action of gravity following a bottom-up scenario \citep{NFW1996ApJ...462..563N, MoWhite2002MNRAS, Springel2005Natur_LSS, Fakhouri2010MNRAS_assemblyhistory}. The massive dark matter haloes we observe today in the nodes of the LSS are the end result of a process of merging and accretion from smaller haloes formed in the early Universe from the collapse of initial perturbations in the density field.
The behaviour of baryonic components such as gas and stars within the distribution of dark matter has been a key scientific puzzle for the last few decades. Feedback from active galactic nuclei \citep[AGN,][]{Padovani2017A&A_agnreview} has been suggested as a solution to a variety of questions, such as the suppression of cooling flows towards the centre of galaxy clusters \citep{McNamara2007ARA&A_gasAGN, Gitti2012_coolingflow, Fabian2012ARA&Areview, Hlavacek-Larrondo2022hxga.book_agnfeedback, Bourne2023Galax_feedback}, the quenching of star formation to reproduce the shape of the observed galaxy stellar mass function in simulations \citep{Silk1998A&A_FEG_SMBH, Pillepich2018MNRAS.473.4077P}, and the origin of scaling relations between galaxy properties and supermassive black hole (SMBH) mass \citep{Magorrian1998AJ....115.2285M, Kauffmann2000MNRAS.311..576K, Kormendy2013ARA&A_BHcoevo, Sahu2019ApJ_BHscalrel}.

Galaxy groups represent a key mass scale in this context, because their gravitational binding energy is comparable to the output energy of the central AGN. Therefore, galaxy groups are very sensitive to the physics of AGN feedback. They are located at the peak of the mass density at the current epoch \citep[][]{Tinker2008}. Groups of galaxies were historically first detected and studied in the optical and infrared bands, as collections of their member galaxies identified as over-densities in angular and redshift distributions \citep{Abell1958ApJS....3..211A, Zwicky1963cgcg.book.....Z, Huchra1982ApJ_grps, Geller1983ApJS...52...61G, Beers1990AJ....100...32B}. Since then, various works presented similar methods to identify groups from galaxy surveys, including phase space methods \citep{Mamon2013MNRAS_mamposst}, the identification of red sequence galaxies \citep{Gladders2000AJ_detRS, Saro2013ApJ_reqseq, Rykoff2014ApJ...785..104R, Licitra2016MNRAS_redgold}, Friends-of-Friends (FoF) algorithms \citep{Trevese2007A&A_fof, Munoz2012MNRAS_fofgrps, Wen2012ApJS_fofgrps, Tempel2012A&A_sdssdr8}, and modified FoF algorithms \citep{Tempel2018A&A...618A..81T, Lambert2020MNRAS.497.2954L}. Groups are often characterized in terms of velocity dispersion of their galaxy members \citep{Mamon2010A&A_veldisp, Gozaliasl2020A&A_grpsKIN}. We refer the reader to \citet{Old2014MNRAS_massrec, Old2015_scatterbias} for a comprehensive description and review about optical detection and mass estimate of galaxy groups. Galaxy groups are also often identified in X-rays thanks to thermal bremsstrahlung and line emission from the hot gas filling their potential wells \citep[e.g.,][]{Boringer2000ApJS..129..435B, Rosati2002ARA&A..40..539R, Gozaliasl2014A&A_XMM, Gozaliasl2019MNRAS_chandra}, reaching temperatures between 10$^6$ and 10$^8$ K \citep{Sanders2023arXiv_spectra}. Wide-field survey instruments in the X-ray band, such as ROSAT \citep{Truemper1982AdSROSAT} and eROSITA \citep{Merloni2012arXiv1209.3114M, Predehl2021A&A_erosita}, are therefore suitable to detect these objects. X-ray observations show prominent features such as cavities in the intra-group medium (IGrM) produced by massive AGN outflows \citep{Birzan2008_jetRadioPow, Gastaldello2009_NGC5044, Randall2015ApJ_NGC5813}. If the outburst is supersonic, shock waves propagating perpendicular to the outflow become visible \citep{Liu2019_3C88}. 

The behaviour of the gaseous atmosphere is strongly connected with both gravitational and non gravitational processes. The physical connection between these two types of properties is therefore related to thermodynamic quantities describing the gas state. For example, AGN feedback tends to disrupt the gas in a more dramatic manner compared to massive clusters, causing an excess entropy in the core \citep{Ponman1999Natur.397..135P} and/or in the outskirts, as observed by \citet{Finoguenov2002ApJ_grps, Ponman2003_entropy}. Combining data from different wavelengths allows tackling these questions from various points of view. The Complete Local Volume Groups Sample \citep[CLoGS,][]{OSullivan2017MNRAS_clogs} started from optical groups from the all-sky Lyon Galaxy
Group catalogue \citep[LGG,][]{Garcia1993A&A_LGG} and combined them with X-ray observations from Chandra and XMM-Newton, and radio data from the Giant Metre wave Radio Telescope (GMRT) and Very Large Array (VLA). Their sample was limited to the local Universe within 80 Mpc and focused on the properties of the brightest group galaxies \citep[BGGs;][]{OSullivan2018A&A...618A.126O, Kolokythas2018MNRAS_clogsRADIOgal, Kolokythas2019MNRAS_clogsRADIO2}. Later studies provided a more systematic analysis of thermodynamical properties for groups out to large radii, even up to R$_{\rm 500c}$\footnote{R$_{\rm 500c}$ is the radius encompassing an average density that is 500 times larger than the critical density of the Universe at the group redshift, $\rho_{\rm c} = 3H(z)/8\pi G$.}, using both XMM-Newton \citep{Johnson2009_xmmgrps} and Chandra \citep{Sun2009ApJ_grpsChandra}. More recently, \citet{Bahar2024_erositagrps} compared a large sample of 1178 groups detected in the first eROSITA all sky survey to various hydrodynamical simulations. They found the entropy profiles to be in good agreement between observations and simulations, whereas the groups core and inner part of the profile show some inconsistencies, meaning that group properties can potentially be a direct tracer of the AGN feedback mechanism.

Currently, hydrodynamical cosmological simulations include the implementation of AGN feedback in various forms and recipes. Some simulations, such as cosmo-OWLS \citep{LeBrun2014MNRAS_cosmowols}, EAGLE \citep{Schaye2015MNRAS.446..521S} and BAHAMAS \citep{McCarthy2017MNRASbahams} rely on an isotropic and thermal response to gas accretion onto the SMBH \citep[see e.g.][]{BoothSchaye2009MNRAS_feedback}. The gas surrounding the central black hole is heated when the AGN turns on, which suppresses gas cooling and hence star formation. Illustris \citep{Vogelsberger2014MNRAS_illustris} has separate radio mode that injects off-set hot bubbles to mimic radio lobes. This concept was also used by the Fable Simulations \citep{Henden2018MNRAS_FABLE}. Other works, including HorizonAGN \citep{Dubois2016MNRAS_horizonAGN}, MAGNETICUM \citep{Dolag2016MNRAS.463.1797D}, IllustrisTNG \citep{Pillepich2018MNRAS.473.4077P}, SIMBA \citep{Dave2019MNRAS_simba}, and Flamingo \citep{Schaye2023MNRAS_flamingo} add a kinetic feedback scheme, where part of the energy injected by the AGN is converted to kinetic energy of the surrounding gas. This allows driving outflows out of the central SMBH, more similar to a standard supernovae feedback approach \citep{Springel2003MNRAS_stellarfeedback}. \\
Most simulations are tuned to reproduce a standard set of observables, mainly the galaxy stellar mass function, but also the gas fraction for clusters in the local Universe (BAHAMAS, Fable, and Flamingo). Although different simulations agree on the prediction of the quantities used to tune them, this is not necessarily the case for inferred quantities. In fact, the predictions of the gas content and radial profiles of thermo-dynamical quantities, such as pressure and entropy, differ significantly between various works in the regime of galaxy groups, as highlighted by the reviews from \citet{Eckert2021_review, Oppenheimer2021Univ....7..209O, Gastaldello2021Univ....7..208G}.

High quality and multi-wavelength observations of galaxy groups are required to inform simulations in this regime. This is the primary goal of the XMM-Newton Group AGN Project
\citep[X-GAP,][]{Eckert2024_xgap}, a large program on XMM-Newton dedicated to 49 galaxy groups that aims at measuring the impact of AGN feedback on the intra-group medium (IGrM) out to R$_{\rm 500c}$. X-GAP is selected from the parent All-sky X-ray Extended Sources project \citep[AXES,][]{Damsted2024_axes, Khalil2024A&A_AXES}, which combines X-ray detection from the ROSAT all-sky survey using wavelet filters \citep{Kaefer2019A&A...628A..43K} and optical FoF groups detected in the Sloan Digital Sky Survey \citep[SDSS,][]{Blanton2017AJsdss} by \citet{Tempel2017A&A_sdssFOF}. On the one hand, the optical detection of galaxy groups using galaxy members is affected by projection effects that typically arise in photometric data \citep[see e.g., redmapper][]{Rykoff2014ApJ...785..104R}. Using spectroscopic information \citep{Robotham2011MNRAS_gamagrps, Tinker2021ApJ_grpssdss} alleviates projection effects, but a few issues still remain, such as low statistic, spectroscopic completeness, or redshift space distortions. On the other hand, X-ray detection of groups is affected by a preferential selection for bright and peaked objects \citep[this is the notion of cool core bias,][]{Eckert2011A&A...526A..79E}, and only at relatively low redshift due to the shallow depth of X-ray surveys such as ROSAT \citep{Ponman1999Natur.397..135P} and eROSITA \citep{Bulbul2024A&A_erass1clu, Bahar2024_erositagrps}. The double selection in X-GAP is devised to obtain a complete and pure sample of galaxy groups, to be compared to hydrodynamical simulations.

In this context, understanding the X-GAP completeness level is key to assess whether any statement about AGN feedback resulting from its analysis, is valid for a fair subsample of the overall group population in the Universe. This concept is encoded in the selection function, i.e. the probability of detection as a function of a given set of properties. In astronomical surveys, it is in fact often fruitful to model the incompleteness and exploit larger amounts of data, instead of restricting to a very complete, high signal to noise sample, at the cost of losing many sources closer to the detection limit \citep{Rix2021AJ_gaiaSelFunc, Clerc2024A&A_eroSF}. The selection function is therefore a key component of scaling relation and cosmological analysis \citep[see e.g,][]{Pacaud2018A&A_xxl, Bahar2022A&A_efedsSR, Ghirardini2024_erass1cosmo, Artis2024_fR}, as it addresses the population of undetected objects in a statistical way. 

In this work we develop a framework to quantitatively model the X-GAP selection function by forward modelling the selection process using end-to-end simulations. Similar approaches were recently dedicated to model the eROSITA X-ray selection function \citep{Seppi2022A&A_erass1twin, Clerc2024A&A_eroSF, Marini2024A&A_erosita}. A multi-wavelength approach to assess the sample properties with simulations is key to obtain observational constraints that are representative of the underlying population \citep{Marini2025A&A_optical, Popesso2024arXiv_simsSTACKS}. We construct a full sky light cone starting from the Uchuu set of N-body simulations \citep{Ishiyama2021MNRAS.506.4210I_UCHUU}. We use dedicated SDSS mocks based on Uchuu \citep{DongPaez2024MNRAS_UchuuSDSS} to develop the optical side of our simulation. We populate our mock sky with X-rays using the AGN model from \citet{Comparat2019MNRAS_agn_model} and a new model for clusters and group informed by hydrodynamical simulations. We forward model the X-ray observations with the Simulation of X-ray Telescopes software \citep[SIXTE,][]{Dauser2019_SIXTE}. We reproduce the detection schemes in X-rays \citep{Damsted2024_axes} and optical bands \citep{Tempel2017A&A_sdssFOF}. We focus on the parent sample selection, as replicating the cuts in source size (angular size of R$_{\rm 500c}$ below 15 arcmin, to fit in the XMM-Newton field of view) and galaxy members (more or equal than 8 spectroscopic members) is then straightforward \citep{Eckert2024_xgap}. We evaluate and model the X-GAP selection function in terms of observables, without directly modelling the mass selection. It makes the selection function more flexible against specific modelling choices, which may impact the relation between mass and luminosity. Finally, we calibrate the scaling relation between galaxy members line of sight velocity dispersion and halo mass.

This paper is organized as follows. In Section \ref{sec:data_concept} we explain the strategy behind our end to end simulation. In Section \ref{sec:clu_model} we describe the neural network model used to assign X-ray profiles and temperature to dark matter haloes. In Section \ref{sec:catalogues} we elaborate on the treatment of each catalogue, for input haloes, X-ray, and optical detections. In Section \ref{sec:probability_detection} we describe sample properties such as purity and completeness, and model the selection function. In Section \ref{sec:veldisp_mass} we calibrate the scaling relation between velocity dispersion and halo mass. Finally, in Section \ref{sec:concl} we elaborate our findings in terms of the dynamical state of dark matter haloes and summarize our results in comparison to other works.
If not otherwise specified, we assume the cosmological parameters from \citet{Planck2020A&A}, the ones used for simulating the Uchuu Universe, i.e. $\Omega_{\rm M}$=0.3089, $\Omega_{\rm B}$=0.0486, $\sigma_{\rm 8}$=0.8159, $H_{\rm 0}$=67.74 km/s/Mpc. We use X-ray luminosities within R$_{\rm 500c}$ in the 0.5-2.0 keV band.

\section{Simulation strategy}
\label{sec:data_concept}

\begin{figure}
    \centering
    \includegraphics[width=\columnwidth]{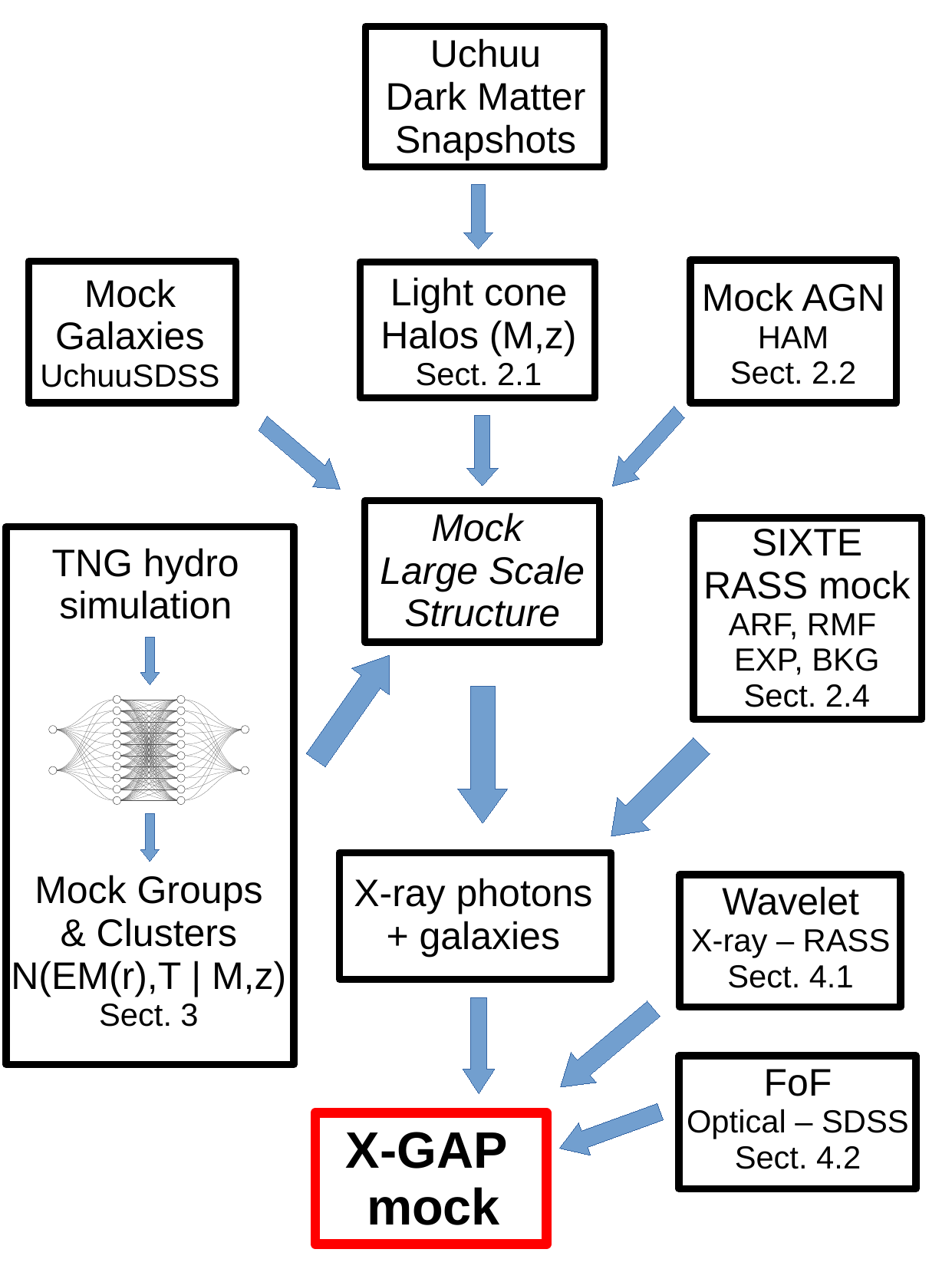}    
    \caption{Forward modelling of the X-GAP selection. We start from a dark matter halo light cone generated from the Uchuu simulation, we build a novel method to assign cluster and groups X-ray profiles and temperatures as a function of halo mass and redshift, we use abundance matching schemes to simulate galaxies and AGN. We generate X-ray events accounting for the telescope response. Finally, we reproduce the detection schemes in X-ray and optical bands to select an X-GAP-like sample from our simulation.}
    \label{fig:model_diagram}
\end{figure}

\begin{figure}
    \centering
    \includegraphics[width=\columnwidth]{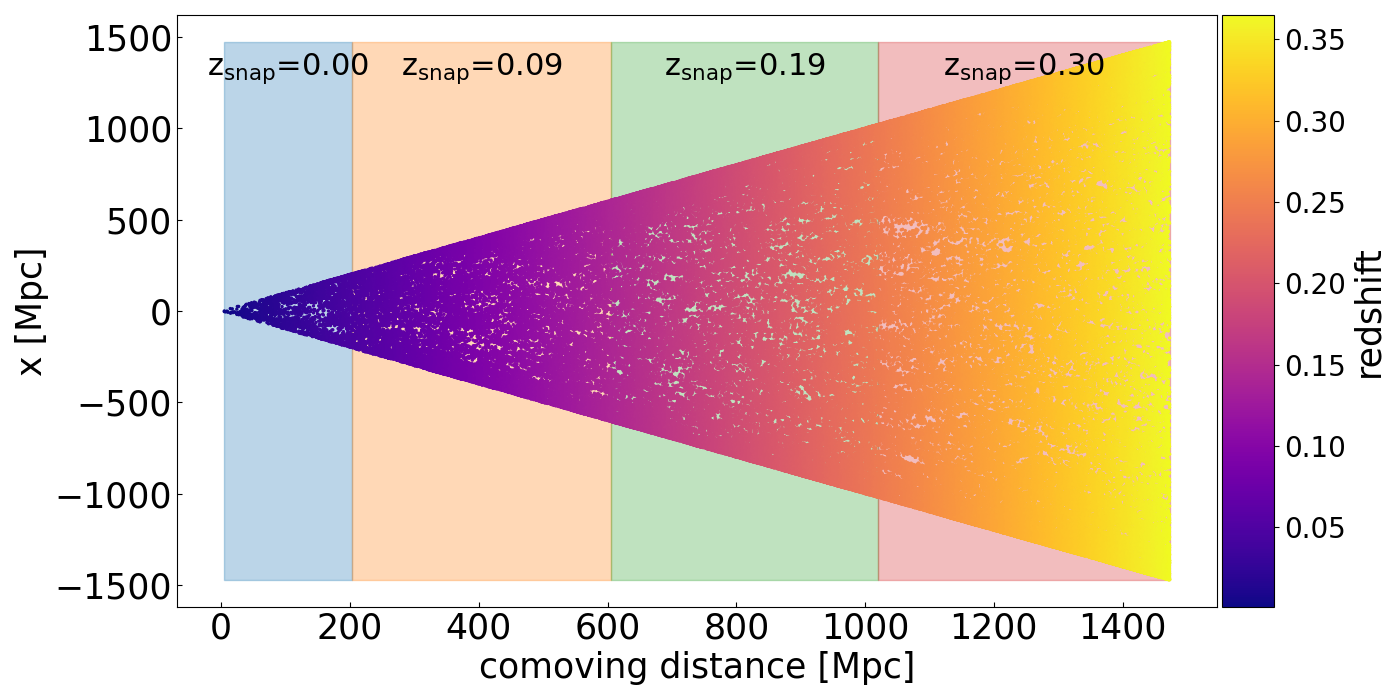}
    \caption{Dark matter halo light cone generated from individual Uchuu snapshots. This panel shows a slice within $\pm$50 Mpc along the z-axis, up to redshift of 0.4. Haloes more massive than 10$^{12.5}$ M$_\odot$ are shown, each one is colour coded by its redshift. The shaded areas denote the various snapshots used to generate the light cone.}
    \label{fig:lightcone}
\end{figure}

Our strategy is to forward model the galaxy groups selection with end-to-end simulations, as shown in Fig. \ref{fig:model_diagram}. We start from individual snapshots of the Uchuu simulations \citep{Ishiyama2021MNRAS.506.4210I_UCHUU} and build a dark matter halo light cone. We use the UchuuSDSS mock from \citet{DongPaez2024MNRAS_UchuuSDSS} to populate our haloes with galaxies. More details are reported in Sect. \ref{subsec:optical_sel}. \\
We develop a novel method to populate haloes with X-rays from galaxy clusters and groups (see Sect. \ref{sec:clu_model}) and implement the AGN model from \citet{Comparat2019MNRAS_agn_model}. We model the diffuse X-ray background following the real ROSAT background maps \citep{Snowden1997ApJ_rass_sxrb}. We generate X-ray events using the SIXTE software \citep{Dauser2019_SIXTE}. We detect sources in the X-ray and optical bands and cross match the output galaxy cluster and groups catalogue to the input dark matter haloes. The whole process is detailed in the next sections.

\subsection{Halo light cone}
\label{subsec:halo_lightcone}

Uchuu is a large dark matter only simulation with high resolution \citep{Ishiyama2021MNRAS.506.4210I_UCHUU}. It is based on a standard Flat $\Lambda$CDM cosmology with parameters from \citet{Planck2020A&A}. Individual snapshots are publicly available\footnote{\url{https://www.skiesanduniverses.org/Simulations/Uchuu}}. The box size is 2 Gpc/h, with a total of 2.1 trillion particles, for a mass resolution of 3.27$\times$ 10$^8$ M$_\odot$/h. The gravitational softening length is 0.4 kpc/h. This is ideal for our purpose to minimize cosmic variance effects due to the all-sky X-ray selection, but also to properly account for faint local galaxies in the SDSS optical selection. Haloes are identified with the \texttt{Rockstar-Consistenttrees} algorithm \citep{Behroozi2013}, based on a FoF approach in six dimensions for positions and velocities. The parallel processing of multiple snapshots provides a consistent halo catalogue as a function of redshift.
We concatenate individual snapshots into a halo light cone by adapting the methodology of \citet{Comparat2020_clumodel} to match the geometry of \citet{DongPaez2024MNRAS_UchuuSDSS}. We combine snapshots at redshift 0.0, 0.09, 0.19, 0.30, 0.43, 0.49, 0.56, 0.70, 0.78, 0.86, 0.94, 1.03, 1.12, 1.22, 1.32, 1.43, 1.54, and 1.65 to obtain a smooth redshift distribution of dark matter haloes. This allows us to include the bulk of the AGN population detected in the ROSAT All-Sky Survey (RASS), which peaks at redshift 0.3 and has a long tail extending just above redshift 1.5 \citep[see e.g.,][]{Anderson2007_rassAGN}. A rotation of 10 deg in the x-z cartesian plane allows us to match our coordinates to the geometry of the UchuuSDSS simulation, for the observer placed in the origin, providing the same conversion between cartesian to angular coordinates. We convert cartesian coordinates into equatorial coordinates according to:

\begin{align}
    &\text{d$_{\rm C}$} = \sqrt{x^2 + y^2 + z^2} \nonumber \\
    &\theta = \arccos{z/dc}  \nonumber \\
    &\phi = \arctan{y/x}  \nonumber \\
    &\text{DEC} = \theta - 90  \nonumber \\
    &\text{RA} = \phi - 360 \left\lfloor \frac{\phi}{360} \right\rfloor,
    \label{eq:cart_to_eq}
\end{align}

where the conversion from $\phi$ to RA is done in python with the \texttt{numpy.mod} function \citep{vanderWalt2011CSE_numpy}. We infer the cosmological redshift $z_{\rm cosmo}$ from the comoving distance $d_{\rm C}$ using \texttt{astropy} \citep{astropy2013A&A...558A..33A} and add the effect of peculiar velocities according to:

\begin{align}
    &v_{\rm pec} = \frac{x\times v_{\rm x} + y\times v_{\rm y} + z\times v_{\rm z}}{ d_{\rm C}} \nonumber \\
    &z_{\rm pec} = \sqrt{\frac{1 + v_{\rm pec}/c}{1 - v_{\rm pec}/c}} \nonumber \\
    &z_{\rm spec} = (1 + z_{\rm cosmo}) \times (1 + z_{\rm pec}) - 1,
    \label{eq:redshift}
\end{align}
where c is the speed of light and $z_{\rm spec}$ is the idealized spectroscopic redshift. Note that in Eq. \ref{eq:cart_to_eq} we place the observer in (0,0,0). The box is replicated 8 times around the origin by shifting all combinations of each cartesian coordinate by one box length. In the construction of the light cone we use the output of a given snapshot at redshift $z_{\rm snap}$ according to $(z_{\rm snap-1}+z_{\rm snap})/2 < z < (z_{\rm snap}+z_{\rm snap + 1})/2$. For z>0.78 the comoving distance to the edge of the snapshot becomes larger than the box length. Therefore, we replicate the snapshots one more time along each direction, obtaining 64 replicas around the observer at the origin for snapshots between z=0.78 and z=1.65. 

We then query subhalo members for each distinct halo and use the individual peculiar velocities to compute the true halo line of sight velocity dispersion, according to:
\begin{align}
    \sigma_{\rm v,T} =& \sqrt{\frac{1}{(N_{\rm mem}-1)}\displaystyle\sum_{\rm i=1}^{N_{\rm mem}} (v_{\rm pec,i} - \overline{v_{\rm pec}})^2 },
    \label{eq:vel_disp}
\end{align}
where $N_{\rm mem}$ is the number of subhaloes for each distinct halo. We restrict our measurement to subhaloes with virial mass larger than 9.3$\times$10$^{9}$ M$_\odot$. It allows resolving these structures with more than 20 particles, so that \texttt{Rockstar} provides secure (sub)halo properties \citep{Knebe2011MNRAShalos, Knebe2013MNRAS_halos2}. In addition, it allows a robust description of SDSS-like galaxies with stellar down to 10$^8$ M$_\sun$, the lower limit in the GALEX-SDSS-WISE Legacy Catalog \citep[GSWLC,][]{Salim2016_galex}. We verified the robustness of the measured true velocity dispersion by increasing the threshold to 50 and 100 particles, hence increasing the subhalo resolution level at the cost of reducing the number of subhaloes used in Eq. \ref{eq:vel_disp} \citep{Onions2012MNRAS_subhalo}, and did not find significant differences in the estimate of velocity dispersion. For computational power reasons, we compute velocity dispersions only for haloes in the first two redshift snapshots, meaning that we can access intrinsic halo velocity dispersion in our light cone up to z$\leq$0.14. Therefore, we miss it for haloes present in the optical mock for z>0.14, but we are not affected by this limitation thanks to the upper limit of X-GAP at z=0.06, allowing us to focus on lower z when modelling the selection function.

Various versions of \texttt{UchuuSDSS} are available, depending on the geometry and the conversion between cartesian and angular coordinates. By placing the observer in the origin of the boxes, we obtain a dark matter halo light cone with coordinates matching the \texttt{UchuuSDSS0} galaxy mock from \citet{DongPaez2024MNRAS_UchuuSDSS}. A slice of the full light cone is shown in Fig. \ref{fig:lightcone}, where the LSS expands from the origin as a function of the comoving distance to the observer. Each halo is colour coded by its redshift (see Eq. \ref{eq:redshift}) and the shaded areas denote the intervals corresponding to the output from different snapshots. In addition, we rotate and flip the coordinates to match the geometry of three additional light cones with the observer at the same place, allowing us to independently reproduce the SDSS sky footprint of 7261 deg$^2$ with four different regions, corresponding to the \texttt{UchuuSDSS} mocks from 0 to 3. In the rest of the article we will refer to the combination of the four light cones when stating our results. When needed, we assign hydrogen column density due to galactic absorption following \citet{HI4PI2016A&A...594A.116H}.

\subsection{Active galactic nuclei}
\label{subsec:AGN}

We populate dark matter haloes from the Uchuu simulation with AGN using the model from \citet{Comparat2019MNRAS_agn_model}. It is based on a halo abundance matching scheme (HAM) between stellar and halo mass. We select haloes with virial mass larger than $10^{11}$ M$_\odot$. The stellar masses are assigned to each halo based on the stellar to halo mass relation from \citet{Moster2013_smhm_rel}, which accounts for the intrinsic scatter in the relation. The model is calibrated to reproduce the AGN duty cycle and the hard X-ray luminosity function \citep[see][]{Georgakakis2017MNRAS_duty, Aird2015MNRAS_AGNLxfunc}. The choice of the 2-10 keV band is particularly useful because the hard band flux is less affected by the AGN spectra and obscuration properties compared to a softer X-ray band. The model produces an AGN population that matches measurements of the evolution of their number density with redshift and luminosity, and the clustering properties \citep{Georgakakis2019MNRAS_model}. Various works in the literature reported the identification of a soft excess feature in AGN spectra \citep{Boissay2016A&A_softexcess, Ricci2017ApJS_AGNswift, Waddell2024A&A_softexcess}, which we include in our modelling. We assume a typical spectrum composed by an absorbed power-law with photon index equal to 1.9, with the addition of a scatter component from cold matter with 2$\%$ normalisation \citep{Yaqoob1997ApJ_plcabs}, and a reflection one with emission lines \citep{Nandra2007MNRAS_pexmon}. We add a phenomenological 0.2 keV thermal bremsstrahlung component to model the soft excess \citep{Boissay2016A&A_softexcess}. We generate model spectra with \texttt{Xspec} \citep[version 12.13.1,][]{Arnaud1996_XSPEC}, our model reads \texttt{TBabs(ztbabs(powerlaw + constant$\times$bremss) + constant$\times$powerlaw + pexmon$\times$constant)}. We assign the spectra by a nearest neighbour search in redshift, intrinsic absorption, and galactic extinction.
We simulate AGN down to an X-ray flux of 10$^{-13.2}$ erg/s/cm$^2$, slightly below the flux limit of the ROSAT all sky survey \citep{Voges2000IAUC_RASSFSC}. This allows us to simulate the AGN population corresponding to the one detected in the real RASS, but without double counting the faint, undetected AGN contributing to the cosmic X-ray background (see next Section). Indeed, in the most recent processing of the ROSAT All-Sky Survey, i.e. the Second ROSAT all-sky survey (2RXS) source catalogue by \citet{Boller2016A&A_2RXS}, we find that the 1st percentile of the cumulative flux distribution calculated assuming a power-law spectrum appropriate for AGN corresponds to a flux of approximately 5.8$\times$10$^{-14}$ erg/s/cm$^2$. Given the uncertainties associated with flux measurements at these faint levels, which are sensitive to the assumed spectral model, we consider this threshold to be broadly consistent with our adopted flux cut of 10$^{-13.2}$ erg/s/cm$^2$.

\subsection{X-ray background}
\label{subsec:bkg}
We merge the individual band 4-7 RASS background maps \citep{Snowden1997ApJ_rass_sxrb}\footnote{\url{https://www.jb.man.ac.uk/research/cosmos/rosat/}} into a single map in the 0.44-2.05 keV band, summing the count rates from each individual band. This strategy accounts for spatial variations of the X-ray background, including the emission of the eROSITA bubble \citep{Predehl2020_bubble} in the sky covered by SDSS, which was already detected in the RASS as the North Polar spur \citep{Egger1995A&A_NPS}. Its brightest part is in the galactic plane \citep{Willingale2003MNRAS.343..995W}. We model the background assuming three main components: unabsorbed plasma emission from the local hot bubble (LHB), absorbed plasma emission from the galactic halo (GH), and the cosmic X-ray background (CXRB) from undetected point sources. In \texttt{Xspec} terms our model reads \texttt{APEC + Tbabs $\times$ (APEC + powerlaw)}. We assume solar abundance ($Z_\odot$) and $kT=0.097$ keV for the LHB, $0.3 Z_\odot$ and $kT=0.22$ keV for the GH, and a photon index $\Gamma=1.46$
 for the CXRB. We stress that the faint AGN population is not simulated as individual sources (see previous Section), because its emission is already present in the background maps. The proportional counter PSPC of ROSAT has low sensitivity to high energy particles and soft protons, therefore we neglect the particle induced background component in our spectral model. Indeed \citet{Eckert2012A&A_outergas} demonstrated that the PSPC instrumental background is more than one order of magnitude lower than the sky background.
 We process the merged map into \texttt{HEALPix}\footnote{\url{https://healpy.readthedocs.io/en/latest/}} format with NSIDE=64, generating 49152 areas of about 0.84 deg$^2$. We convert the count rate to flux using the spectral model defined above folded with the ROSAT PSPC response file and integrate it in the individual areas to obtain the total flux from each pixel in the energy band of interest.

\subsection{Event generation}

To generate X-ray photons we use the SIXTE software \citep{Dauser2019_SIXTE}. It is an end-to-end X-ray simulator, which allows forward modelling observations accounting for vignetting, energy-dependent PSF, ancillary response file (ARF), and redistribution matrix file (RMF). We build an ad-hoc SIXTE module to simulate RASS data.
We use publicly available response files from ROSAT\footnote{\url{https://heasarc.gsfc.nasa.gov/FTP/caldb/data/rosat/pspc/cpf/matrices}}. We build an analytical PSF model following the work from \citet{Boese2000A&A_rosatPSF} (see Eq. 6, 7 therein). The ROSAT PSPC has a sensitive area of 8 cm diameter \citep{Pfeffermann1986SPIE..733..519P}.
The total area is $A = \pi \times (d/2)^2$, corresponding to a mock pixel with a size of $s = \sqrt{A} = 70.898\,mm$ on a side. 
During the readout, we ignore any correction between pulse height amplitude and channels, as the time resolution of proportional counters is very good. We account for a dead time interval after each event of $180\,\mu s$, when the detector is not sensitive to radiation. 
We collect information about the telescope pointing from the publicly available ancillary ROSAT data \footnote{\url{https://heasarc.gsfc.nasa.gov/FTP/rosat/data/pspc/processed_data/rass/release}}. We concatenate each file to generate a single RASS attitude file describing the coordinate pointing and the roll angle of the telescope as a function of time. We ignore time spans with operational problems \citep[see Table 2 in][]{Voges1999A&A_RASS_BSC}.
For processing purposes, we divide the area covered by the X-GAP sample into 627 pixels of about 53 square degrees using \texttt{healpix} with NSIDE=8. For clusters and groups, we generate idealized input images following the ellipticity of each dark matter halo, which SIXTE used to simulate events from extended sources on the plane of the sky.

\section{Clusters and groups model}
\label{sec:clu_model}
Since the Uchuu simulations contain only dark matter, we need a prescription to populate the haloes with baryons. Various implementations exist in the literature, such as semi-analytical models \citep{Shaw2010_analytical, Osato2023MNRAS_baryonpaste}, or phenomenological approaches based on real observations \citep{Zandanel2018MNRAS.480..987Z}, also accounting for covariances between different observables \citep{Comparat2020_clumodel}.
We proceed in this direction and develop a new model to predict the X-ray emissivity profile and temperature as a function of halo mass and redshift using a machine learning approach. Our method builds on \citet{Comparat2020_clumodel} and aims to predict cluster and group properties with the correct covariances. The \citet{Comparat2020_clumodel} model is based on high signal to noise observation of massive galaxy clusters and required ad-hoc corrections in the galaxy group regime \citep[see discussion in][]{Seppi2022A&A_erass1twin}. Although hydrodynamical simulations are not a perfect reconstruction of the real Universe due to various assumptions, e.g in the feedback and star formation prescriptions, they still provide a complete view of the galaxy group population under those assumptions. On the one hand, our new model is more reliable at low masses <10$^{14}$ M$_\odot$, because it is informed by hydrodynamical simulations, which do not lack objects in this mass range. On the other hand, hydrodynamical simulations can not consistently predict hot gas properties for galaxy groups \citep{Eckert2021_review}. However, our definition of the selection function in terms of observables makes our modelling less dependent on specific models assumed in the implementation of baryonic physics (see also Appendix \ref{appendix:rescale_test}). Similarly, various prescriptions for the hot X-ray gas may change the total number of sources as a function of X-ray observables at fixed optical properties in our framework. However, the selection function is a ratio, and it is not dependent on a given X-ray model if it also accounts for optical properties in its definition (see Sect. \ref{sec:probability_detection}). We introduce the concepts, the model, and show the results about inferred observables in this section.

\subsection{Emission measure profiles}

We use the emission measure integrated along the line of sight to model the profiles. This quantity is also known as emission integral and should not be confused with the volume integrated emission measure \citep{Eckert2016A&A_XXL}. When mentioning the emission measure, we refer to the one integrated along the line of sight. At radius $x=r/r_{\rm 500c}$ it can generally be deduced from the X-ray surface brightness \citep{Neumann1999A&Arosat_cluprofiles, Arnaud2002A&A_profiles}.  It is defined as follows:

\begin{align}
    EM(x) &= \int n_e n_{H} dl \nonumber \\
    EM(x) & = \frac{4\pi (1+z)^4 S(x)}{\epsilon(T,z)} \nonumber \\
    \epsilon(T,z) & = \int_{\rm E1}^{\rm E2} S(E)e^{\rm -\sigma(E)N_{\rm H}} f_{\rm T}((1+z)E)(1+z)^2 dE,
    \label{eq:EM_profiledef}
\end{align}
where S(E) is the detector effective area at energy E, $\sigma(E)$ is the absorption cross section, $f_{\rm T}((1+z)E)$ is the emissivity in cts/s/keV$\times$cm$^3$ for a plasma of temperature T.

In practice, various works in the literature use a conversion factor between count rate and APEC normalisation with \texttt{xspec} to obtain the emission measure profile from surface brightness. It allows accounting for the response of the instrument \citep{Pratt2009A&A_rexcess, Eckert2012A&A_outergas, Bartalucci2023A&A_chexmate_SB}. More details are given in Appendix \ref{appendix:em_prof}. %We then convert the apec normalisation to emissivity in units of Mpc/cm$^6$. 
The self similar scaled emissivity profile is finally obtained as follows, see e.g. \citet{Arnaud2002A&A_profiles, Eckert2012A&A_outergas}:
\begin{equation}
    EM_{\rm SS}(x) = EM(x) \Big[\frac{kT}{10 keV}\Big]^{-1/2} E(z)^{-3}.
    \label{eq:EM_selfsimilar}
\end{equation}

\subsection{Profile extraction from TNG}

We train a neural network on galaxy clusters from the hydrodynamical TNG300 simulation\footnote{\url{https://www.tng-project.org}} \citep{Nelson2019ComAC_TNGdatarelease}. The box size is 205 Mpc/h and the dark matter particle mass is 5.9$\times$10$^7$ M$_\odot$/h, therefore galaxy groups and clusters are simulated with extremely high resolution. We use groups and clusters with M$_{\rm 500c}$>8$\times$10$^{12}$ M$_\odot$ at snapshots corresponding to z = 0.01, 0.03, 0.06, 0.1, 0.2, 0.3, 0.5, 1.0, 1.5. In the snapshot at z=0.03, which is within the X-GAP window, we model 4101 objects.

Following a methodology similar to \citet{Shreeram2025A&A_CGM_forwmod}, we retrieve the data stored for each gas cell and model its emission measure using \texttt{pyxsim} \citep{Zuhone2015ApJ_visco_pyxsim}, which provides a Python interface to the \texttt{PHOX} code \citep{Biffi2012MNRAS_phoxintro, Biffi2013MNRAS_phox}. We consider gas cells with temperature between 0.1 and 20 keV, with gas density below 5$\times$10$^{-25}$ g/cm$^3$. We project the 3D model of each cell along the x cartesian direction and integrate them within circular apertures to retrieve the emission measure profile integrated along the line of sight (as defined in Eq. \ref{eq:EM_profiledef}).

Next, we need an estimate for the source temperature in the simulation. Although computing the local temperature of a gas cell or particle in hydrodynamical simulations is possible using the internal energy, translating a large amount of individual temperatures into a global halo temperature is not trivial, as it often requires the assumptions of weights for different gas properties, which may result in disagreements between outputs from hydro simulations and observations \citep{Mazzotta2004MNRAS_SL, Rasia2005ApJ_SL, Rasia2014_Txstruct}. In observations, the temperature measurements comes rather from fitting the source spectrum with a single global temperature. Therefore, to be as close as possible to an ideal X-ray temperature from the observations' perspective, we generate an ideal spectrum starting from the emission measure profile. We determine the photon X-ray emissivity using an \texttt{APEC} model \citep{Smith2001ApJapec} as a function of density and temperature of each gas cell, assuming a metallicity of Z=0.3 Z$_\odot$ and the abundance table from \citet{Asplund2009ARA&Aabund}, using the emissivity of each cell as a distribution function for photon energy, assuming a large collective area of 10\,000 cm$^2$ and a long exposure time of 500 ks. We infer the ideal X-ray halo temperature by fitting the resulting global spectrum with an APEC model in \texttt{Xspec}, fixing the redshift to the redshift of the TNG snapshot and metallicity to 0.3, while leaving temperature and normalisation free to vary. The result is a collection of clusters and groups with mass, redshift, emission measure profile, and temperature. The assumption of Z=0.3 Z$_\odot$ is reasonable for the average metallicity in galaxy groups, especially outside the core \citep{Sun2009ApJ_grpsChandra, Mernier2017A&A_metalprof, Bahar2024_erositagrps}. For detailed comparisons between the prediction of hydrodynamical simulations and the real X-GAP, accounting for the metallicity distribution along the group profile may also help in shedding light on AGN feedback prescriptions and we will consider it in future work.

\subsection{Profile generation with machine learning}

The field of machine learning and neural networks has enormously grown in recent years, with a multitude of implementations for various algorithms, such as variational auto encoders \citep{Kingma2013arXiv_VAE}, and generative adversarial networks \citep{Goodfellow2014_GAN}. Different works in the literature have successfully applied machine learning techniques in astronomy, for example to study galaxy cluster morphologies \citep{Benyas2024ApJ_morph}, predict galaxy cluster masses \citep{Ntampaka2019ApJ_clumass, Krippendorf2024A&A_clumasses}, infer the gravitational potential of the Milky Way \citep{Green_2023_MWpotential}, and study dark matter simulations \citep{Rose_2025ApJ_DREAMS}.

In this context, normalising flows have emerged as a popular way to model and generate data \citep{Hahn2022ApJ_SEDFlow, Li2023arXiv_PopSED, Crenshaw2024AJ_NF_galaxies}.
normalising flows constitute a machine learning approach to construct a complex probability distribution from a set of transformations of simple distributions \citep[see][for a review]{Kobyzev2019_NFreview}. Given a set of observed variables $\overline{x}$ the goal is to model its probability distribution $p_{\rm x}(\overline{x})$, by starting from a continuous random variable $\overline{y}$ that follows a simple probability distribution $p_{\rm y}(\overline{y})$, e.g. a Gaussian one. The idea is to transform the simple $p_{\rm y}(\overline{y})$ into a more complex one through a collection of $N$ invertible and differentiable functions (bijectors) $f_{\rm T}(\overline{y}) = f_{\rm N} \circ ... f_{\rm k} \circ ... f_{\rm 1}(\overline{y})$. The transformed probability distribution is:
\begin{align}
    p_{\rm x}(\overline{x}) &= p_{\rm y}(\overline{y}) \Big | \det \Big ( \frac{\partial f_{\rm T}^{-1}(\overline{x})}{\partial \overline{x}} \Big ) \Big | \nonumber \\
    & = p_{\rm y}(\overline{y}) \prod_{\rm i=0}^{\rm N} \Big | \det \Big ( \frac{\partial f_{\rm i}^{-1}(\overline{x})}{\partial \overline{z_{\rm i}}} \Big ) \Big |,
    \label{eq:probdistr_bijectors}
\end{align}
where $\overline{y} = f_{\rm T}^{-1}(\overline{x})$ and $\det \Big ( \dfrac{\partial f_{\rm i}^{-1}(\overline{x})}{\partial \overline{y_{\rm i}}} \Big )$ is the Jacobian of each transformation $f_{\rm i}$. The variable $y$ flows through the bijectors chain, while the Jacobians normalise the probability distributions of the transformed variable. The training process of a normalising flow consists of optimizing the parameters of the bijectors in the chain to best reproduce the probability distribution of the observable $p_{\rm x}(\overline{x})$. During training we minimize the negative log likelihood defined from Eq. \ref{eq:probdistr_bijectors} as follows:
\begin{equation}
    \log \mathcal{L} = \sum_{\rm i=0}^{\rm N} \log p_{\rm y}[f^{-1}_{\rm i}(\overline{x})] + \sum_{\rm i=0}^{\rm N}\log | \det Df_{\rm i}(\overline{x}) |,
\end{equation}
because for bijectors the determinant of the product of the jacobian matrices of each transformation is the product of the individual determinants.

Our model is based on a multivariate normal distribution with the same dimension as our training dataset. Then we create an autoregressive normalising flow \citep[MAF,][]{Papamakarios2019MAF} using autoregressive models for density estimation \citep[MADE,][]{Germain2015MADE}. They describe the final probability distribution as the product of conditional probabilities:
\begin{equation}
    p_{\rm x}(\overline{x}) = \prod_{\rm i=0}^{D} p_{\rm x}(x_{\rm i} | \overline{x_{\rm <i}}),
\end{equation}
where $D$ is the number of dimensions of a given dataset. 
In particular, each component of the observable $\overline{x}$ is predicted by transforming the latent variable $\overline{y}$ with a shift $\mu_{\rm i}$ and a rescale factor $\sigma_{\rm i}$ that depend on the previous components:
\begin{equation}
    x_i = \mu_i(x_{\rm <i}) + y_{\rm i}\times \exp{\log \sigma_{\rm i}(x_{\rm <i})}.
    \label{eq:shift_scale}
\end{equation}
This setup makes the calculation of the Jacobian in Eq. \ref{eq:probdistr_bijectors} simple, which reduces for each component to 
\begin{equation}
    \det Df_{\rm i}(\overline{x}) = \sigma_{\rm i}(x_{\rm <i}).
    \label{eq:jacobian_MAF}
\end{equation}
Finally, we minimize the negative of the log likelihood in Eq. \ref{eq:probdistr_bijectors} by the gradient descent method using the \texttt{Adam} optimizer with a learning rate of $10^{\rm -3}$.
One of the caveats in the field of computer vision is that the MAF architecture is slow in generating new data, because the generation of each variable needs all the previous inputs in the flow. Given the size of the dataset we need to generate we are not affected by this limitation.

In this study, we develop a network to predict the self similar emission measure profile (see Eq. \ref{eq:EM_selfsimilar}) and X-ray temperature, conditionally on mass and redshift. This allows us to generate X-ray quantities (i.e. profiles and temperatures) using prior information of mass and redshift from our Uchuu halo light cone. Together with the profiles extracted from TNG, in the training set we add the HIFLUGCS sample \citep{Reiprich2002ApJ_hiflugcs}, and Chandra observations of SPT-selected clusters from \citet{Sanders2018MNRAS_chandraSPT}. They make the prediction of our model more robust at high mass, where only a few haloes are available in TNG. If the X-ray data is not sufficient to reach the halo outskirts, we follow the approach of \citet{Comparat2020_clumodel} and extrapolate the profile using a power law whose slope is fitted on the three outermost bins. The input dimension of our network is equal to 2 (M,z), while the output dimension is equal to 21, i.e. 20 radial bins logarithmically spaced between 0.02 and 3 R$_{\rm 500c}$ plus one value for temperature. In contrast to temperature, the values of mass and surface brightness span many orders of magnitudes. Therefore, we model their log-value and normalise each array in our training data, so that the network only needs to work with numbers between 0 and 1. We then apply the inverse transformations to convert the output values of the network to predictions in physical units. We apply a median smoothing to the radial profiles. This technique replaces points that exhibit strong fluctuations relative to their neighbors with the median of the two preceding and two following points along the profile, effectively preserving the overall shape and slope of the profile.
We build model with \texttt{TensorFlow} \citep{tensorflow2015-whitepaper}, it is composed of three MAF. Each autoregressive network has two output parameters ($\mu_{\rm i}, \sigma_{\rm i}$), two layers of 128 units each, and a sigmoid activation function. We train the model for 200 epochs, with a batch size of 16. We find that this set-up reproduces well the distributions of observed clusters and group properties compared to existing observations, as shown in the next section. This model does not introduce a correlation between the X-ray morphology (cool core or non cool core) and the dynamical state of the dark matter haloes \citep[see e.g.,][]{Seppi2021A&A_MF}. The X-ray detection scheme is designed to be insensitive to the relaxation state of the hot gas and we verified that this is the case in Sect. \ref{subsec:SBshape}.

\subsection{Results}
\label{subsec:clumodel_results}

\begin{figure}
    \centering
    \includegraphics[width=\columnwidth]{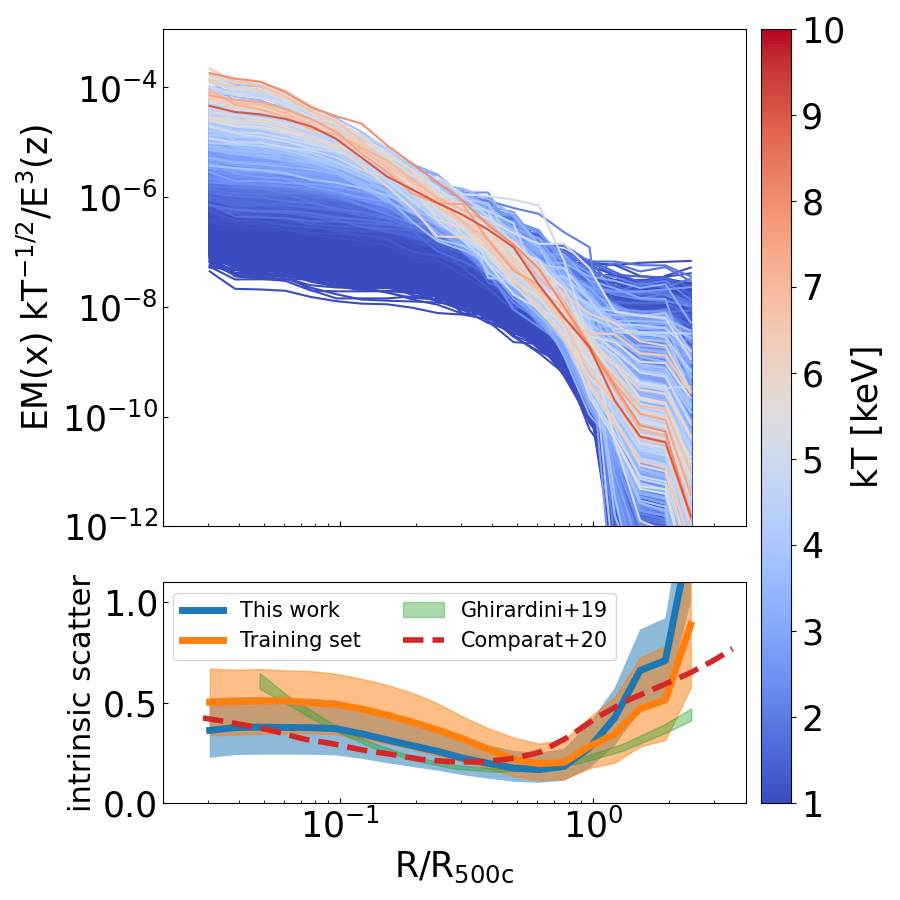}
    \caption{Emission measure profiles with self similar scaling, colour coded for temperature, as a function of radius in units of R$_{\rm 500c}$. The units on the y-axis are Mpc keV$^{-1/2}$ cm$^{-6}$. The bottom part of the panel shows the evolution of the profile intrinsic scatter at different radii. For reference it is compared to the model from \citet{Comparat2020_clumodel} and profiles of X-COP clusters \citep{Ghirardini2019A&A_xcop}.}
    \label{fig:EM_profiles}
\end{figure}

Starting from the halo masses and redshifts in the Uchuu light cone, we use the trained neural network to generate profiles and temperatures for each dark matter halo. The result is shown in Fig. \ref{fig:EM_profiles}. It displays the self similar scaled emission measured profiles, colour coded by temperature. The scaling of the profiles with temperature is clear and as expected: hot clusters exhibit high surface brightness with steep profiles, while cooler systems show lower normalization and flatter slopes. Specifically, we observe a correlation between emission measure and temperature in the central region within 0.5$\times$R$_{\rm 500c}$, and a more uniform distribution of the profiles towards the outskirts \citep{Eckert2012A&A_outergas}. Our result is in agreement with \citet{Comparat2020_clumodel}, which was built to reproduce observations of massive clusters by construction, and the profiles of X-COP clusters \citep{Ghirardini2019A&A_xcop}. The bottom part of the panel shows the intrinsic scatter of the profiles at different radii. It is defined as the 16th-84th percentile distribution of the profiles around their median. It is the only scatter component, as the ideal model has no noise for individual systems. It is compared to the model from \citet{Comparat2020_clumodel}. The intrinsic scatter is computed on the overall population, so the halo selection is different between the lines shown in the bottom panel of Fig. \ref{fig:EM_profiles}. Nonetheless, we find overall a good agreement between the prediction of our neural network to previous models and observations.
Our model predicts a slightly smaller intrinsic scatter compared to the training set in inner region, with a value of 0.39 against 0.48 at 0.1$\times$R$_{\rm 500c}$. The opposite holds in the outskirts, with values of 0.7 and 0.55 at 2$\times$R$_{\rm 500c}$. However, the model prediction and the training set are always compatible within 1$\sigma$ throughout the whole profile.

We compute the value of X-ray luminosity with a cylindrical integral of the emissivity profile as follows:

\begin{equation}
    L_{\rm 500c} = \int_0^{\rm R_{\rm 500c}} 2\pi  EM(r) r dr\ \Lambda(kT),
    \label{eq:Lx_calc}
\end{equation}
where $r$ is the radius and $\Lambda(kT)$ is the temperature dependent cooling function in the 0.5-2.0 keV band in units of cm$^3$erg s$^{-1}$ \citep{Sutherland1993_coolfunc}. For convenience, we tabulate the cooling function on a fine grid of temperature (every 0.075 keV) and interpolate. We apply a 2D K-correction to convert intrinsic luminosities to the observer frame. We tabulate conversion factors as a function of cluster temperature and redshift. Finally, we additionally account for galactic absorption. We compute the galactic gas column density ($N_H$) at the angular position of each cluster using the maps from \citet{HI4PI2016A&A...594A.116H}. Similarly to the cooling function case in Eq. \ref{eq:Lx_calc}, we tabulate the conversion factors and then interpolate at the exact values of luminosity, redshift, and $N_H$ of each cluster.

\begin{figure}
    \centering
    \includegraphics[width=\columnwidth]{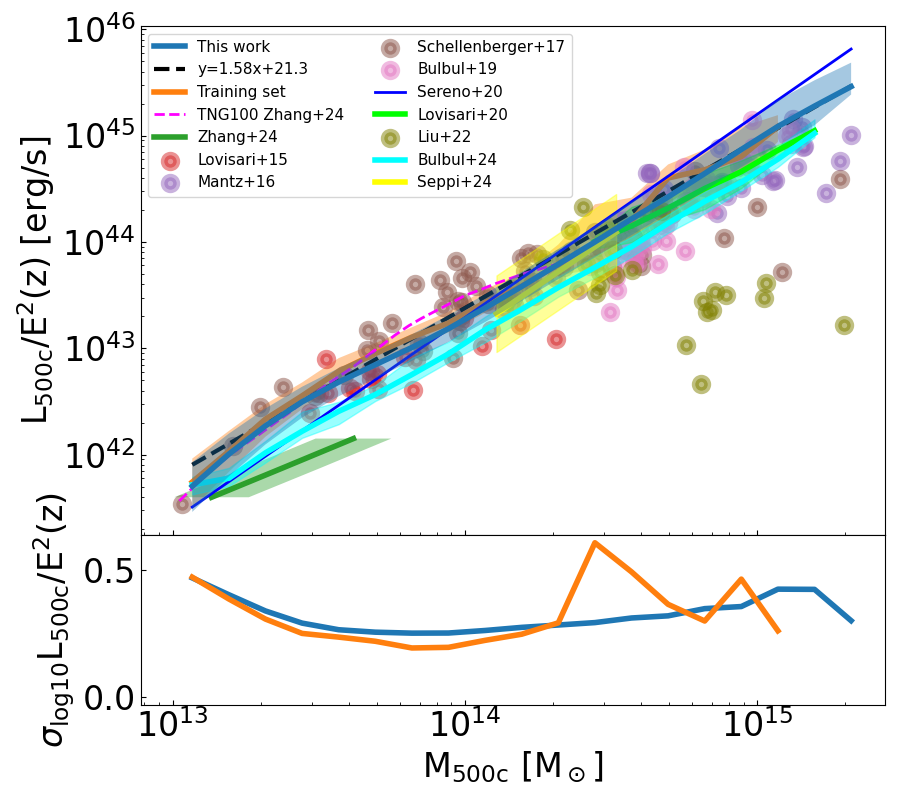}
    \includegraphics[width=\columnwidth]{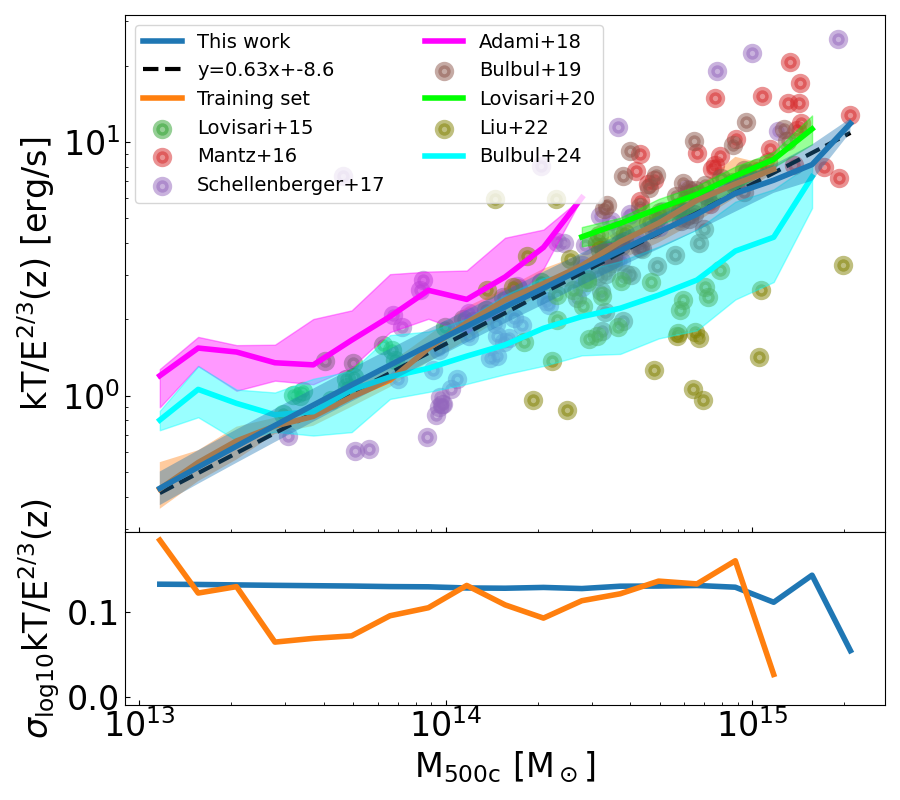}
    \caption{X-ray scaling relation prediction of the cluster and group model. \textbf{Top panel}: Scaling relation between X-ray luminosity obtained by integrating the emission measure profiles (see Eq. \ref{eq:Lx_calc}) and halo mass. \textbf{Bottom panel}: Scaling relation between temperature and halo mass. The blue shaded area shows the 16th-84th percentile distribution for the model applied to the Uchuu light cone in each mass bin. Our model is compared to a collection of real groups and cluster samples. The bottom part of both panels shows the intrinsic scatter evolution as a function of halo mass.}
    \label{fig:Clumodel_SR_validation}
\end{figure}

To assess the robustness of the model, we compare the scaling relations between observables and halo mass to literature results. The results are shown in Fig. \ref{fig:Clumodel_SR_validation}, displaying the X-ray luminosity to mass and temperature to mass relations. Overall, we find good agreement between our result and observations. The model predicted by the neural network applied to the Uchuu light cone is in blue, and it is compared to a collection of real clusters and groups, including a RASS selected groups sample \citep{Lovisari2015A&A_grps}, some ROSAT selected sample of massive clusters \citep{Mantz2016_scalrel, Schellenberger2017MNRAS_highflugs}, the XMM-XXL survey \citep{Adami2018A&A_XXL}, some Sunyaev-Zeldovich (SZ) selected samples in the radio/millimeter bands from SPT \citep{Bulbul2019ApJ_SPT_SR} and Planck \citep{Lovisari2020ApJ_scalrel}, and the eROSITA catalogues from eFEDS \citep{Liu2022A&A_angEFEDS} and eRASS1 \citep{Bulbul2024A&A_erass1clu, Seppi2024A&A_clustering}. The bottom part of both panels shows the intrinsic scatter evolution as a function of halo mass. For the $L_{\rm X}-M_{\rm 500c}$ relation we find a slope of 1.58$\pm$0.02. It is steeper than the self-similar model expectation of 4/3, because the gas fraction in the galaxy group regime is smaller than the cosmic one, which reduces luminosity at fixed mass. Our model aligns well with previous observations. The luminosity at fixed mass is slightly higher than the eRASS1 sample, likely due to software and calibration differences \citep[see discussion in][]{Bulbul2024A&A_erass1clu}. In any case, an accurate and precise comparison is not possible since for some of these observational samples a full scaling relation model including systematics and selection function is not available. In addition the masses have been computed using different techniques, such as weak lensing calibration, hydrostatic equilibrium assumption, or scaling relations, which means that the observational samples have different accuracy and precision along the x-axis in Fig. \ref{fig:Clumodel_SR_validation}. The scatter in luminosity predicted by the neural network evolves for different masses, from about 0.4 dex at about 2$\times$10$^{\rm 13}$ M$_\odot$ to 0.3 dex at 8$\times 10^{\rm 14}$ M$_\odot$, with a median value of 0.31 dex. This is in agreement with expectations when comparing these values with observations, given the fact that our model does not include measurement uncertainties that affect the latter. Literature values span between 0.2 and 0.4 \citep{Lovisari2015A&A_grps, Bulbul2019ApJ_SPT_SR, Sereno2019A&A_xxl, Seppi2024A&A_clustering}. The relation is flatter in the mass range between 1$\times$10$^{13}$ and 5$\times$10$^{13}$ M$_\odot$. This is intrinsic to the TNG simulation, as indicated by the dashed magenta line, which reports the TNG100 prediction as shown in \citet{Zhang2024A&A_LxM}. This effect could be due to line cooling. At fixed density, the X-ray emissivity is enhanced for temperatures around 1 keV due to the addition of line cooling on top of the self-similar expectation of thermal bremsstrahlung. Indeed \citet{Lovisari2021Univ_review} showed that the emissivity quickly increases by up to a factor of about two in the temperature range of 1 keV, this holds for different metallicities and energy bands. Disentangling such an effect from the reduced gas fraction in groups compared to clusters and the variation of the gaseous atmosphere due to AGN feedback is not trivial. In any case, we verify that this is not a limitation for our purpose of formulating the selection function in terms of observables. We reduce the luminosities extracted from the cluster model by a constant factor of 10$\%$, i.e. the fluxes are rescaled by a factor of 0.9. We re-generate the SIMPUT cluster files for the mock number three and process it with our end-to-end approach. We find that the final detection probability as a function of flux is in excellent agreement with the main work combining the four light cones. In addition, the X-GAP flux limit is about 5$\times$10$^{-13}$ erg/s/cm$^2$, where we find an excellent agreement between different mocks and the test with rescaled luminosities. We further elaborate in Appendix \ref{appendix:rescale_test}.

For the $T_{\rm X}-M_{\rm 500c}$ relation we find a slope of 0.63$\pm$0.02, close to the self similar model expectation of 2/3. The scatter is fairly constant as a function of mass, with values around 0.13. This is larger compared to the model from \citet{Comparat2020_clumodel}, reporting values around 0.07. Results from observational studies span from 0.05 \citep{Lovisari2020ApJ_scalrel}, 0.064 \citep{Sereno2019A&A_xxl}, 0.069 \citep{Chiu2022A&A_efedsSR}, to 0.18 \citep{Bulbul2019ApJ_SPT_SR} . 

\section{Catalogue creation}
\label{sec:catalogues}

We follow the selection scheme devised by \citet{Damsted2024_axes} for AXES, the successor of CODEX \citep{Finoguenov2020A&A_codex}, and the parent sample of X-GAP \citep{Eckert2024_xgap}. The optical FoF detection follows \citet{Tempel2017A&A_sdssFOF}. On top of the positional matching between X-ray and optical detections, we add a matching to the input clusters and groups, which allows us to quantify completeness and purity levels in the AXES catalogues.
The whole procedure is detailed in this section.

\subsection{X-ray detection}

We run a wavelet source detection algorithm, as introduced by \citet{Vikhlinin1998ApJ_wvdet}. It consists in convolving the image with a kernel with a positive core and a negative outer ring, allowing the isolation of objects with a given angular size. The kernel is a Mexican hat function. This allows an accurate background subtraction, because the convolution of the kernel with any local linear function is zero \citep[see][for details]{Vikhlinin1998ApJ_wvdet}. The wavelet scale $n$ is defined such that the outer scale corresponds to a Gaussian with size of $2^{n-1}$ pixels. We follow the implementation described in \citet{Kaefer2019A&A...628A..43K}. We divide the detection of point-like emission using wavelet scales of 2, 3, and 4 pixels; from the detection of extended-like emission using wavelet scales of 5 and 6 pixels. A key step at this point is using images with the same pixel size as the real RASS maps, because this is the scale at the base of the wavelet filters. Given the pixel size of 45 arcsec, these scales correspond to 1.5, 3, and 6 arcmin for point-like emission, and 12, 24 arcmin for the extended-like case.
Given the redshift range of X-GAP, using such large angular scales allows the creation of a source catalogue that is only sensitive to the baryonic content of groups in the outskirts. For a typical 5$\times$10$^{13}$ M$_\odot$ group at z=0.04, $R_{\rm 500c}$ covers about 12 arcmin. This minimizes the impact of AGN feedback, mostly evident in the central region, on the sample selection. The centre of X-ray emission is located by running \texttt{SExtractor} \citep{Bertin1996_sextractor} on the sum of the extended source wavelet images. 

\subsection{Optical detection}
\label{subsec:optical_sel}

Our optical mock is based on the \texttt{UchuuSDSS} simulation \citep{DongPaez2024MNRAS_UchuuSDSS}. The UchuuSDSS catalogues are publicly available\footnote{\url{https://skun.iaa.csic.es/SUsimulations/UchuuDR2/Uchuu_SDSS/uchuusdss_lightcones}}. They are based on a subhalo abundance matching (SHAM) between the maximum value of the halo circular velocity, serving as a stellar mass proxy, and a target luminosity to match the SDSS galaxy luminosity function. They add a smoothing to reproduce the observed redshift trend of the luminosity function and assign colours based on empirical models. Apparent r-band magnitude (mag-r) are computed accounting for colour-dependent k-corrections. Finally, they assign additional galaxy properties, such as stellar mass and star formation rate, by a nearest neighbour search within SDSS. The authors demonstrated that this approach recovers SDSS population properties by construction, such as the stellar mass function, galaxy clustering, redshift distribution, and colour magnitude diagrams \citep[see][for more details]{DongPaez2024MNRAS_UchuuSDSS}. \\
The optical FoF algorithm from \citet{Tempel2017A&A_sdssFOF}, used in \citet{Damsted2024_axes} and in this work, only requires angular coordinates, redshift, and mag-r. We treat their catalogue as the mock corresponding to the real galaxy SDSS data used in FoF search of galaxy groups. We select galaxies with redshift $z<0.2$ and r-band magnitude $\text{r-mag}<17.77$ to reproduce the same selection as \citet{Tempel2017A&A_sdssFOF}. The authors used a redshift dependent linking length according to:
\begin{equation}
    d_{\rm LL}(z) = d_{\rm LL,0}[1 + \arctan(z/z_\ast)],
\end{equation}
with $d_{\rm LL,0} = 0.34$ Mpc and $z_\ast = 0.09$.
The original SDSS catalogue contains $584\,449$ galaxies and $88\,662$ groups with at least two members. 
In the four simulations we used there are $624\,639$, $610\,845$, $595\,183$, and $626\,333$ galaxies. Differences between the mocks are attributed to cosmic variance.
These numbers are higher compared to the real Universe observed in SDSS. This is due to an overestimation of the galaxy density in the optical mock compared to the real SDSS at redshift close to 0.2, as pointed out in \citet{DongPaez2024MNRAS_UchuuSDSS}. This is not a limitation in our redshift range of interest $z<0.14$, where we also have access to velocity dispersions (see Sect. \ref{subsec:halo_lightcone}). In fact, in this range the mocks contain $484\,960$, $473\,439$, $456\,510$, $492\,000$ galaxies compared to $464\,978$ in the real SDSS. We discard groups with less than five galaxy members, as their properties are hard to measure quantitatively \citep[replicating the selection of][]{Damsted2024_axes}. In addition to the catalogue of FoF groups, the algorithm from \citet{Tempel2017A&A_sdssFOF} provides the catalogue of member galaxies assigned to each FoF group. We use the latter for matching the FoF groups to input dark matter haloes, as explained in the next section.

We further clean our optical catalogue following the same approach as \citet{Damsted2024_axes} and apply the \texttt{CLEAN} algorithm \citep{Mamon2013MNRAS_mamposst}. It minimizes the impact of interlopers, i.e. galaxies falling within the projected virial radius but that are actually located outside the main halo, by selecting member galaxies based on their position in the phase space of radial distance from the centre and rest frame velocity. An iterative process based on an NFW model estimates $r_{\rm 200c}$ from velocity dispersion. Then the algorithm selects galaxies within the $\pm K \sigma_{\rm los}(R)$, with $K=2.7$. For more details about the theoretical formalism of \texttt{CLEAN}, we refer the reader to Appendix B in \citet{Mamon2013MNRAS_mamposst}. Finally, using the cleaned members, we compute observed velocity dispersion using the \texttt{Gapper} method \citep{Beers1990AJ....100...32B}, an estimator based on the gaps between ordered measurements: it sorts the values of individual velocities and uses the difference between velocity intervals as weights to compute the final velocity dispersion \citep{Wetzell2022MNRAS_desy3veldisp}. The uncertainty on the measurement is computed as the standard deviation of 1000 bootstrap resamples.

\begin{figure}
    \centering
    \includegraphics[width=\columnwidth]{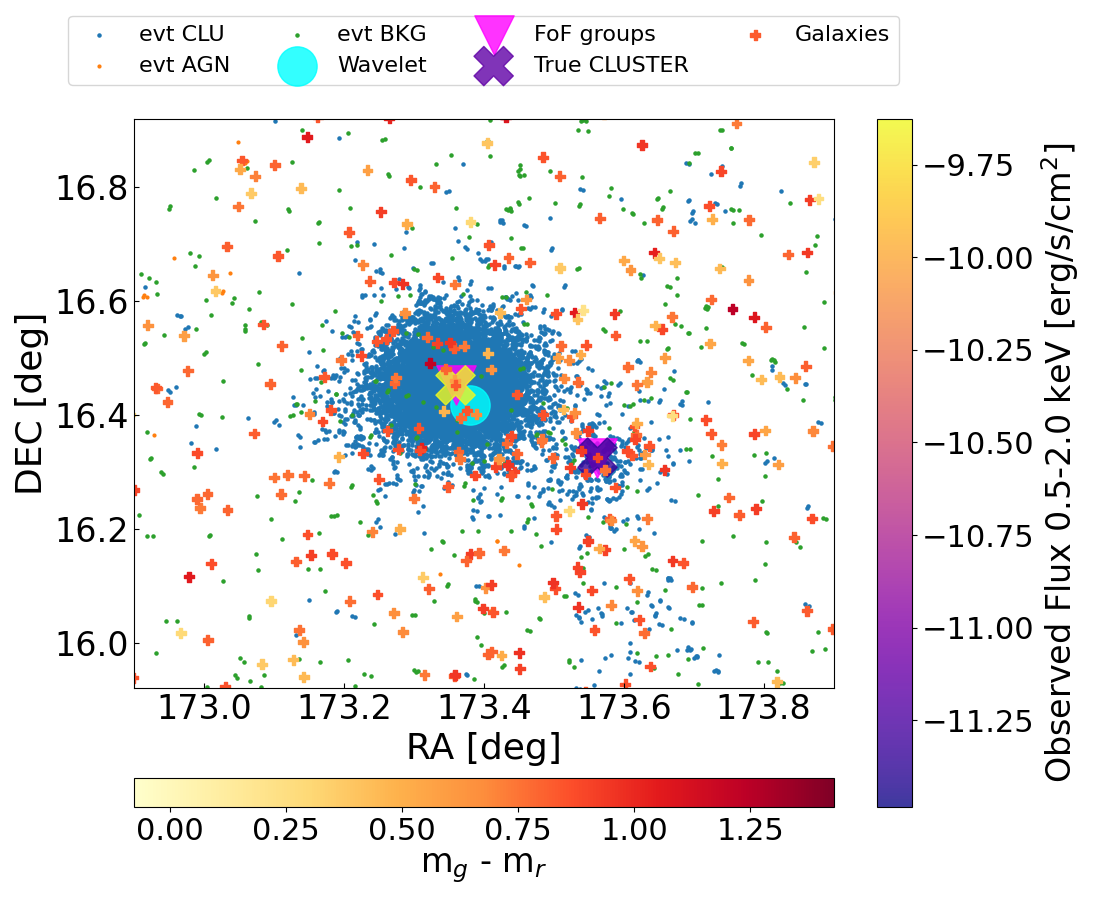} 
    \caption{Image example showing one of the brightest clusters in our mock with the additional components included in the simulation. A secondary halo is located in the south-east region compared to the brightest one. The events generated by clusters (AGN, the X-ray background) are displayed as small dots in blue (orange, green). The true halo position is shown with large crosses, colour coded by the X-ray flux. The position of the X-ray wavelet (optical FoF) detection is located at the cyan circle (magenta triangle). Each galaxy is marked by the small plus signs, according to their g-r colour. The vertical colour bar refers to input groups and clusters (the large crosses), the bottom one to individual galaxies (small plus signs). }
    \label{fig:image_example}
\end{figure}

\begin{table*}[]
    \centering
    \caption{Summary statistic about our mocks.}
    
    \begin{tabular}{|c|c|c|c|c|}
    \hline
    \hline
       \textbf{Sample} & \textbf{Total Number}  & \textbf{FoF Groups} & \textbf{X-ray Wavelet} & \textbf{X-ray + Optical} \\
       \hline
       \rule{0pt}{2ex}  
       Full Sample & 62$\,$868 & 17$\,$669 (28.1) & 5589 (8.9) & 3975 (6.3) \\
        z < 0.08 & 14$\,$366 & 6990 (48.6) & 2804 (19.5) & 2144 (14.9) \\
        z < 0.08, M$_{\rm 500c}$ > 5$\times$10$^{13}$ M$_\odot$ & 1962 & 1614 (82.2) & 1352 (68.9) & 1166 (59.4) \\  
        0.02 < z < 0.06 & 5768 & 3230 (56.0) & 1582 (27.4) & 1206 (20.9) \\
        \hline
    \end{tabular}
    \tablefoot{The table reports the total number of simulated clusters and groups, together with the number and fraction of detected objects in the optical (FoF) and X-ray (Wavelet) mocks, for different redshift and mass references. The last line refers to the X-GAP redshift range of interest. The numbers between brackets is the detection probability expressed as percentage.}
    \label{tab:compl_summary}
\end{table*}

\subsection{Matching input and output}

Given the setup of our simulation, we need to perform a three-way matching, between dark matter haloes in the light cone, detections in the X-ray mock, and FoF optical groups. For a given dark matter halo, we can ask whether it is detected in the X-ray, optical, or both mocks. For the matching procedure we focus on haloes at $z \leq 0.14$, where we measured input velocity dispersions. We ignore optical detections in the range between 0.14 and 0.2, which is nonetheless outside the X-GAP range of interest.

\subsubsection{X-ray to haloes}
We follow a similar scheme to \citet{Seppi2022A&A_erass1twin} to match input and output catalogues from the X-ray point of view, and use the information stored in the unique ID of each photon. For each event, the ID encodes the source responsible for the emission, either a cluster, an AGN, or the background. For each entry in the wavelet catalogue, we query all the events within a radius of 6 arcmin. We assign such detection to the simulated source emitting the majority of the photons. If different sources provide the same exact amount of events, we give priority to the input cluster. We only account for input sources providing more than two events on the mock detector, and sources providing an amount of events larger than the 0.8 percentile point of the Poisson distribution with mean value equal to the total number of counts provided by the background within the given aperture of 6 arcmin \citep[see also][for a similar implementation]{Liu2022A&A_teng_efedsSIM}.

The main difference compared to previous work about eROSITA simulations cited above is that in this case we are particularly interested in the detection of extended sources rather than contamination. Indeed, it is clear from the X-GAP data that contamination from AGN is not an issue \citep{Eckert2024_xgap}, with only one false detection out of 49. Therefore, if a detection contains cluster emission but is not assigned to the cluster directly, i.e. due to the presence of a bright AGN, we still consider the cluster as detected. Indeed, the cross-match with the optical mock allows to clean the otherwise contaminated X-ray classification of such sources. However, we track all cases of sources whose emission is contaminated by a secondary object. To do this, we require that the amount of counts provided by the secondary source within the given aperture is larger than the square root of the counts produced by the primary match.

\subsubsection{Optical to haloes}

We now search for an optical FoF counterpart to each simulated halo. We start from the input halo catalogue and search for optical matches in radial angular apertures of $3 \times R_{\rm 500c}$ in the RA-DEC plane. We then query the candidate matches, if present, and keep only the ones with a relative redshift error below $1\%$, i.e. $\dfrac{|z_{\rm FoF} - z_{\rm true}|}{1+z_{\rm true}}<0.005$. This threshold is comparable to state of the art results on cluster photometric data \citep[see e.g.,][]{Kluge2024A&A_eromapper}, and thus serves as a reasonable upper limit that also accounts for additional observational redshift uncertainties. We perform the matching in redshift space, i.e. $z_{\rm true}$ includes peculiar velocities (see Eq. \ref{eq:redshift}), which help to disentangle close pairs or mergers due to the fact that the optical detection uses observed redshifts as well. For each candidate we evaluate the matching quality by computing a matching statistic accounting for a mass probability, encoded in the estimate of velocity dispersion, and a distance probability, related to the 3D cartesian distance from the true halo centre. The matching statistic is therefore higher if the optical candidate is well centred on the true position of the halo, and if velocity dispersion is accurately estimated. The final equation reads:
\begin{align}
    r =& \frac{d_{\rm 3D}}{R_{\rm 500c}} \nonumber \\
    P_{\rm dist}(r) =& \frac{1}{1+r^2} \nonumber \\
    P_{\rm match}(\sigma_{\rm v,M},\sigma_{\rm v,T},r) =& \mathcal{LN}(\sigma_{\rm v,M}|\sigma_{\rm v,T}, \sigma_{\rm LN}) \times P_{\rm dist}(r), 
    \label{eq:matching_probability}
\end{align}
where $\mathcal{LN}(\sigma_{\rm v,M}|\sigma_{\rm v,T})$ is a log-normal function describing the measured velocity dispersion $\sigma_{\rm v,M}$ that is centred around the true value $\sigma_{\rm v,T}$. We use a log-normal scatter of $\sigma_{\rm LN} = 0.08$ dex. An accurate refinement of this quantity is provided in Sect. \ref{sec:veldisp_mass}.

We consider the optical FoF object with the highest matching statistic as the primary match to an input halo, if more than one candidate is present.
Although this setup easily allows us to state whether a halo is detected or not, it may assign the same optical FoF object to different input haloes. In such cases, we assign the primary match to the halo whose FoF object has the highest matching probability. For the other haloes, we then check if they have secondary matches. If the secondary matches have not been assigned as primary matches to another halo, we upgrade them to primary matches. Otherwise, we mark the cluster as blended with another primary detection. With this method we now have a unique way of mapping input haloes into FoF optical groups and clusters.

An example of the end result of the processing detailed in this section is shown in Fig. \ref{fig:image_example}. It is a field of one square degree with one of the brightest clusters in our simulation. X-ray photons generated by various components (clusters and groups, AGN, background) are shown as small dots (in blue, orange, green). Galaxies are displayed by the plus signs and represented according to their g-r colour. True halo positions (wavelet detections, optical FoF detections) are located at the large crosses (cyan circles, magenta triangles). In this case, the bright system in the centre of the image is properly detected both in X-ray and optical bands. In addition, a fainter nearby halo is found by the optical FoF algorithm, but not in the wavelet processing.

We generate X-ray photons for a total of 62$\,$868 clusters and groups below redshift 0.14, with flux larger than 10$^{-14}$ erg/s/cm$^2$. This limit is much below the expected RASS detection limit \citep{Boringer2000ApJS..129..435B, Ebeling2001ApJ_macs}. Between them, 17$\,$669 are detected in the optical mock, for an optical completeness of 28.1$\%$. In the X-ray mock, the wavelet processing identifies 5589 sources, for an X-ray completeness of 8.8$\%$. After combining X-ray and optical detections, we obtain 3975 sources, for a global completeness of 6.3$\%$. When focusing on a lower redshift slice, or more massive systems, the detection probability increases. In the redshift range of interest for X-GAP, between 0.02 and 0.06, the global probability of detection is 20.9$\%$. A summary is reported in Table \ref{tab:compl_summary}.

\begin{figure}
    \centering
    \includegraphics[width=\columnwidth]{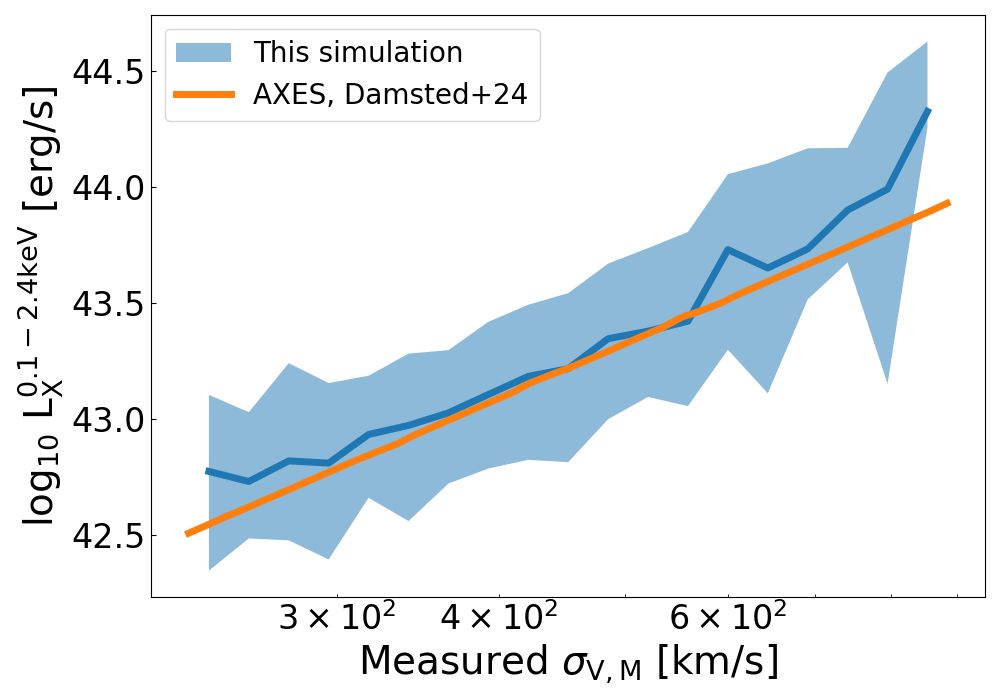}    
    \caption{Scaling relation between X-ray luminosity and measured velocity dispersion. The result from our simulation (the real AXES) is shown in blue (orange). We find excellent agreement between the mock and real data in \citet{Damsted2024_axes}.}
    \label{fig:Lx_sigma}
\end{figure}

Comparing the output sample obtained as described above to the real AXES presented in \citet{Damsted2024_axes} is a test for our full workflow, from the individual models of clusters and groups, the X-ray simulation, the detection algorithm, and the matching procedure. We do so by comparing the X-ray luminosity as a function of the measured galaxy member velocity dispersion. In this case we use the X-ray luminosity in the 0.1-2.4 keV band to match the values measured in the real AXES. The procedure is the same as Eq. \ref{eq:Lx_calc}, but in this case we use the cooling function in the 0.1-2.4 keV band computed using \texttt{pyatomdb} \citep{Foster2020Atoms_pyatomdb}. We focus on the X-GAP redshift range 0.02-0.06. The result is shown in Fig. \ref{fig:Lx_sigma}. It displays the relation for our mock in blue and for the real AXES in orange. The shaded area accounts for 16th-84th percentiles. We find excellent agreement between our simulation and the results from \citet{Damsted2024_axes}. Therefore, our end to end simulation provides a robust sample in comparison to real data and is suitable to study its selection effects.

\section{Selection function}
\label{sec:probability_detection}

In this section we report our results about the sample completeness, directly encoded in the selection function, and purity. These results include the combination of the four light cones described in Sect. \ref{subsec:halo_lightcone} into a single summary catalogue. We express the true X-ray flux in the 0.5-2.0 keV band within R$_{\rm 500c}$.

\subsection{Sample completeness}
\label{subsec:completeness}

\begin{figure}
    \centering
    \includegraphics[width=0.98\columnwidth]{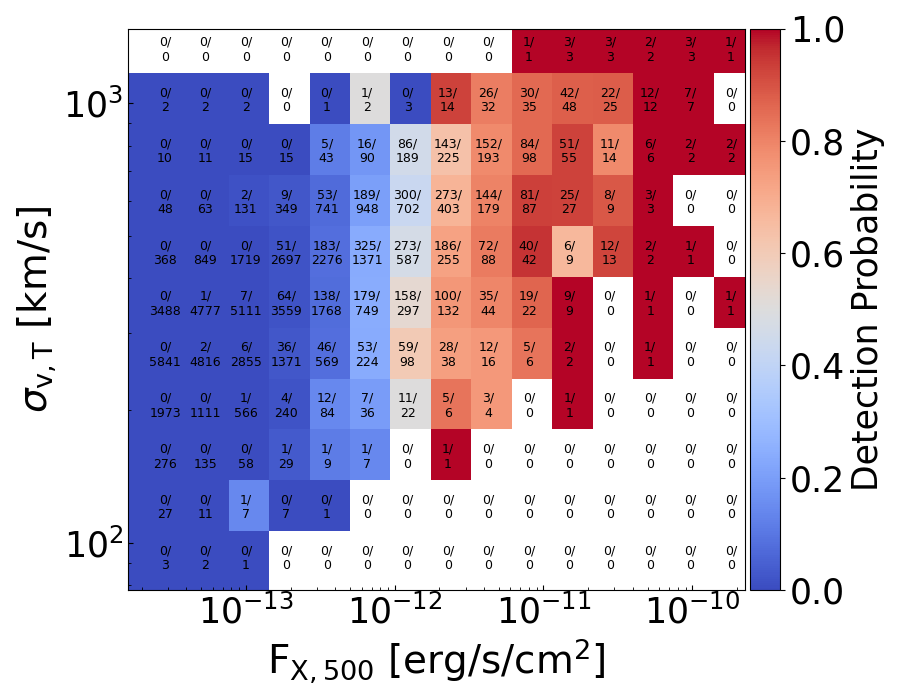}
    \includegraphics[width=0.98\columnwidth]{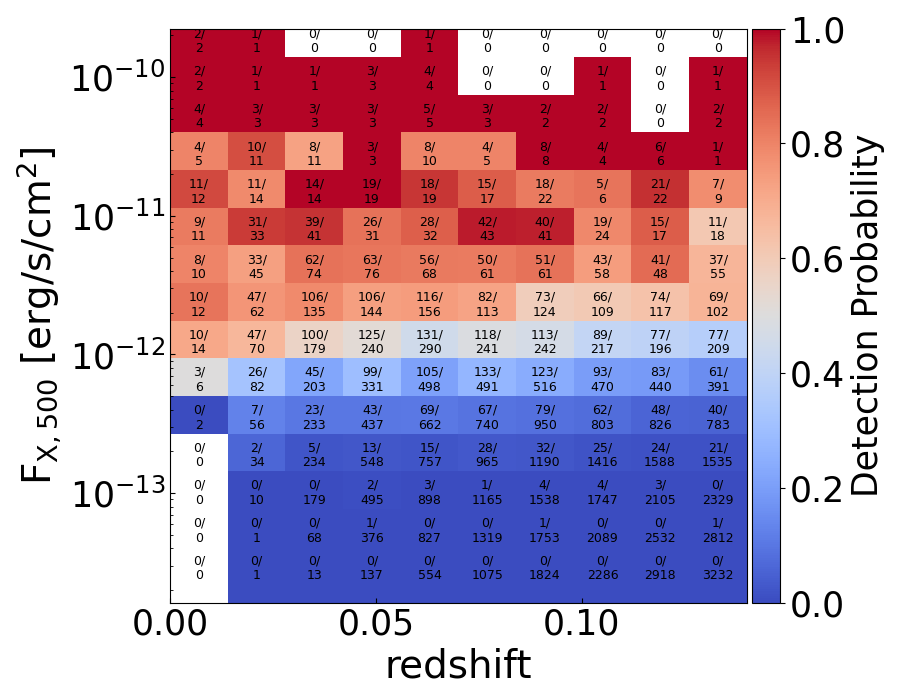}
    \includegraphics[width=0.98\columnwidth]{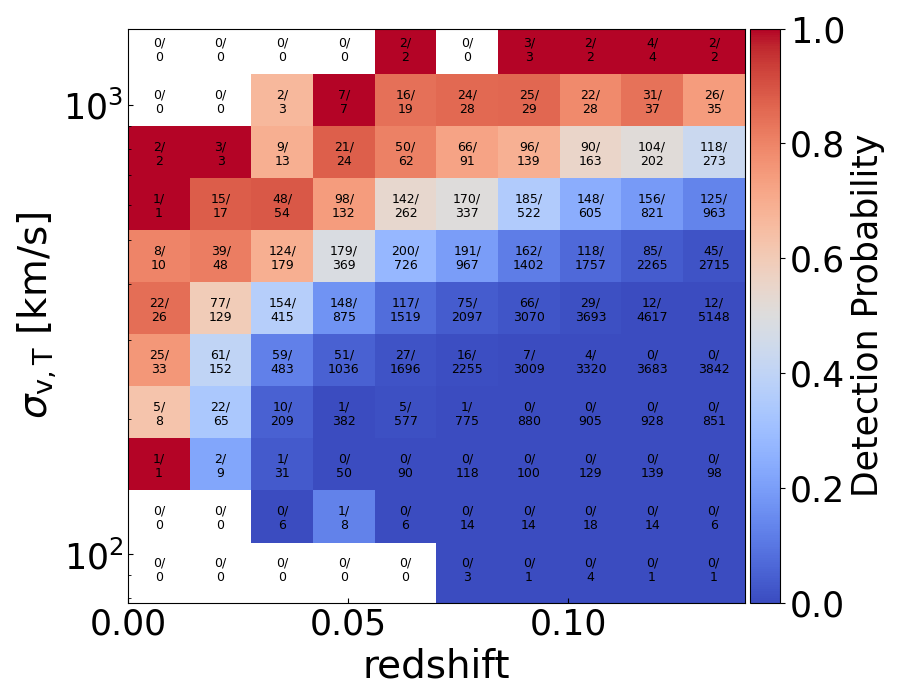}    
    \caption{Two dimensional probability of detection for X-ray plus optically selected haloes in our simulation (see Sect. \ref{subsec:completeness}). The panels shows the completeness fraction as a function of three combinations of X-ray flux, velocity dispersion, and true velocity dispersion.}
    \label{fig:Pdet2d}
\end{figure}

\begin{figure}
    \centering
    \includegraphics[width=\columnwidth]{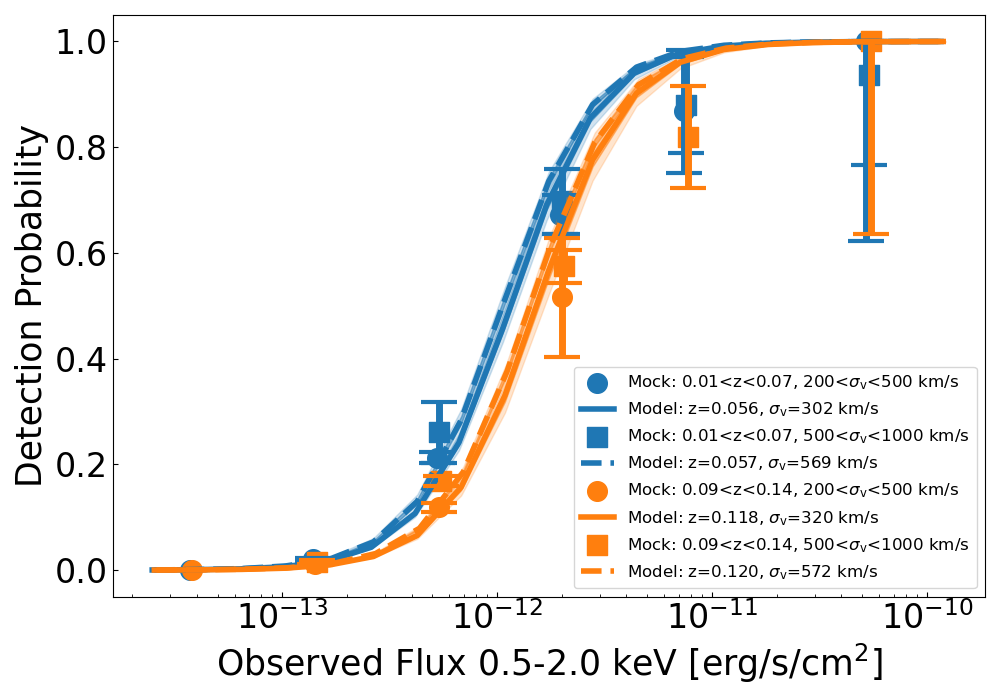}
    \includegraphics[width=\columnwidth]{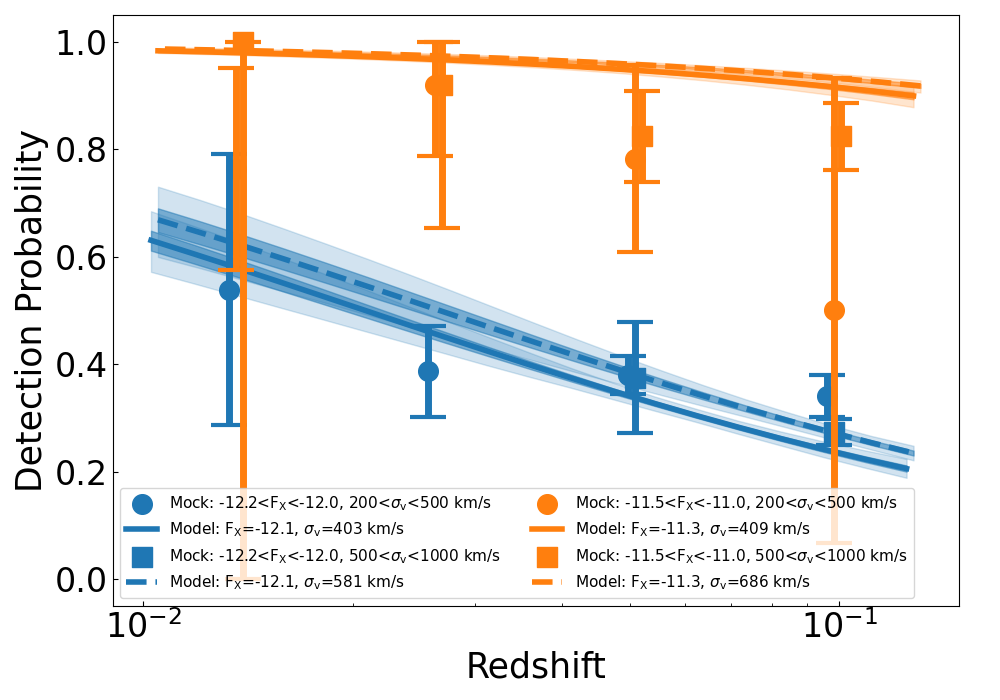}
    \includegraphics[width=\columnwidth]{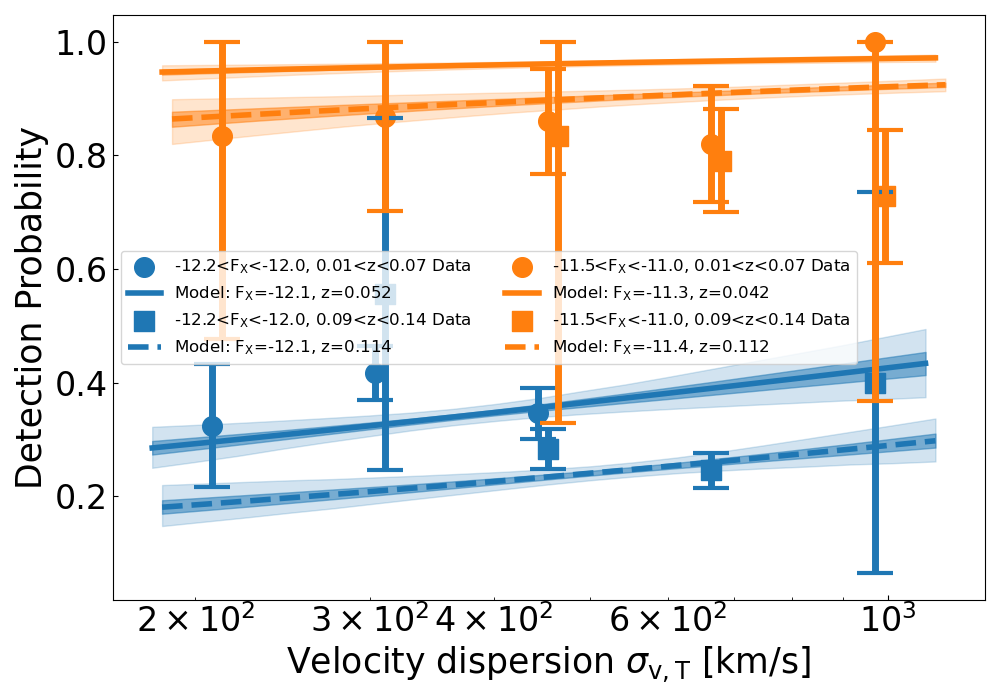}
    \caption{Probability of detection for X-GAP selected haloes as a function of observables. The three panels show different combination of flux measured within R$_{\rm 500c}$, redshift, and velocity dispersion on the x-axis, in different colours, and symbols. The various line styles denote the selection function model in Eq. \ref{eq:selfunc_model}. The error bars account for the Poisson error on the number of total and detected sources. The lines in each panel were shifted respectively by -0.02, -0.01, 0, and 0.01 dex for clarity.}
    \label{fig:Pdet_selfunc}
\end{figure}

Information and knowledge about the sample completeness level is necessary to characterize the population of galaxy clusters and groups from a sample such as X-GAP. Our end to end mock allows a direct comparison between the halo population selected in the optical and X-ray bands to the global one in the full light cone. We start by analysing the fraction of selected sources as a function of X-ray flux and velocity dispersion. Just like for the luminosity, we use fluxes within R$_{\rm 500c}$ in the 0.5-2.0 keV band. The result is shown in Fig. \ref{fig:Pdet2d}. We see that the probability of detection is higher (in red) for bright systems with high velocity dispersion, and therefore high mass. However, we notice that the detection depends primarily on flux, rather than velocity dispersion. In the top panel of Fig. \ref{fig:Pdet2d}, at fixed velocity dispersion $\sigma_{\rm v,T}$=400 km/s, the detection probability changes from about 53$\%$ at X-ray flux of 2$\times 10^{-12}$ erg/s/cm$^2$ to about 86$\%$ at 8$\times 10^{-12}$ erg/s/cm$^2$. Conversely, it varies only from 75$\%$ at fixed flux of 5$\times 10^{-12}$ erg/s/cm$^2$ for $\sigma_{\rm v,T}$ of 250 km/s to 81$\%$ at 1000 km/s. Although flux and velocity dispersion are inevitably intrinsically correlated, their impact on detection is not strongly linked. The same holds for redshift in comparison to X-ray flux. In the middle panel of Fig. \ref{fig:Pdet2d} the detection probability is one for flux above 4$\times 10^{-11}$ erg/s/cm$^2$ at each redshift, whereas faint sources at about 1$\times 10^{-13}$ erg/s/cm$^2$ are not detected even if they are nearby. This result is in agreement with \citet{Damsted2024_axes}, who find a small redshift evolution of the 10$\%$ completeness limit of AXES, from about 300 km/s at z=0.05 to 480 km/s at z=0.15. The bottom panel of Fig. \ref{fig:Pdet2d} shows the detection probability as a function of redshift and velocity dispersion. In this case we do see a combined evolution. At z=0.1, the completeness fraction increases from about 0.1$\%$ at $\sigma_{\rm v,T}$=300 km/s to 78$\%$ at $\sigma_{\rm v,T}$=1000 km/s. Similarly, at z=0.02
it increases from 40$\%$ to 1. This trend is encoded in the correlation between more massive systems with larger velocity dispersion and X-ray brightness, meaning that the X-ray sensitivity is driving the variation of detection probability with redshift and velocity dispersion. We stress that this is true for our particular survey set up and does not necessarily hold for other selection criteria. Indeed, this is the result of the ROSAT selection being the limiting factor compared to the SDSS one. In an opposite case, with a deep X-ray coverage and a shallow optical one, the main driver of the selection method would likely be the optical proxy. Finally, we notice that the average R$_{\rm 500c}$ of the simulated systems in the X-GAP redshift range of 0.02 to 0.06 is equal to about 8 arcminutes and the average size of the detected systems is 11 arcminutes. This confirms that the size of the wavelet scales equal to 12 and 24 arcminutes in \citet{Damsted2024_axes} is suitable to detect X-GAP groups using emission from their outskirts.

\subsection{Selection function model}

\begin{figure}
    \centering
    \includegraphics[width=\columnwidth]{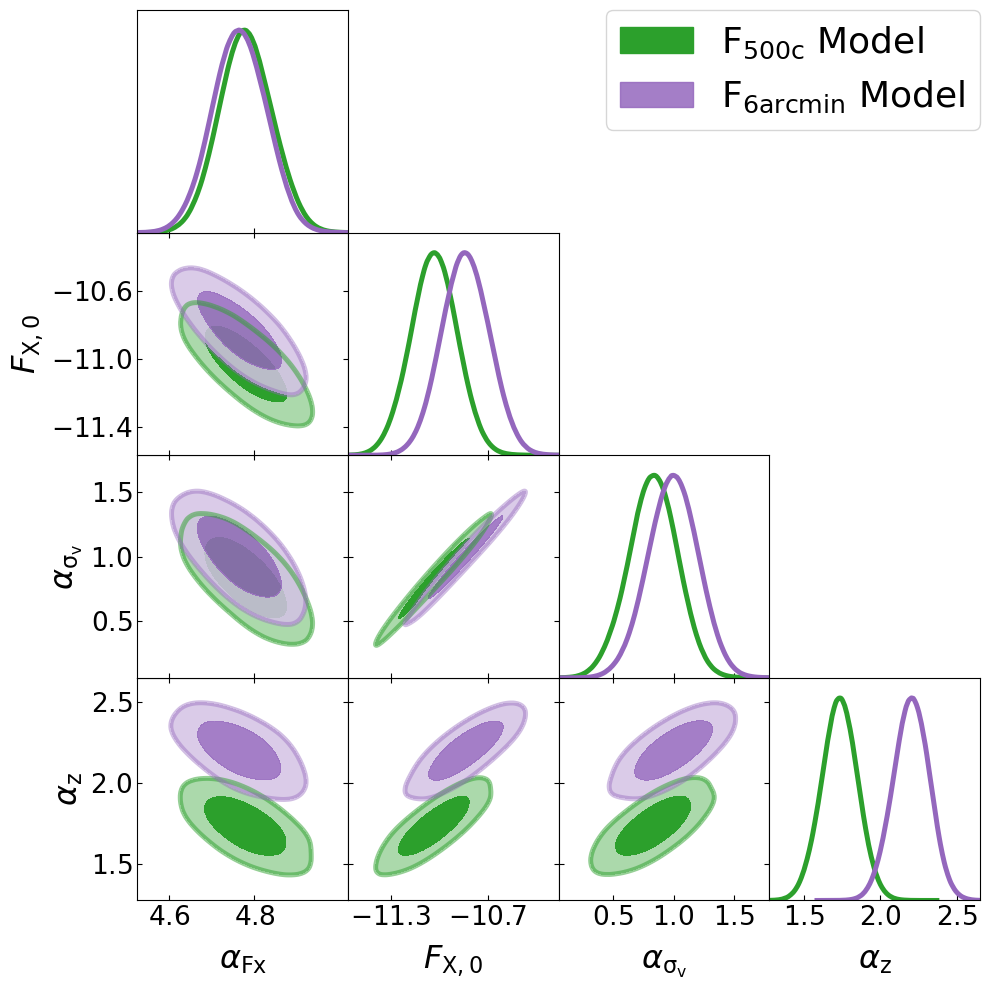}
    \caption{Marginalized posterior distribution of the best fit model of the X-ray plus optical selection function (see Eq. \ref{eq:selfunc_model}). The filled 2D contours show the 1-$\sigma$ and 2-$\sigma$ confidence levels of the posteriors after convolution with the uniform priors. The green contours denote the base selection function where flux is measured within R$_{\rm 500c}$, the violet contours show the effect of measuring flux within an angular aperture of six arcminutes. The parameter values are reported in Table \ref{tab:selfunc_model_pars}.}
    \label{fig:corner_selfunc}
\end{figure}

The selection function encodes the probability for a source to be detected as a function of a given set of parameters. We compute the ratio between the detected and the simulated sources:
\begin{equation}
    P_{\rm det}(F_{\rm X}, z, \sigma_{\rm v,T}) = \frac{N_{\rm DET}}{N_{\rm ALL}}(F_{\rm X}, z, \sigma_{\rm v,T}).
    \label{eq:pdet}
\end{equation}

In particular, we evaluate the ratio in Eq. \ref{eq:pdet} as a function of observable properties. This makes our selection function directly related to observations, bypassing intrinsic halo properties (e.g. halo mass), that are not directly measurable. When using multi wavelength data for the identification of clusters and groups, accounting for mass tracers using different observables is key to model selection effects in different surveys \citep[e.g.][]{Finoguenov2020A&A_codex}. We use the X-ray flux in the 0.5-2.0 keV band within an aperture equal to R$_{\rm 500c}$ (computed following Sect. \ref{subsec:clumodel_results}), velocity dispersion (see Eq. \ref{eq:vel_disp}), and redshift. The result is shown in Fig. \ref{fig:Pdet2d}. This combination is particularly suitable in the case where the selection function is needed for forward modelling a population. One caveat is that one would need to account for an aperture correction since the radius encompassing an average density that is 500 times larger than the critical density depends on cosmology. It is possible to account for it by modelling its evolution with $E(z)=\dfrac{H(z)}{H_0}$. In the opposite case, where one wants to start from the data and does not necessarily have access to M$_{\rm 500c}$, a selection function expressed in terms of flux measured within a fixed angular aperture is more convenient. Therefore, we add a second model on top of the base one, where we model the detection probability as a function of observed flux within an angular aperture of six arcminutes, velocity dispersion, and redshift. In addition, the latter model does not depend on any mass assumptions to estimate the radial aperture, which is also makes it useful for forward modelling.

\begin{table}[]
    \centering
    \caption{Priors and posteriors of the parameters describing the X-GAP selection function model in Eq. \ref{eq:selfunc_model}.}
    
    \begin{tabular}{c|c|c|c}
    \hline
    \hline
       \textbf{Parameter}  & \textbf{Prior} & \textbf{Posterior F$_{\rm X,500c}$} & \textbf{Posterior F$_{\rm X,6'}$} \\
       \hline
       \rule{0pt}{2ex}    
        $\alpha_{\rm Fx}$ & $\mathcal{U}$(0 , 10) &  4.78 $\pm$ 0.08 & 4.76 $\pm$ 0.08 \\
        F$_{\rm X,0}$ & $\mathcal{U}$(-14, -5) & -11.03 $\pm$ 0.14 & -10.84 $\pm$ 0.15 \\
        $\alpha_{\rm v}$ & $\mathcal{U}$(0, 5) & 0.83 $\pm$ 0.19 & 1.01 $\pm$ 0.21  \\
        $\alpha_{\rm \sigma z}$ & $\mathcal{U}$(0, 5) & 1.73 $\pm$ 0.12 & 2.19 $\pm$ 0.12 \\
        \hline
    \end{tabular}
    \tablefoot{Posteriors are reported for both cases of X-ray flux measured within R$_{\rm 500c}$ or in an aperture of six arcminutes. The symbol $\mathcal{U}(M,N)$ denotes a uniform prior between the values M and N.}
    \label{tab:selfunc_model_pars}
\end{table}

For both cases, we expect massive, bright, and nearby sources to be easier to detect compared to lighter, fainter, further ones. Therefore, the completeness is directly proportional to flux and velocity dispersion, and inversely proportional to redshift. In the previous paragraph describing Fig. \ref{fig:Pdet2d}, we notice that our selection is primarily driven by X-ray flux, which is the main variable of our model. We combine individual sigmoid functions into a comprehensive detection probability model with four free parameters, that reads:

\begin{align}
    P_{\rm det}(F_{\rm X}, z, \sigma_{\rm v,T}) =\ (1 &+ \exp[-\alpha_{\rm Fx}(\log_{\rm 10}F_{\rm X} - F_{\rm X,0}) \ + \nonumber \\
    & - \alpha_{\rm \sigma v} \times \log_{\rm 10}\sigma_{\rm v,T} + \alpha_{\rm z} \times \log_{10}z ])^{-1},
    \label{eq:selfunc_model}
\end{align}

where the parameters $\alpha_{\rm Fx}$, $\alpha_{\rm z}$, $\alpha_{\rm \sigma v}$ regulate the slope of the global sigmoid function, and F$_{\rm X,0}$ sets the flux scale to centre its zero point along the flux axis. A similar approach was followed by \citet{Clerc2018A&A...617A..92C}, who modelled the detection of extended sources in eROSITA simulations with an error function. We find that a sigmoid allows us to better capture the completeness trend in our simulations. To test the performance of our fit, we compute a reduced $\chi^2_r = \dfrac{1}{F}\sum\dfrac{(D-M)^2}{\delta D^2}$, where $F$ is the number of degrees of freedom, i.e. the number of bins minus the number of free model parameters, $D$ is the measured completeness, $\delta D$ is its uncertainty, and $M$ is the best fit model evaluated at the median flux, velocity dispersion, and redshift of the full population in each 3D bin. We obtain $\chi^2_{r}=0.97$, therefore we conclude that our fit is adequate. We derive posterior probability distributions and the Bayesian evidence for the parameters in Eq. \ref{eq:selfunc_model} with the nested sampling Monte Carlo algorithm MLFriends \citep{Buchner2016S&C_mlfriends, Buchner2019PASP_mlfriends} using the
UltraNest\footnote{\url{https://johannesbuchner.github.io/UltraNest/}} package \citep{Buchner2021JOSS_ultranest}.

The result for the base model using F$_{\rm 500c}$ is shown in Fig. \ref{fig:Pdet_selfunc}. The three panels show the detection probability with all combinations of flux, velocity dispersion, and redshift. The circles and squares represent the measurements from the end to end mocks, while the lines and shaded areas denote the best fit model with 68$\%$ and 95$\%$ confidence intervals. The error bars account for the Poisson uncertainty in counting the total and the detected number of sources when computing the completeness (see Eq. \ref{eq:pdet}). The model favours a selection driven by flux, as noted in the previous section, with a secondary dependence on redshift and velocity dispersion. We find good agreement between the measured completeness and the model of the selection function. The full marginalized posterior distribution is shown by the green contours in Fig. \ref{fig:corner_selfunc}. There is a negative correlation between the parameters describing the slope and flux normalisation, with a Pearson correlation coefficient of -0.75. This is unexpected because if the zero point is set to fainter fluxes, the slope needs to be shallower to describe the bulk of the population. However, this effect is mitigated by the positive correlation of F$_{\rm X,0}$ to the slopes related to velocity dispersion and redshift, with Pearson coefficients respectively equal to 0.98 and 0.79. We find also a positive correlation between the slopes related to redshift and velocity dispersion, with a coefficient of 0.69. This is expected by construction, because even if the detection probability increases with velocity dispersion and decreases with redshift, the sign in slope definition is opposite in Eq. \ref{eq:selfunc_model}. The results for the additional model using the flux within six arcminutes are qualitatively very similar to the base model using flux within R$_{\rm 500c}$. The flux and the velocity dispersion slopes are compatible within 1$\sigma$. The flux normalisation is slightly larger at -10.84 $\pm$ 0.15 compared to -11.03 $\pm$ 0.14. One would naively expect the opposite, because at the average redshift (equal to 0.075) of our detected population, the mean R$_{\rm 500c}$ covers about 7 arcminutes, meaning that the integrated flux within a smaller aperture of six arcminutes should end up in a lower normalisation of the sigmoid function. However, this effect is counter balanced by the positive correlation between the flux normalisation and redshift slope, which is steeper in this second case at 2.19 $\pm$ 0.12 compared to 1.73 $\pm$ 0.12 for the base model. The larger slope values push the flux normalisation to higher values as well, but at fixed flux the completeness is higher for the model using flux within six arcminutes compared to the base one, as expected. All the parameters are reported in Table \ref{tab:selfunc_model_pars}.

\begin{figure*}
    \centering
    \includegraphics[width=0.66\columnwidth]{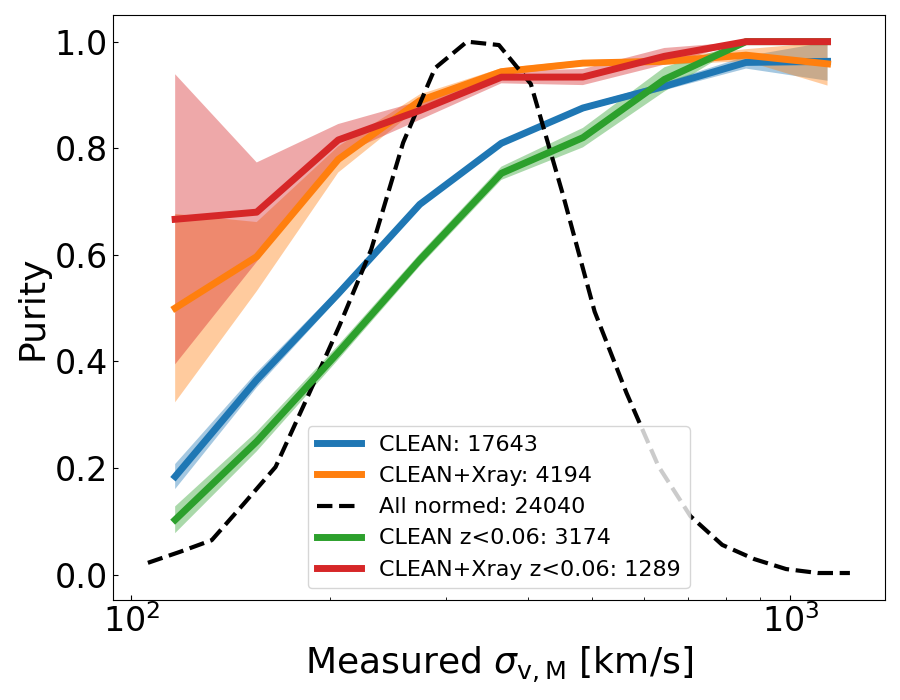}    \includegraphics[width=0.66\columnwidth]{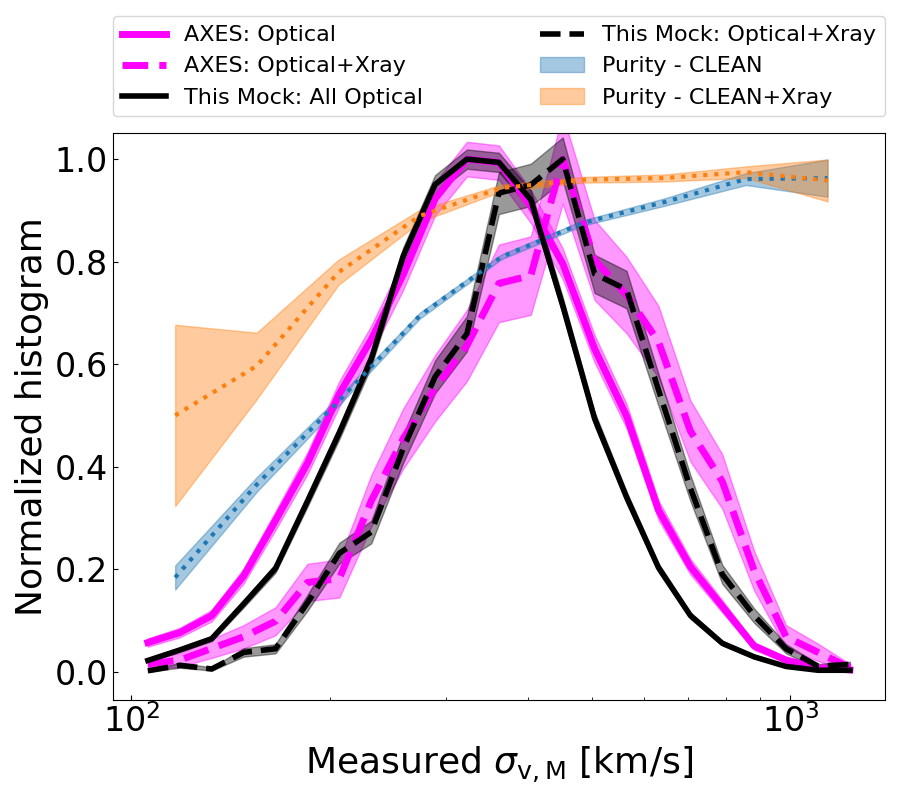}
    \includegraphics[width=0.66\columnwidth]{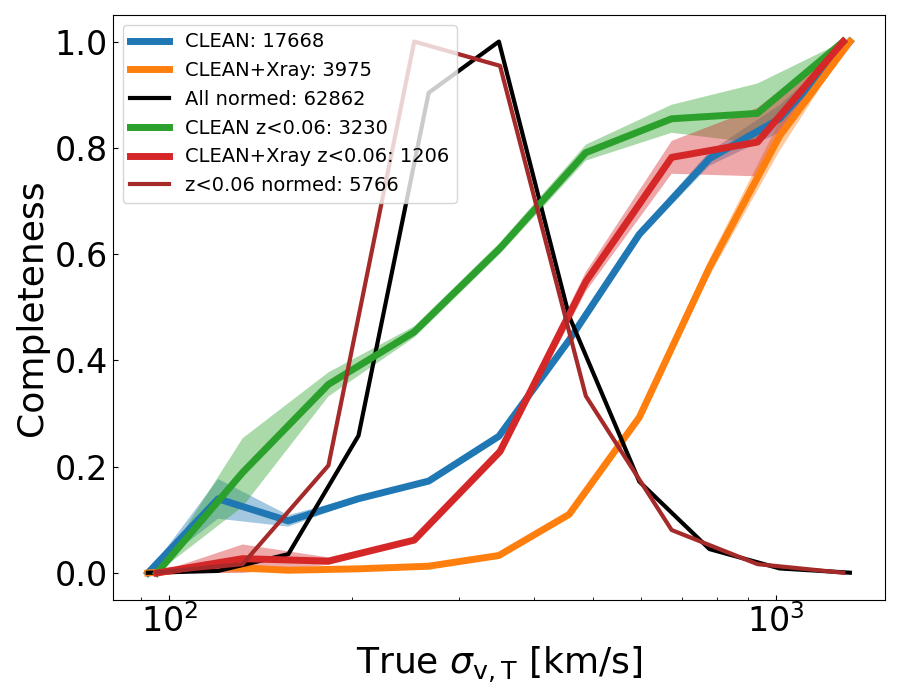}    
    \caption{Trade-off between completeness and purity as a function of velocity dispersion. \textbf{Left hand panel}: Fraction of objects with a corresponding real input dark matter halo as a function of measured velocity dispersion. The blue line shows the fraction of objects detected in the optical mock that are matched to a real halo in the light cone. The orange line
    denotes objects with a corresponding X-ray detection. The green and
    red lines refer to the low redshift population, in the interval of interest
    for X-GAP, between 0.02 and 0.06. The black line shows the normalised distribution of measured velocity dispersion in the simulation. \textbf{Central panel}: Distribution of the measured velocity dispersion in our mock and in the real data. The blue and orange lines are left as a reference from the previous panel. The pink lines show the normalised distributions of the real AXES sample from \citet{Damsted2024_axes}, while the black ones refer to the simulation. The solid (dashed) lines denote the full optical (optical plus X-ray) detections. \textbf{Right hand panel}: Fraction of real input dark matter haloes with a corresponding detection in the mock. The coloured lines represent the same populations as the top panel, but starting from the true input dark matter haloes. }
    \label{fig:purity-veldisp}
\end{figure*}

\subsection{Purity-completeness trade off}
\label{subsec:purity}

In the previous subsection we analysed the sample completeness after cross matching the input dark matter haloes to the X-ray and optical mocks. Here instead we do the opposite, i.e. we start from the output mock catalogues and query the matched haloes from the procedure outlined in Sect. \ref{sec:catalogues}. To compare the X-ray and optical catalogues we follow a prescription similar to \citet{Damsted2024_axes}. For each optical detection, we estimate R$_{\rm 200c}$ from a halo mass inferred from the measured galaxy velocity dispersion, based on the scaling relation calibrated in the next section. We stress the fact that this is not necessarily a precise and accurate measure of groups' radii, but we simply use it as an aperture to search for X-ray matches. We consider an optical group to have a corresponding X-ray detection if there is a wavelet detection within the $R_{\rm 200c}$ estimated from velocity dispersion. This process allows us to measure the purity of our sample, i.e. the fraction of sources detected in the simulation that correspond to real input objects.

Since defining an input property for a source that is not matched to an input halo is impossible, we analyse purity as a function of a measured quantity: velocity dispersion. The result is shown in the left hand panel of Fig. \ref{fig:purity-veldisp}. The blue line denotes the global population detected in the optical mock, after the application of \texttt{CLEAN}. The orange line considers only optical detections with a corresponding X-ray detection. The dashed black line shows the global distribution of the measured velocity dispersion in our mocks. The central panel shows the comparison to the full distribution in the real AXES sample (in pink), where the dashed lines denote the population with an X-ray match. We find that distributions in our mock and in the real AXES peak at the same velocity dispersion, and have a similar distribution, especially around the peak. This is an additional confirmation that our end to end simulations produce a high fidelity sample that matches the properties of the real AXES. In addition, it enables a reliable estimate of the purity level in AXES using the results from the mock. We find that the sample with an X-ray detection is very pure, reaching a purity of 90$\%$ already at $\sigma_{\rm v,M} = 280$ km/s. The full optical mock reaches a similar purity level at about $\sigma_{\rm v,M} = 600$ km/s. This means that the cross correlation with X-ray practically serves as a cleaning of the optical catalogue, confirming the finding of \citet{Damsted2024_axes}. In fact, the overall distribution of measured velocity dispersion for all detections with an X-ray match peaks at higher values compared to only optical detections, with values of about 450 km/s and 350 km/s respectively. Using the full AXES population and our purity estimate from the simulation, we can estimate the purity level in the real data:
\begin{equation}
    P_{\rm avg} = \sum_{\sigma_{\rm v,M}} \frac{f_{\rm obs}(\sigma_{\rm v,M}) P_{\rm sim}(\sigma_{\rm v,M})}{f_{\rm obs}(\sigma_{\rm v,M})},
    \label{eq:purity_level}
\end{equation}
where $f_{\rm obs}$ is the fraction of real sources at a given velocity dispersion, and $P_{\rm sim}$ is the purity estimate from the simulation. We estimate a purity level of 76$\%$ in the optical-only detection, and a 93$\%$ purity level in the AXES sample after adding the X-ray matching.

We also study the detection performance in the lower redshift range between 0.02 and 0.06, where X-GAP is defined. The result is denoted by the green and red lines, that are the analogue of the blue and orange ones respectively, but for the low redshift population. When accounting only for the optical detection, the purity for the latter is lower than the global population by about 5-10$\%$ up to 500 km/s, where the two distributions start to agree within error bars. This is due to the fact that in the nearby Universe the SDSS galaxy sample contains a larger fraction of galaxies with low stellar mass. In the 0.02-0.06 range, the average stellar mass is about 2$\times$10$^{10}$ M$_\odot$, whereas at higher redshift between 0.06-0.14, it increases to about 5$\times$10$^{10}$ M$_\odot$. Therefore, the fraction of isolated galaxies at low redshift is higher, and the chance of merging them together into a fake detection is higher. However, there is excellent agreement between the orange and the red lines, meaning that after adding the X-ray match to the optical detections, the purity level is very stable for different redshift. 

The effect of the X-ray correlation is also clear in the right hand panel of Fig. \ref{fig:purity-veldisp}. It shows the completeness fraction as a function of true velocity dispersion, the same concept analysed in the previous section, in this case marginalized on flux and redshift. The optical catalogue reaches a 50$\%$ completeness at velocity dispersion equal to about 500 km/s, while the optical plus X-ray case requires about 730 km/s. This is due to the relatively shallow and low resolution coverage of the ROSAT all sky survey, which allows detecting only massive, bright systems in large fractions. Conversely to purity, we see a significant improvement of the detection probability as a function of velocity dispersion at low redshift. In the 0.02-0.06 range, the 50$\%$ completeness is reached at 300 km/s. When adding the X-ray match, the same completeness level is reached at about 450 km/s. Both cases are an improvement by a factor of about 1.6 compared to the full population. The higher completeness at low redshift is also evident in the global distribution of measured velocity dispersion, the brown lines in the top panel of Fig. \ref{fig:purity-veldisp}. In fact, such distribution peaks at 250 km/s and at 350 km/s when adding the X-ray match. Both are about 100 km/s lower than the peak of the distribution for the full population, even if the underlying true velocity dispersion follows a very similar trend (brown line in the bottom panel of Fig. \ref{fig:purity-veldisp}): at low redshift we are able to probe the groups population down to lower velocity dispersion. We stress that this redshift trend is encoded in the selection function defined in terms of flux. We conclude that combining X-ray detected clusters and groups in the RASS with optical FoF detections provides pure samples even at low velocity dispersions, at the cost of reducing the sample completeness. However, the latter can be accounted for via the selection function calibrated in the previous section.

\section{Mass -- velocity dispersion}
\label{sec:veldisp_mass}

\begin{figure}
    \centering
    \includegraphics[width=\columnwidth]{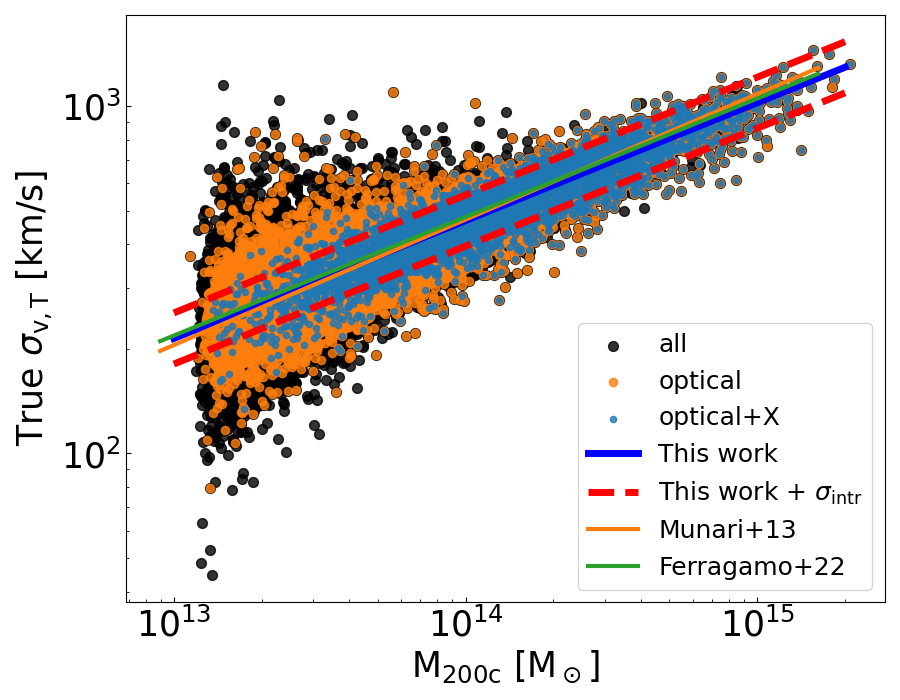}
    \includegraphics[width=\columnwidth]{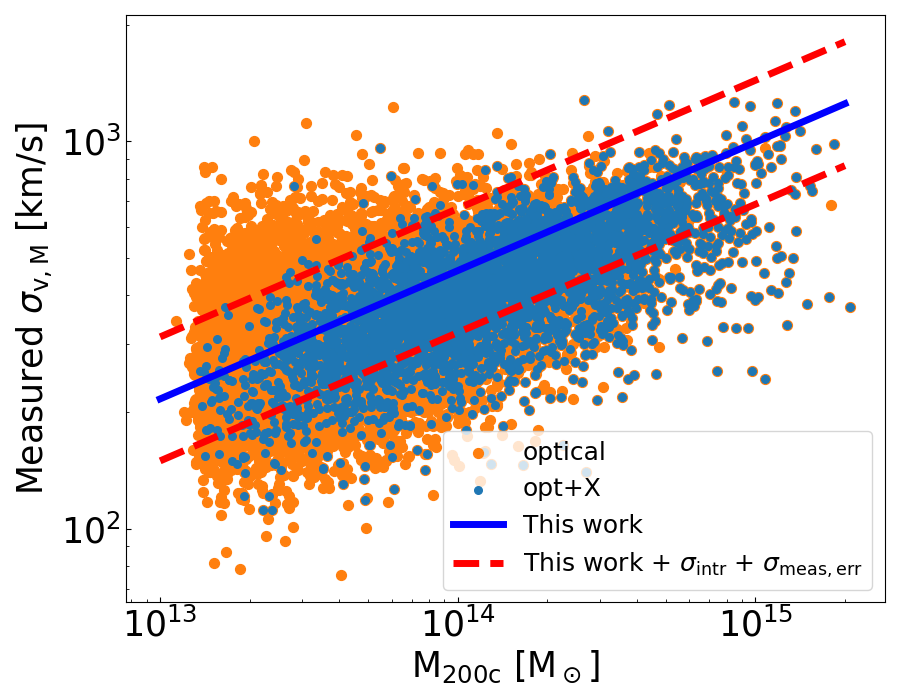}    
    \caption{Scaling relation between velocity dispersion and halo mass. \textbf{Top panel}: Scaling relation for the true velocity dispersion (Eq. \ref{eq:vel_disp}). The orange and green lines show the results from \citet{Munari2013MNRAS_veldisp} and \citet{Ferragamo2022EPJWC_veldispmass}. \textbf{Bottom panel}: Scaling relation for the measured velocity dispersion (Sect. \ref{subsec:optical_sel}). The blue line denotes the best fit relation, and the red lines include the contribution of the intrinsic and measurement scatter summed in quadrature (see Eq. \ref{eq:scal_rel}).}
    \label{fig:veldisp_mass}
\end{figure}

In the section, we use our mock to calibrate the scaling relation between halo mass and velocity dispersion. Within the scaling relation we account for two distributions. The first one is a log-normal distribution of the true velocity dispersion $\sigma_{\rm v,T}$ (see Eq. \ref{eq:vel_disp}) around the linear relation with halo mass. The second one is another log-normal distribution of the measured velocity dispersion $\sigma_{\rm v,M}$ (see Sect. \ref{subsec:optical_sel}) around the true one.

\begin{figure}
    \centering
    \includegraphics[width=\columnwidth]{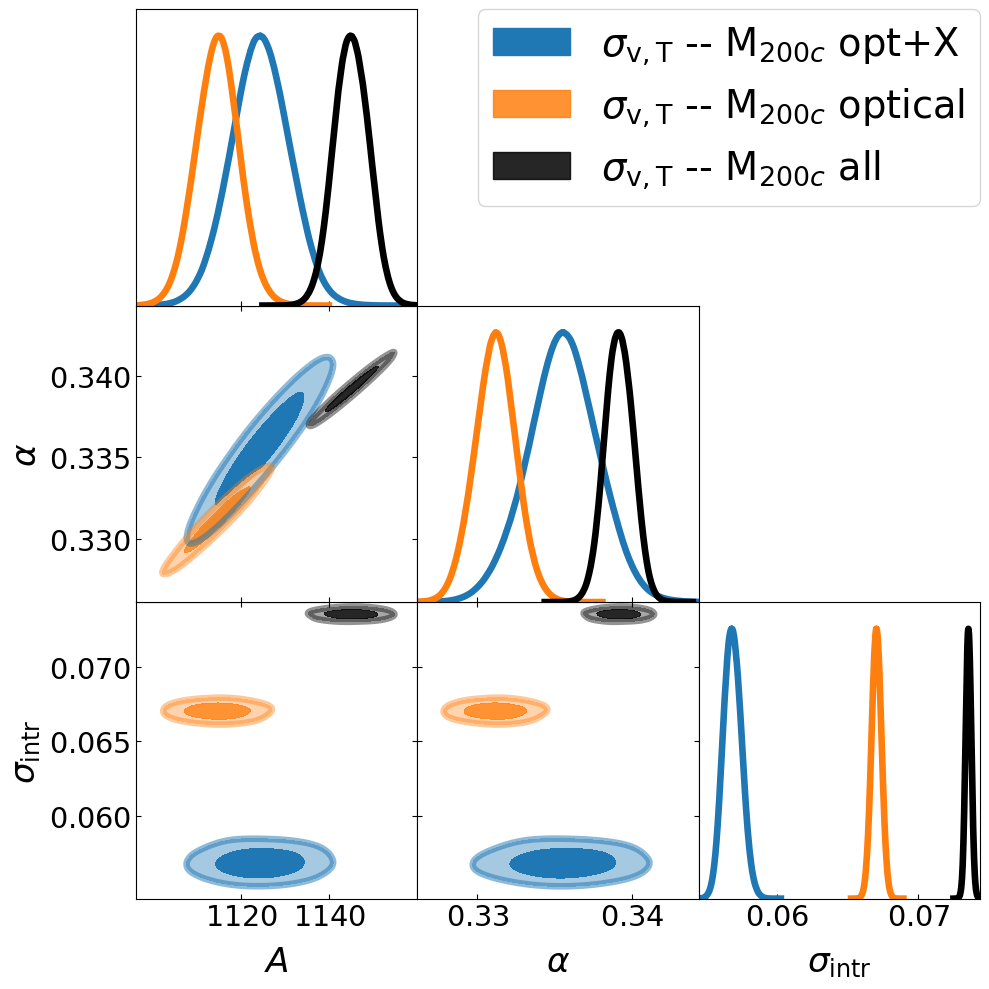}
    \includegraphics[width=\columnwidth]{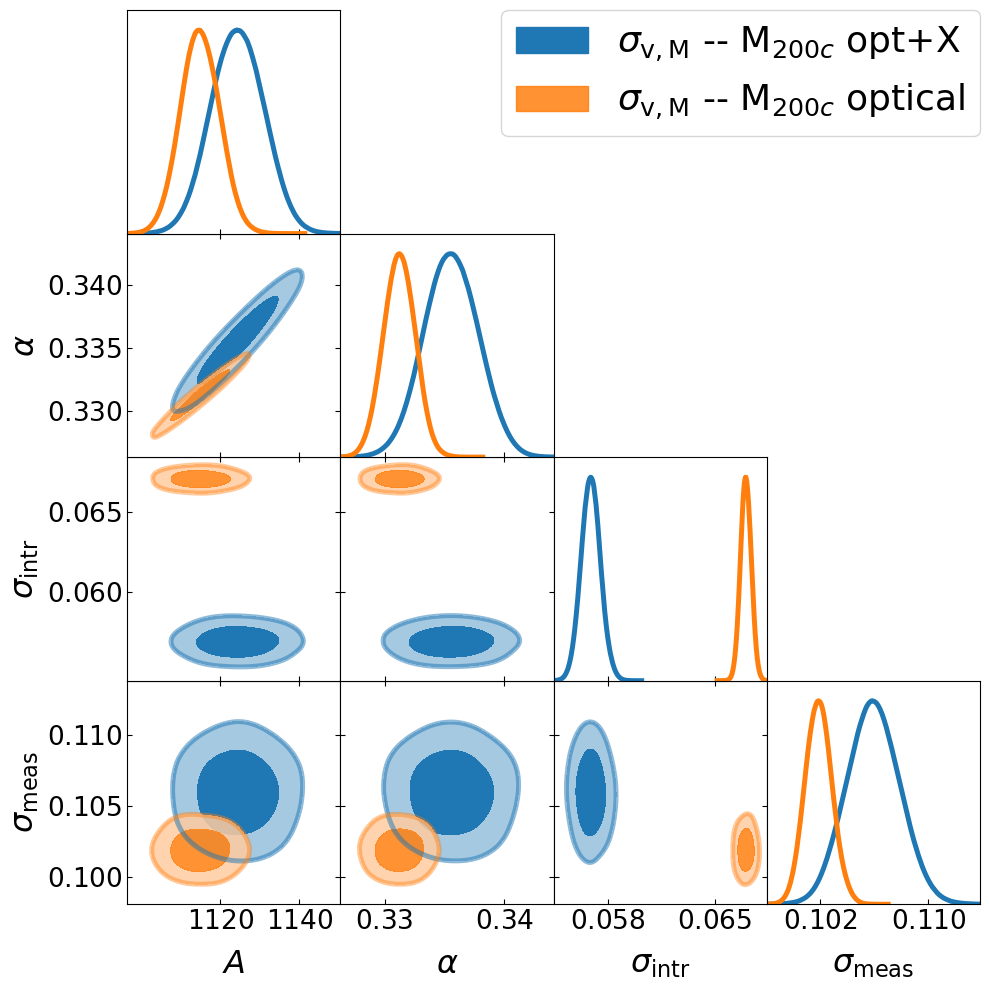}
    \caption{Marginalized posterior distributions of the best fit scaling relation parameters between velocity dispersion and M$_{\rm 200c}$. The filled 2D contours show the 1-$\sigma$ and 2-$\sigma$ confidence levels of the posteriors after convolution with the uniform priors. The model is given by Eq. \ref{eq:scal_rel}. The corresponding 1D parameter constraints are reported in Table \ref{tab:scaling_rel_pars}. The top panel refers to direct relation between true velocity dispersion and halo mass, with black contours referring to the full sample, orange ones to clusters and groups detected in the optical mock, and blue ones with the addition of the X-ray detection. The bottom panel refers to the full model of the scaling relation accounting for the measured velocity dispersion.}
    \label{fig:scalrel_posterior}
\end{figure}

Following the work from \citet{Munari2013MNRAS_veldisp} and \citet{Ferragamo2022EPJWC_veldispmass}, we model the relation between the true velocity dispersion and halo mass using a power law model with normalisation $A$ and slope $\alpha$. In addition, we add the intrinsic scatter of the true velocity dispersion at fixed halo mass $\sigma_{\rm intr}$. Finally, we consider the measured velocity dispersion to be distributed around the true value in a log-normal way with a scatter of $\sigma_{\rm meas}$. We report the two scatters $\sigma_{\rm intr}$ and $\sigma_{\rm meas}$ in units of dex, they are the scatters of the base 10 logarithm of velocity dispersion. We also account for the measurement uncertainty of velocity dispersion on top of the scatter between measured and true velocity dispersion. This mitigates the lower precision of the velocity dispersion measurement for systems with a lower amount of members. 
%\newpage
Overall, our hierarchical formalism reads as follows:

\begin{align}
    P(\sigma_{\rm v,T} | M,z) =& \mathcal{LN} ( \mu\, =\, A\, \Bigg( \frac{h(z)M}{10^{15} M_\odot} \Bigg)^\alpha,  \sigma\, =\, \sigma_{\rm intr}) \nonumber \\
    \sigma_{\rm meas, err} =& \sqrt{\sigma_{\rm meas}^2 + \Bigg( \frac{\delta \sigma_{\rm v, M}}{\sigma_{\rm v, M}\log(10)}\Bigg) ^2} \nonumber \\
    P(\sigma_{\rm v,M}|\sigma_{\rm v,T}) =& \mathcal{LN}(\mu\, =\, \sigma_{\rm v,T},\, \sigma\, =\, \sigma_{\rm meas,err}) \nonumber \\  
    P(\sigma_{\rm v,M}|M,z, \theta) =& P(\sigma_{\rm v,M}|\sigma_{\rm v,T}) P(\sigma_{\rm v,T}|M,z, \theta) P(I|\sigma_{\rm v,T}, z),
    \label{eq:scal_rel}
\end{align}
where $\theta$ is the collection of parameters describing the scaling relation, in our case $A$, $\alpha$, $\sigma_{\rm intr}$, $\sigma_{\rm meas}$. The normalisation $A$ is always expressed in units of km/s. The last term $P(I|\sigma_{\rm v,T}, z)$ is the detection probability. We model it similarly to Eq. \ref{eq:selfunc_model}, but with a single sigmoid function for velocity dispersion. It depends on two parameters, the slope and normalisation, similarly to the corresponding flux terms in Eq. \ref{eq:selfunc_model}. We find a slope of 6.4 (10.8) and a normalisation of 2.64 (2.68) for the optically (optically plus X-ray) detected systems. Accounting for the detection probability allows down-weighting the low mass groups that are less likely to be detected in the forward model of the selection process within the likelihood formalism. We perform five different fits in total. For the first three we focus on the true velocity dispersion. For the last two we account for the measured velocity dispersion and use the full calibration described above. In this section we focus on M$_{\rm 200c}$, for consistency with the literature about velocity dispersion in clusters and groups. In Appendix \ref{appendix:sigmav_M500} we calibrate the relation as a function of M$_{\rm 500c}$, which is the mass measured for X-GAP galaxy groups \citep{Eckert2024_xgap}. 

We note that when studying the scaling relation with real data, disentangling the two components of true and measured velocity dispersion is not feasible. However, this is possible within our framework, as we have access to $\sigma_{\rm v,T}$ in the simulations. This allows us to separate the selection function, defined in terms of true observable halo properties, and the measurement of such observables, that in this case is $\sigma_{v,M}$.

\begin{table*}[]
    \centering
    \caption{Priors and posteriors of the scaling relation between halo mass M$_{\rm 200c}$ and velocity dispersion.}
    
    \begin{tabular}{|c|c|c|c|c|c|c|c|}
    \hline
    \hline
       \textbf{Parameter}  & \textbf{Prior} & \textbf{$\sigma_{\rm v,T}$ All} & \textbf{$\sigma_{\rm v,T}$ Optical} & \textbf{$\sigma_{\rm v,T}$ Opt+X} & \textbf{$\sigma_{\rm v,M}$ Optical} & \textbf{$\sigma_{\rm v,M}$ Opt+X} \\
       \hline
       \rule{0pt}{2ex}    
        $A$ & $\mathcal{U}$(100, 2000) & 1145.1 $\pm$ 3.9 & 1114.6 $\pm$ 4.9 & 1124.3 $\pm$ 6.6 & 1114.9 $\pm$ 4.9 & 1124.4 $\pm$ 6.6\\
        $\alpha$ & $\mathcal{U}$(0.1, 0.5) & 0.330 $\pm$ 0.001 & 0.331 $\pm$ 0.001 & 0.335 $\pm$ 0.002 & 0.331 $\pm$ 0.01 & 0.335 $\pm$ 0.002 \\
        $\sigma_{\rm intr}$ & $\mathcal{U}$(0.01, 1.0) & 0.073 $\pm$ 0.001 & 0.067 $\pm$ 0.001 & 0.057 $\pm$ 0.01 & 0.067 $\pm$ 0.001 & 0.057 $\pm$ 0.001 \\
        $\sigma_{\rm meas}$ & $\mathcal{U}$(0.01, 1.0) & - & - & - & 0.102 $\pm$ 0.001 & 0.106 $\pm$ 0.002 \\
        \hline
    \end{tabular}
    \tablefoot{The symbol $\mathcal{U}(M,N)$ denotes a uniform prior between the values M and N. Posterior values are reported from the third column onward, for each case labelled in the top row.}
    \label{tab:scaling_rel_pars}
\end{table*}

\subsection{True scaling relation}
\label{subsec:true_scal_rel}

We first analyse the three cases focusing on the true velocity dispersion with different selections.
The first case describes the full halo population in our light cone and does not account for any selection effect. The second one adds the optical selection, and the third one combines the optical selection with the X-ray one. Any discrepancy between these cases would allows us to understand the impact of selection effects on the true scaling relation. 

We derive posterior probability distributions and the Bayesian evidence with the MLFriends algorithm using the
UltraNest package. The full marginalized posterior distributions are shown in Fig. \ref{fig:scalrel_posterior}. We find a positive degeneracy between the normalisation and the slope of the scaling relation. This is expected, because the normalisation fixes the amplitude of the relation at high mass, therefore, a shallower slope is required to model the bulk of the population for a smaller normalisation.
For the M$_{\rm 200c}$ case, the 1-D marginalized posteriors are $A=1145.1 \pm 3.9$ and $\alpha = 0.330 \pm 0.001$. We find a scatter for the true velocity dispersion of $\sigma_{\rm intr} = 0.073 \pm 0.001$, confirming the tight distributions presented by \citet{Munari2013MNRAS_veldisp} and \citet{Ferragamo2022EPJWC_veldispmass}. The end result is shown in the top panel of Fig. \ref{fig:veldisp_mass}, displaying the scaling relation between the true velocity dispersion computed using subhalo velocities (see Eq. \ref{eq:vel_disp}) and halo mass. The blue line denotes the best fit model, the red lines include the intrinsic scatter $\sigma_{\rm intr}$. Our result is compatible with the orange and green lines, referring to \citet{Munari2013MNRAS_veldisp} and \citet{Ferragamo2022EPJWC_veldispmass}, respectively. When adding the optical and the X-ray selections, we find a scaling relation with lower slope and normalisation. However, all three best fits are in excellent agreement within the intrinsic scatter of the scaling relation. The intrinsic scatter decreases from 0.073 for the full halo sample to 0.067 when adding the optical selection, and to 0.057 when including also the X-ray selection. This suggests that confirming the presence of a halo with one or multiple proxies gets rid of outliers, with velocity dispersions that are far from the mean scaling relation of fixed halo mass. All the parameters are reported in Table \ref{tab:scaling_rel_pars}.

\subsection{Full calibration}
We perform a full calibration of the velocity dispersion-halo mass relation by accounting for the distribution of the measured velocity dispersion around the true value.
For the fourth and fifth case we add the measured velocity dispersion and apply the full formalism in Eq. \ref{eq:scal_rel}. In the first place, we only study haloes detected in the optical mock, then secondly we add the X-ray selection. 
The bottom panel of Fig. \ref{fig:veldisp_mass} shows $\sigma_{\rm v,M}$ as a function of halo mass. In this case, the red lines include the contribution of the two scatters $\sigma_{\rm intr}$ and $\sigma_{\rm meas}$ summed in quadrature, with the inclusion of measurement uncertainty as in Eq. \ref{eq:scal_rel}. The posterior distribution is shown in the bottom panel of Fig. \ref{fig:scalrel_posterior}. The scatter for the measured velocity dispersion assumes larger values of $\sigma_{\rm meas} = 0.102 \pm 0.001$. The recovery of the normalisation, slope, and intrinsic scatter is in excellent agreement with the true velocity dispersion case analysed in the previous section. This is key, because it means that our formalism in Eq. \ref{eq:scal_rel} allows recovering the base scaling relation also when accounting for the measurement of velocity dispersion. The same holds for the addition of the X-ray selection on top of the optical one, with a similar measurement scatter of $\sigma_{\rm meas} = 0.102 \pm 0.001$. We emphasize that it is not possible to directly disentangle intrinsic and measurement scatter in observational data, as the true velocity dispersion is not accessible. However, this distinction is valuable in the context of forward modelling, where true velocity dispersions can be generated as a function of halo mass including intrinsic scatter, and the observed values are subsequently modelled by incorporating measurement uncertainties around the true values. This approach enables a more accurate connection between observations and halo mass by appropriately accounting for both sources of scatter.

\begin{figure}
    \centering
    \includegraphics[width=\columnwidth]{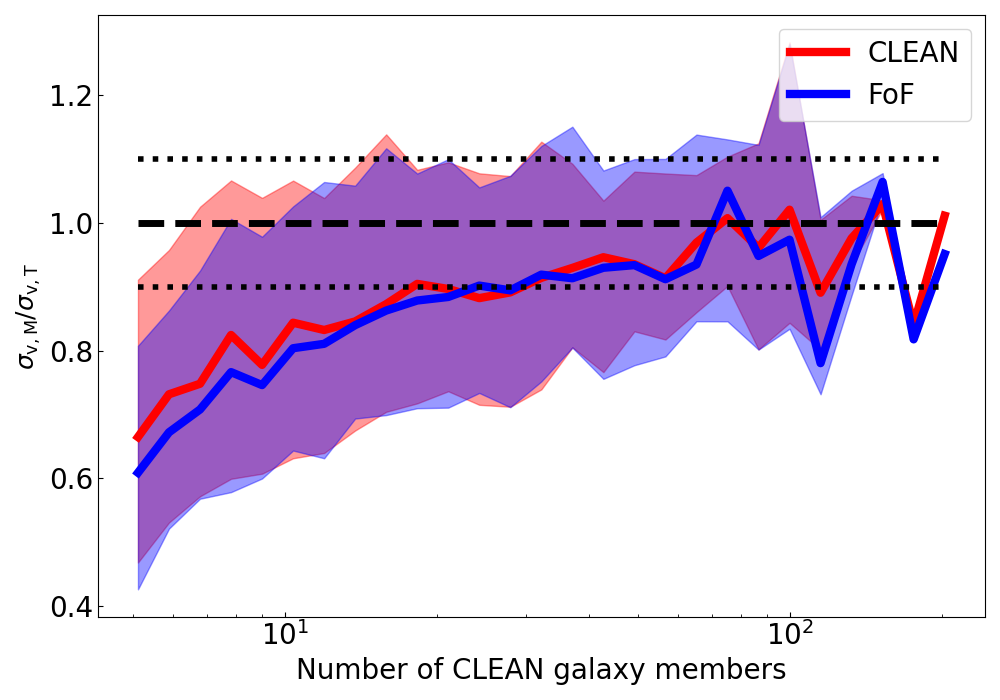}
    \caption{Ratio between measured and true velocity dispersion as a function of number of recovered galaxy members. The red line shows the gapper velocity dispersion measured by the \texttt{CLEAN} algorithm, while the blue one refers to the direct estimate from the FoF finder. The black dashed line denotes the one-to-one ratio, while the dotted ones include a 10$\%$ deviation from unity.}
    \label{fig:sigmaratio_Nmem}
\end{figure}

Finally, we test whether our assumption that the measured velocity dispersion is log-normally distributed around the true value is justified. We measure the ratio between measured and true velocity dispersion as a function of recovered number of \texttt{CLEAN} members. The result is shown in Fig. \ref{fig:sigmaratio_Nmem}. The red line shows the gapper velocity dispersion measured by the \texttt{CLEAN} algorithm, the blue one denotes the estimate of the FoF finder. We do not see significant differences between them here.

The ratio is systematically low for poor systems. About 20 members are required for a 10$\%$ accuracy in the measurement of velocity dispersion. We notice that this does not mean that the members have to be true group members. The mis-identification of nearby galaxies that are not however gravitationally bound to the main halo from the dark matter point of view likely contributes to the scatter around the 1:1 ratio in Fig. \ref{fig:sigmaratio_Nmem}. Our result is consistent with \citet{Marini2025A&A_optical}, who find a large scatter between true halo mass and mass inferred from velocity dispersion spanning values between 25$\%$ and 40$\%$, based on different optical finders for a GAMA-like survey. To assess whether the discrepancy between true and measured velocity dispersion for poor groups and clusters had an impact in our calibration of the total scaling relation, we tested a similar run by allowing the measured velocity dispersion to be biased compared to the true value, i.e.:

\begin{equation}
    P(\sigma_{\rm v,M}|\sigma_{\rm v,T}) = \mathcal{LN}(\mu\, =\, m\times \sigma_{\rm v,T}+q,\, \sigma\, =\, \sigma_{\rm meas,err}),
    \label{eq:sigmav_bias}
\end{equation}

where $m$, $q$ describe the linear relation between the logarithmic values of measured and true velocity dispersion. On the one hand, we obtain $m=0.85 \pm 0.01$ and $q=0.06 \pm 0.01$, in agreement with the fact that their ratio does not follow a perfect 1:1 relation, as suggested by Fig. \ref{fig:sigmaratio_Nmem}. This also means that estimating mass via velocity dispersion can be problematic also for massive galaxy clusters if the cluster is poor. This is a matter of galaxy population rather than halo mass. The majority of X-GAP groups are rich \citep{Eckert2024_xgap}, so accounting for velocity dispersion in the mass calibration is justified for large part of the sample. On the other hand, the scaling relation parameters are unbiased compared to the standard case (see Fig. \ref{fig:scalrel_posterior}, \ref{fig:veldisp_mass}). We obtain a normalisation of 1124.3 $\pm$ 6.5, a slope of 0.335 $\pm$ 0.01, and an intrinsic scatter of 0.056$\pm$ 0.001, all in great compatibility with the last column of Table \ref{tab:scaling_rel_pars}. The additional uncertainties in the measurement of velocity dispersion just contributes to the measurement scatter, in fact we obtain a value of 0.078 $\pm$ 0.002 compared to 0.106 $\pm$ 0.02. We note that the linear model in Eq. \ref{eq:sigmav_bias} may not perfectly capture an effect that vanishes for rich systems. Nevertheless, given the results presented above, we expect that a more complex model would yield an even lower measurement scatter while preserving the same scaling relation parameters. We conclude that when using velocity dispersion as a mass proxy, accounting for measurement uncertainties and modelling the distribution around its true value, even with a simple lognormal function, is a key step towards retrieving an unbiased relation to halo mass.

\section{Discussion and conclusions}
\label{sec:concl}
Here we summarize our work and further discuss the results obtained in the previous sections in terms of the group's surface brightness and dynamical state.

\subsection{Surface brightness shape}
\label{subsec:SBshape}

\begin{figure}
    \centering
    \includegraphics[width=\columnwidth]{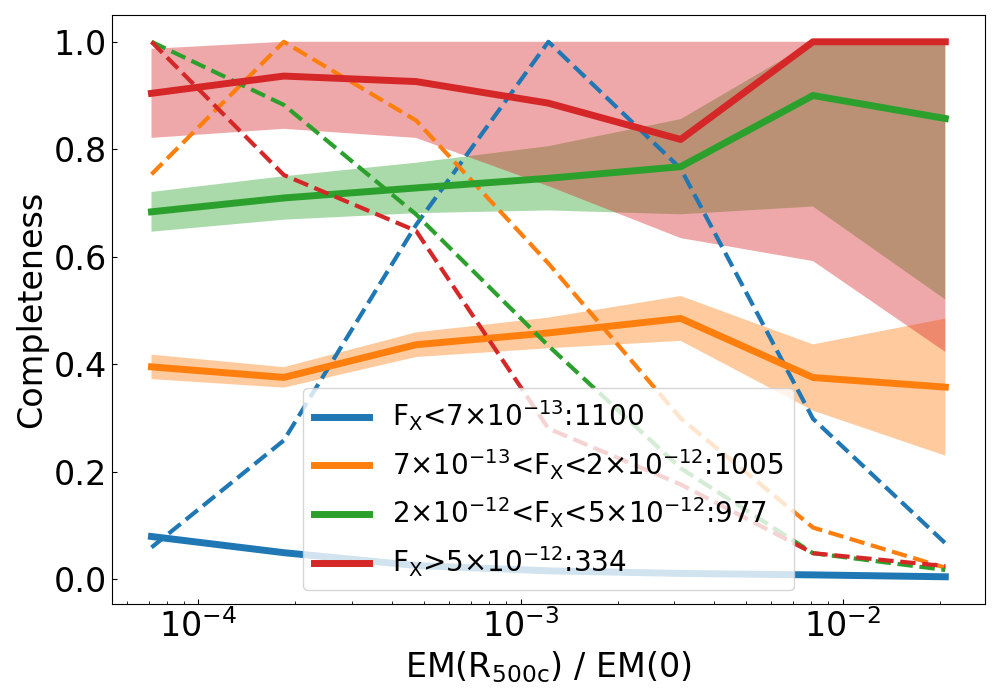}
    \includegraphics[width=\columnwidth]{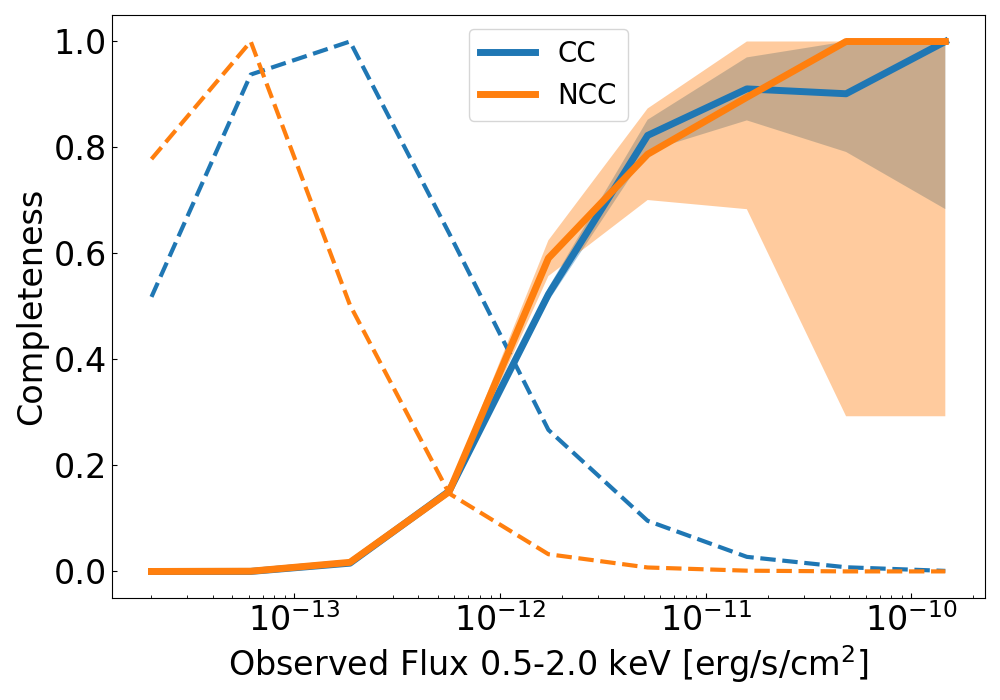}
    \caption{Probability of detection as a function of core properties. The dashed lines represent the normalised distribution of the full population. \textbf{Top panel}: completeness fraction as a function of emissivity profile ratio measured at R$_{\rm 500c}$ and in the inner most bin. Different flux intervals are encoded in different colours. \textbf{Bottom panel}: completeness fraction as a function of flux for the cool core (non cool core) systems with profile ratio smaller (larger) than the median values of 0.001.} 
    \label{fig:Pdet_SB}
\end{figure}

The wavelet detection scheme is designed to detect groups and clusters using the emission from the outskirts. It is therefore not sensitive to the core properties, such as peaked surface brightness profile in relaxed systems with an efficient cooling \citep{Eckert2011A&A...526A..79E}. Because of this, we did not introduce a direct correlation between the dynamical state of dark matter haloes and X-ray core properties. 

Here we test whether this assumption is valid by studying the detection probability as a function of the shape of the surface brightness profiles generated from the neural network in Sect. \ref{sec:clu_model}. We encode the surface brightness shape in the ratio between the emissivity at R$_{\rm 500c}$ and in the inner most bin. If the ratio is low, the system is more similar to a cool core with a steep profile. On the opposite, a large ratio means that the profile is flatter and the system is closer to a non cool core. The result is shown in Fig. \ref{fig:Pdet_SB}. The top panel shows the completeness fraction as a function of the profile ratio, for different flux intervals identified by the various colours. We do not see a clear trend of the detection probability as a function of the profile ratio, meaning the detection process is not highly sensitive to the surface brightness shape. This holds from faint systems with X-ray flux below 7$\times$10$^{-13}$ erg/s/cm$^2$ where the completeness is close to zero, up to bright systems with flux larger than 5$\times$10$^{-12}$ erg/s/cm$^2$, with detection probability larger than 80$\%$. In the bottom panel of Fig. \ref{fig:Pdet_SB} we split our population in cool core and non cool core by selecting systems with profile ratio respectively below and above the median ratio equal to 0.55. We show the completeness fraction as a function of flux for these two populations and do not find a significant difference between them. Both panels in Fig. \ref{fig:Pdet_SB} confirm that the wavelet detection scheme is not sensitive to the dynamical state and the core properties of clusters and groups in our simulation.
Finally, assuming a constant metallicity profile is justified, as variations in metallicity would primarily affect the shape of the surface brightness profile, to which our selection method is largely insensitive.

\subsection{Dynamical state}

\begin{figure}
    \centering
    \includegraphics[width=\columnwidth]{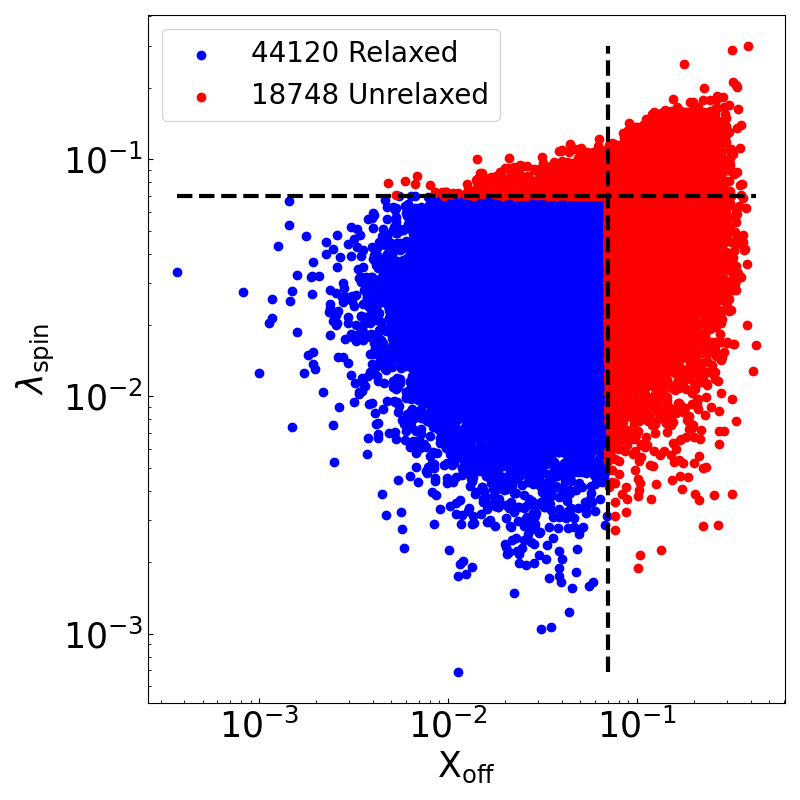}
    \caption{Selection of relaxed and unrelaxed groups from the dark matter halo point of view, using the spin and offset parameters.}
    \label{fig:relunrel}
\end{figure}

The dynamical state of dark matter haloes hosting galaxy clusters and groups has been vastly studied in past decades, both in theory and simulations \citep{Neto2007MNRAS.381.1450N, Prada2012MNRAS.423.3018P, Klypin2016_MD, Seppi2021A&A_MF, Deluca2021_300dynstate}, and observations \citep{Wojtak2010_orbits, Eckert2011A&A...526A..79E, Ota2020PASJ...72....1O,  Ghirardini2022A&A_efeds, Seppi2023A&A_offset, Cerini2023ApJ_dynstate}. From the purely dark matter point of view, various parameters linked to the halo rotational properties and/or merger history have been used to characterize the dynamical state of haloes. A popular choice is a combination of the spin parameter and the offset parameter. They are defined as:

\begin{align}
    \lambda_{\rm P} &= \frac{J\sqrt{E}}{GM^{5/2}} \nonumber \\
    X_{\rm off} &= \frac{|R_{\rm peak} - R_{\rm CM}|}{R_{\rm vir}},
    \label{eq:dynstate_pars}
\end{align}

where $J$ is the halo angular momentum, $E$ its total energy, $R_{\rm peak}$ the position of the deepest point in the potential well, and $R_{\rm CM}$ its centre of mass. A disturbed or recently merged halo has a high ratio of rotational kinetic to gravitational energy, and a large offset between the peak and centre of mass positions. Therefore, both the spin and offset parameter values will be higher compared to a relaxed halo, according to the definitions in Eq. \ref{eq:dynstate_pars}. Following the popular thresholds for these parameters of 0.07 \citep[see e.g.,][]{Neto2007MNRAS.381.1450N, Klypin2016_MD, Seppi2021A&A_MF}, we divide our halo population detected in the optical mock in a relaxed and an unrelaxed sample. The sample selection is shown in Fig. \ref{fig:relunrel}. It displays the halo distribution in the spin-offset parameter plane, with relaxed haloes shown in blue and disturbed ones in red. 

\begin{figure}
    \centering
    \includegraphics[width=\columnwidth]{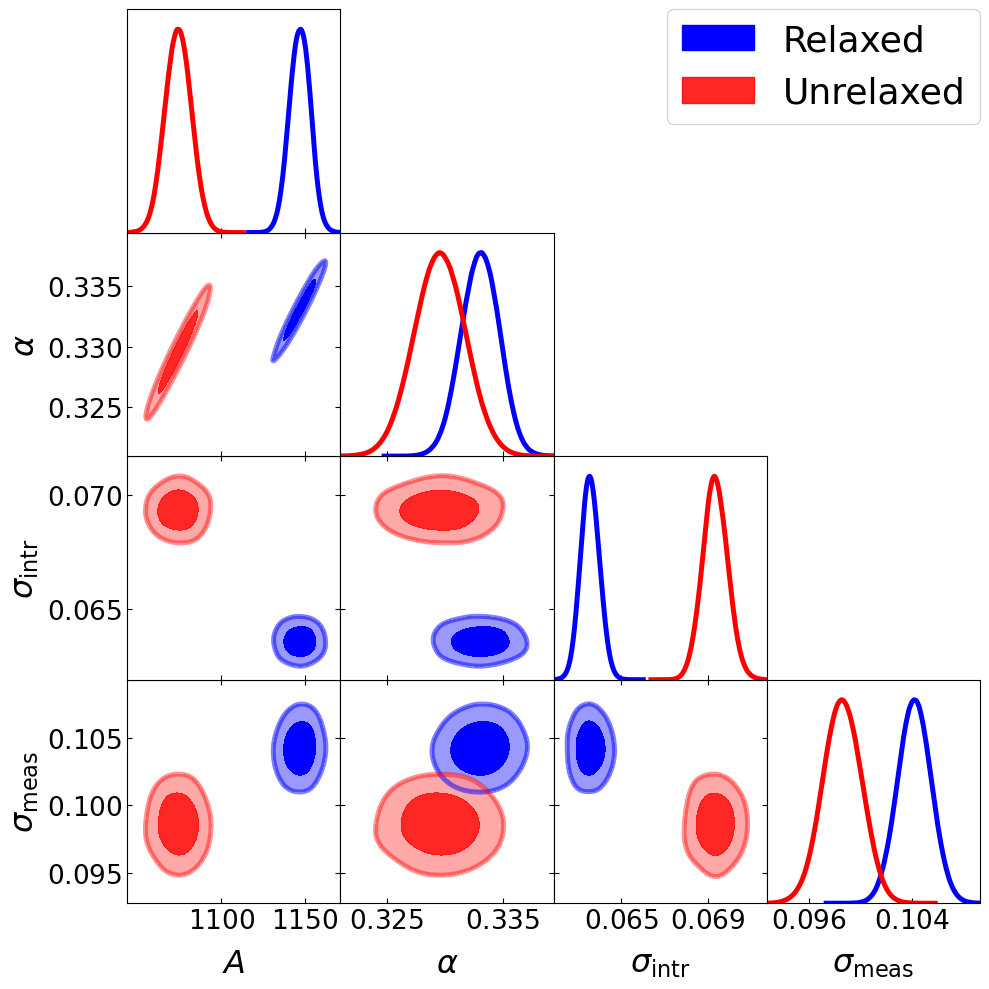}
    \caption{Full marginalized posterior distribution for the scaling relation between halo mass and velocity dispersion for relaxed and unrelaxed haloes.}
    \label{fig:corner-relunrel}
\end{figure}

We repeat the scaling relation analysis explained in Sect. \ref{sec:veldisp_mass} to the individual samples selected from haloes matched with an optical detection.
We use the same priors as the ones for the primary analysis in Table \ref{tab:scaling_rel_pars}. The results are qualitatively similar, the posterior distributions are shown in Fig. \ref{fig:corner-relunrel}. The slope of the scaling relation is compatible within 1$\sigma$ for the two populations, assuming values of 0.333 $\pm$ 0.002 and 0.329 $\pm$ 0.002. However, the normalisation is significantly different, meaning that at fixed halo mass, a relaxed halo tends to host a group or cluster with higher velocity dispersion. The normalisation is 1147.1 $\pm$ 6.3 for the relaxed population, compared to 1074.0 $\pm$ 7.8 for the disturbed one. One may expect the opposite, with unrelaxed groups that are in the evolutionary stage of undergoing tidal interactions and mergers hosting a galaxy population with irregular velocities, ultimately increasing velocity dispersion at fixed mass.

\begin{figure}
    \centering
    \includegraphics[width=\columnwidth]{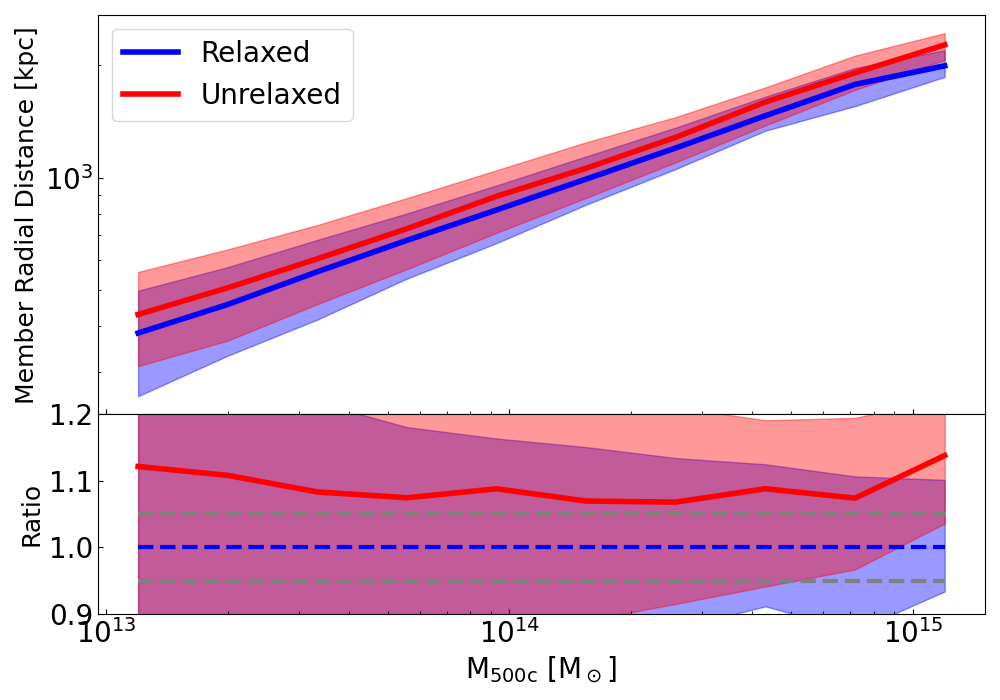}
    \caption{Average galaxy members' distance from the halo center as a function of halo mass. The blue line denoted relaxed haloes, the red line refers to disturbed ones.}
    \label{fig:Member_distance}
\end{figure}

To investigate this discrepancy, we measure the average distance of the recovered galaxy members from the halo centre as a function of halo mass. The result is shown in Fig. \ref{fig:Member_distance}. In particular, the bottom panel shows the ratio of the average member distance between the disturbed and the relaxed population. It is always higher than 1 throughout the whole mass range, spanning between an increase of at least 5$\%$, up to 10$\%$ at different masses. This means that for relaxed haloes, the member galaxies are probing a more central region of the gravitational potential compared to the disturbed case. Velocity dispersion tends to increase radially up to about the scale radius \citep{Hoeft2004ApJ_veldisp_prof, Faltenbacher2006MNRAS_veldispprof}, and tends to decrease toward the outer regions. Since both radii used to describe clusters and groups, such as R$_{\rm 200c}$ or R$_{\rm 500c}$, are significantly larger than the scale radius \citep[depending on halo concentration, see][]{Klypin2016_MD}, the members of relaxed groups and clusters are tracing a region of the potential with larger velocity dispersion, explaining the higher normalisation of the scaling relation measured in Fig. \ref{fig:corner-relunrel}.

We stress that our formalism (see Eq. \ref{eq:scal_rel}) within the full calibration allows us to probe the true velocity dispersion, which is unaffected by interlopers. This implies that the observed difference in normalisation is encoded in the radial distribution of member galaxies, which varies with the dynamical state of the halo. This effect plays a more significant role than the presence of galaxies with irregular velocities in disturbed systems. The latter effect rather contributes to the intrinsic scatter: the relaxed haloes trace a population with a velocity dispersion that lies closer to the mean relation to halo mass compared to the disturbed ones, in fact the intrinsic scatter is lower at 0.063 $\pm$ 0.001 compared to 0.069 $\pm$ 0.001. Because of projection, it is more likely to observe a merger closer to the 2D sky plane than along the 1D line of sight. Nonetheless, the line of sight velocity dispersion is able to capture the larger scatter in disturbed haloes. Finally, one might expect disturbed haloes to contain a higher fraction of interlopers, which would introduce additional scatter in the measured velocity dispersion. However, we find that the measured dispersions for disturbed ($\sigma_{\rm meas}$=0.099$\pm$0.002) and relaxed ($\sigma_{\rm meas}$=0.104$\pm$0.002) haloes are compatible within 1.7$\sigma$, indicating no statistically significant difference. This suggests that \texttt{CLEAN} is effectively removing interlopers in both relaxed and disturbed systems, ensuring robust velocity dispersion measurements.

\subsection{Summary and comparison to other works}
In this work, we produced and analysed a full end-to-end mock sky in the optical and X-ray wavelengths, with the goal of forward modelling the selection function of the X-GAP galaxy groups sample \citep{Eckert2024_xgap}, a collection of 49 galaxy groups selected by cross-matching extended sources in the Rosat All Sky Survey (RASS) and FoF galaxy groups and clusters in the SDSS area \citep{Tempel2017A&A_sdssFOF, Damsted2024_axes}. The workflow is summarized in Fig. \ref{fig:model_diagram}. We built a dark matter halo light cone using individual snapshots of the Uchuu simulation \citep{Ishiyama2021MNRAS.506.4210I_UCHUU} (see Fig. \ref{fig:lightcone}) and adapted dedicated simulations of the SDSS sky \citep{DongPaez2024MNRAS_UchuuSDSS}. 
We developed a novel method to populate dark matter haloes with X-ray emission using an optimized neural network informed by hydrodynamical simulations, to simulate emissivity profiles and temperatures with the proper covariances in terms of mass and redshift. We demonstrated that our model is able to predict the observed distributions of clusters and groups profiles and observables, recovering scaling relations with halo mass for X-ray luminosity and temperature with observations (see Sect. \ref{sec:clu_model}, Fig. \ref{fig:EM_profiles} and \ref{fig:Clumodel_SR_validation}). We used the AGN model from \citet{Comparat2019MNRAS_agn_model} and generated a mock X-ray diffuse background by re-sampling the real RASS background maps. We generated X-ray photons using SIXTE \citep{Dauser2019_SIXTE}, accounting for the instrumental response and observing strategy of the spacecraft. \\
We run a wavelet based detection scheme \citep{Vikhlinin1998ApJ_wvdet, Kaefer2019A&A...628A..43K} on the mock X-ray images and a FoF finder on the mock galaxies to obtain galaxy group candidates. We cross-matched the output catalogues with the dark matter halo light cone using the information stored in the source ID generating each X-ray event and the parent IDs stored for each galaxy member of the optical groups. An example for a 1 square degree field covering a massive cluster in the mock is shown in Fig. \ref{fig:image_example}.

We measured the detection probability by comparing the histogram of detected haloes to the complete population as a function of flux, velocity dispersion, and redshift. The result is shown in Fig. \ref{fig:Pdet2d}. In total, we simulated 62$\,$868 clusters and groups at z<0.14. The global completeness is 6.3$\%$, with the X-ray mock being the limiting factor, with a completeness of 8.8$\%$, compared to 28.1$\%$ for the optical FoF catalogue alone. We found that the detection probability mainly depends on flux, while it weakly depends on velocity dispersion and redshift. This confirms that the X-GAP selection is mainly X-ray flux limited, due to the relatively shallow coverage of the RASS. We found a 50$\%$ detection probability at about 1.5$\times$10$^{-12}$ erg/s/cm$^2$. This 50$\%$ flux limit is about five times larger compared to the estimate for the first eROSITA all sky survey (eRASS1), carried out by \citet{Seppi2022A&A_erass1twin}, at about 3$\times$10$^{-13}$ erg/s/cm$^2$. Since the average exposure time is comparable between eRASS1 and the X-GAP area covered by the RASS, the difference resides in the higher eROSITA sensitivity compared to ROSAT, thanks to its efficient grasp, i.e. the effective collecting area multiplied by the field of view, that is larger than the ROSAT one by about three at 1 keV. Previous work by \citet{Clerc2018A&A...617A..92C} estimated an average 50$\%$ completeness at 6$\times$10$^{-15}$ erg/s/cm$^2$ for the cumulative sum of eight eROSITA all sky surveys. In this case, the much lower flux limit is also due to the increment in exposure time, reaching average values of more than 1.5 ks compared to our average coverage of about 400 s. Indeed, pointed programs such as the XMM-XXL survey \citep[][]{Pierre2016A&A_XXL}, covering about 50 square degrees with an exposure time of about 10 ks, allows reaching faint fluxes of 2$\times$10$^{-14}$ erg/s/cm$^2$ at z=0.2 \citep{Pacaud2006MNRASXXLselfunc}, with an effective area that is comparable to eROSITA. The different flux limits also depend on the set up of a detection algorithm and source classification, which makes them dependent on the scientific goal of specific project.

Various approaches have been adopted in the literature to model the selection function, from empirical methods \citep{Bocquet2019ApJ_SPTcosmo}, to using analytical models \citep{Clerc2018A&A...617A..92C, IderChitham2020MNRAS_codex}, to more complex interpolation schemes \citep{Clerc2024A&A_eroSF}. The latter is preferable if one needs to model the selection on a very large parameter space, in such case the effort becomes computationally very expensive. For example, \citet{Clerc2024A&A_eroSF} combined 100 different realizations of a single mock to model the eRASS1 cluster detection probability, accounting for the very large variations of exposure time and local background across the eROSITA sky, in addition to count rate, redshift, and various combinations of morphological parameters.
In comparison, the X-GAP area is covered with a stable background and more uniform exposure due to marginal overlap between the deep region around the ecliptic pole and the SDSS area. We modelled the selection function with a sigmoid distribution accounting for a main flux component with secondary slopes describing the redshift and velocity dispersion trends. We provided two models for the selection function, one based on the observed flux within R$_{\rm 500c}$, ideal for forward modelling population studies, and a second one using flux within a fixed angular aperture of 6 arcminutes, useful for studies starting from the perspective of the observations. Both models capture the smooth increment of the detection probability as a function of flux and its secondary trend with increasing velocity dispersion and decreasing redshift (see Fig. \ref{fig:Pdet_selfunc}). The expression for our model is in Eq. \ref{eq:selfunc_model}, the best fit parameters are reported in Table \ref{tab:selfunc_model_pars}, and their posterior distribution is shown in Fig. \ref{fig:corner_selfunc}.

We analysed the purity of the selection method by comparing the fraction of detections that is not matched to an input halo. We studied this fraction as a function of velocity dispersion in Fig. \ref{fig:purity-veldisp}. We found that the X-ray data serves as a cleaning of the optical catalogue, reaching 90$\%$ purity already at about 280 km/s, compared to 600 km/s for the optical FoF group catalogue. A similar comparison is performed by \citet{Marini2025A&A_optical} and \citet{Popesso2024arXiv_simsSTACKS}. Together with the FoF algorithm from \citet{Tempel2017A&A_sdssFOF}, they test the optical finders from \citet{Yang2005MNRAS_optfind} and \citet{Robotham2011MNRAS_gamagrps}. They study completeness and purity as a function of halo mass instead of velocity dispersion, however our calibration of the mass to velocity dispersion scaling relation allows a comparison with their work. At M$_{\rm 200c}$=2$\times$10$^{13}$ M$_\odot$ they report a contamination level between about 6$\%$ and 15$\%$ depending on the optical finder. Although there is a larger scatter at low masses, at a corresponding velocity dispersion of about 250 km/s, we obtain a higher contamination of about 35$\%$. The discrepancy is reduced after the X-ray cleaning, where our purity increases to about 85$\%$. There are multiple effects contributing to such difference. \citet{Marini2025A&A_optical} focused on a GAMA-like spectroscopic survey instead of SDSS, allowing them to reach fainter magnitudes down to 19.8 \citep{Popesso2024arXiv_simsSTACKS}. Their mock extends to z=0.2, therefore they are probing fainter magnitudes compared to our work but not necessarily a galaxy population with lower stellar mass. This is consistent with our result in Fig. \ref{fig:purity-veldisp}, with purity decreasing at low redshift. In addition, our matching procedure is different, as we directly involve velocity dispersion in the matching (see Sect. \ref{sec:catalogues}). One additional difference is that we do not include X-ray binaries (XRB) in our mock, which lack baryons in the first place. \citet{Popesso2024arXiv_simsSTACKS} showed that the XRB contribution to the total X-ray emission is at most 5$\%$ for small groups around 10$^{13}$ M$_\odot$, and it drastically drops moving to the cluster scale. A similar result using observations comes from \citet{Anderson2015MNRAS.449.3806A}, who report an X-ray luminosity from XRBs around 10$^{40}$ erg/s at stellar mass M$_\star$=2$\times$10$^{11}$ M$_\odot$, while the total luminosity is closer to 10$^{40}$ erg/s. According to \citet{Tinker2021ApJ_grpssdss} this range corresponds to an halo mass around 8$\times$10$^{12}$ M$_\odot$. Therefore, we consider this a negligible limitation for the purpose of this work. Nonetheless, accounting for XRBs will be relevant in the future model comparison of thermodynamical profiles to X-GAP. Finally and most importantly, the selection of the parent optical catalogue is different, as they push down to galaxy pairs, while we focus on groups with at least five members. \\
The latter point also boosts completeness in \citet{Popesso2024arXiv_simsSTACKS}, reaching levels higher than 80$\%$ in the group regime, compared to our result around 20$\%$ at a velocity dispersion of about 300 km/s. This is expected, because the majority of these groups are detected with a low number of members, which are naturally filtered out by our selection of at least five members. Finally, the construction of the optical simulation also plays a role. On the one hand, we use the abundance matching prescription from \citet{DongPaez2024MNRAS_UchuuSDSS}, which recovers the SDSS optical properties by construction. On the other hand, \citet{Marini2025A&A_optical} built an optical mock with a stellar mass function skewed towards massive centrals, and tested their analysis with a different simulation built using the UniverseMachine software \citep{Behroozi2019MNRAS_UNIVERSEMACHINE}. They showed that their completeness is reduced to about 60$\%$ for galaxy groups, therefore reducing the difference with our work. Estimating completeness and purity levels using simulations is strongly sensitive to the specific survey setup and assumptions in the detection scheme. A careful modelling of the whole process is key to properly assess the systematics for different works. 

Finally, we calibrate the scaling relation between halo mass and velocity dispersion. We account for two different types of scatter, the first is the intrinsic one, and the second is due to the uncertainty in the measurement of velocity dispersion. The scaling relation is shown in Fig. \ref{fig:veldisp_mass}, and the posterior distribution of the best fits parameters is in Fig. \ref{fig:scalrel_posterior}. Our results are in agreement with previous work on simulations \citep{Munari2013MNRAS_veldisp, Ferragamo2022EPJWC_veldispmass}. We find that the scatter is dominated by measurement uncertainty, assuming values of 0.102 and 0.106 for groups detected in the optical mock and with the addition of the cross correlation to X-ray. These are about 60$\%$ larger than the respective intrinsic scatters of 0.067 and 0.057. We also find that the measurement of velocity dispersion is accurate compared to its true value only for rich systems, with a bias smaller than 10$\%$ for systems with at least twenty members (see Fig. \ref{fig:sigmaratio_Nmem}). For smaller groups the velocity dispersion is biased towards lower values. This is true also for more massive systems if they are poor. Our result is in agreement with \citet{Marini2025A&A_optical}, who obtain halo masses inferred from velocity dispersion that are biased low compared to true mass from groups to clusters using the FoF finder from \citet{Tempel2017A&A_sdssFOF}. In general, accounting for the distribution of measured velocity dispersion around its true value is key to recover an unbiased scaling relation to halo mass. In the context of forward modelling, this is usually done for mass proxies, especially in cluster counts cosmological studies in the X-ray or millimeter bands, using X-ray count rate \citep{Ghirardini2024_erass1cosmo} or Y-SZ \citep{Bocquet2025PhRvD.111f3533B}.

The mocks and the selection function developed in this article will be used a benchmark when comparing the real X-GAP data to hydrodynamical simulations in future work.

\begin{acknowledgements}

The authors thank the anonymous referee for their helpful comments that improved the quality of the article. RS and DE are supported by Swiss National Science Foundation project grant \#200021\_212576.
RS thanks S. Krippendorf and W. Hartley for useful discussions about neural networks for this project. RS thanks F. Prada for discussions about the Uchuu simulations. ET acknowledges funding from the HTM (grant TK202), ETAg (grant PRG1006) and the EU Horizon Europe (EXCOSM, grant No. 101159513). KK acknowledges support from the South African Radio Astronomy Observatory (SARAO; www.sarao.ac.za). MAB acknowledges support from a UKRI Stephen Hawking Fellowship (EP/X04257X/1).
We thank Instituto de Astrofisica de Andalucia (IAA-CSIC), Centro de Supercomputacion de Galicia (CESGA) and the Spanish academic and research network (RedIRIS) in Spain for hosting Uchuu DR1, DR2 and DR3 in the Skies and Universes site for cosmological simulations. The Uchuu simulations were carried out on Aterui II supercomputer at Center for Computational Astrophysics, CfCA, of National Astronomical Observatory of Japan, and the K computer at the RIKEN Advanced Institute for Computational Science. The Uchuu Data Releases efforts have made use of the skun@IAA\_RedIRIS and skun6@IAA computer facilities managed by the IAA-CSIC in Spain (MICINN EU-Feder grant EQC2018-004366-P).

\end{acknowledgements}

\bibliographystyle{aa}
\bibliography{biblio} 

\appendix

\section{Rescaled Luminosity Test}
\label{appendix:rescale_test}

Our cluster and group model, used to assign X-ray properties to dark matter haloes, is based on TNG. Compared to a stacking analysis of X-ray data with eROSITA from \citet{Zhang2024A&A_LxM}, it over predicts X-ray luminosities around 10$^{13}$ M$_\odot$ by about a factor 2 (see Sect. \ref{subsec:clumodel_results}). By construction, our selection function for the X-ray plus optically selected clusters and groups is not affected by this particular prescription because it is formulated in terms of observables. A given model prescription will change the relation between luminosity and halo mass. On the one hand, at a fixed mass, the detection rate could change based on a given set up for the X-ray model. On the other hand, different models will map different luminosities at fixed mass. This does certainly impacts the detection rate at a given halo mass, but the same does not necessarily hold at a given flux. Nonetheless, we test such hypothesis by rescaling the input X-ray luminosities by 10$\%$ in the mock number 3. We regenerate the X-ray events and process this rescaled version of the simulation through our end to end pipeline. We finally compare the detection rate as a function of X-ray flux to the standard processing. 

\begin{figure}[h]
    \centering
    \includegraphics[width=\columnwidth]{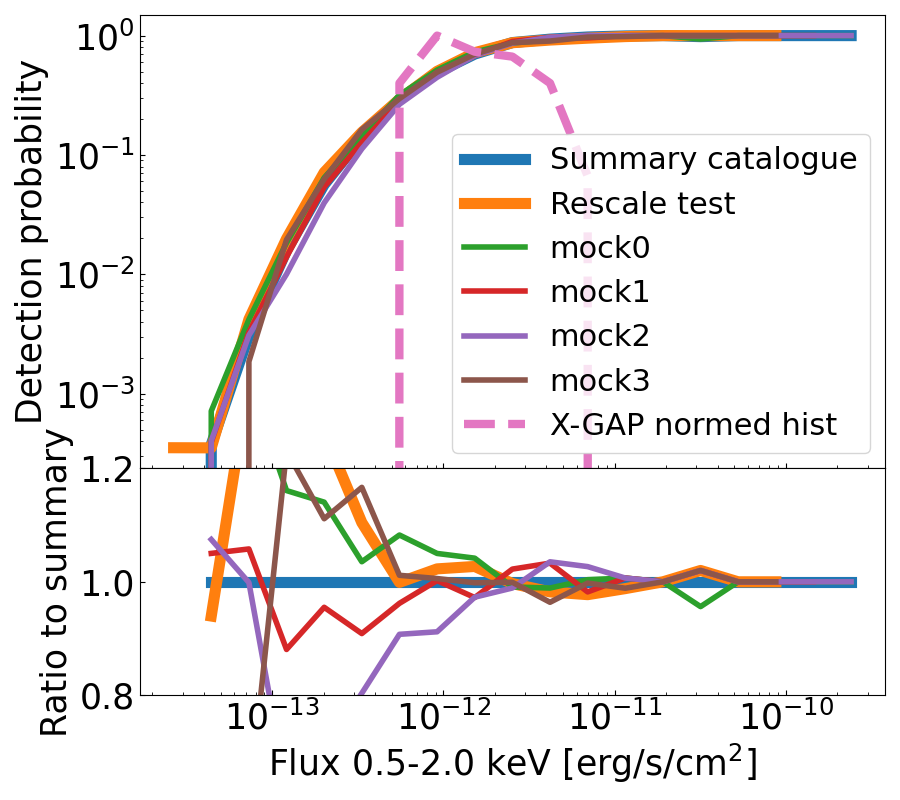}
    \caption{Detection probability as a function of X-ray flux. The collection of the four mocks processed with the standard pipeline is shown in blue, the rescale test is in orange. The individual mocks are displayed with thinner coloured lines. The pink line denotes the normalised histogram of the X-GAP fluxes. The bottom panel shows the ratio between each catalogue and the collection of the four mocks.}
    \label{fig:Rescale_test}
\end{figure}

The result is shown in Fig. \ref{fig:Rescale_test}. The collection of the four mocks is shown in blue. The detection probability in the rescaled simulation is in orange. The results from the standard processing of each individual mock is shown by the thinner lines. The bottom panel displays the ratio between each experiment and the reference collection used for the analysis in the main body of this paper. There is almost a perfect agreement between the rescaled version of the mock and the summary catalogue until 5$\times$10$^{-13}$ erg/s/cm$^2$, where the total completeness is about 20$\%$. Moreover, the agreement between the rescaled experiment and mock number 3, the one used to generate the rescaling in the first place, reaches even fainter fluxes at about 3$\times$10$^{-13}$ erg/s/cm$^2$, where the detection probability is low, at about 5$\%$. At fainter fluxes, both distributions are noisy, with less than 250 detected groups in total. We conclude that our selection method is not affected by assumptions included in the X-ray model for cluster and groups. In addition, the pink line shows the normalised flux distribution of the X-GAP groups presented in \citet{Eckert2024_xgap}. There is no object fainter than about 5$\times$10$^{-13}$ erg/s/cm$^2$. For this flux threshold we have an excellent agreement between various mocks and the rescaled test. The selection function formulated in this work is robust for the X-GAP sample.

\section{Velocity dispersion and M$_{\rm 500c}$}
\label{appendix:sigmav_M500}

\begin{figure}
    \centering
    \includegraphics[width=0.8\columnwidth]{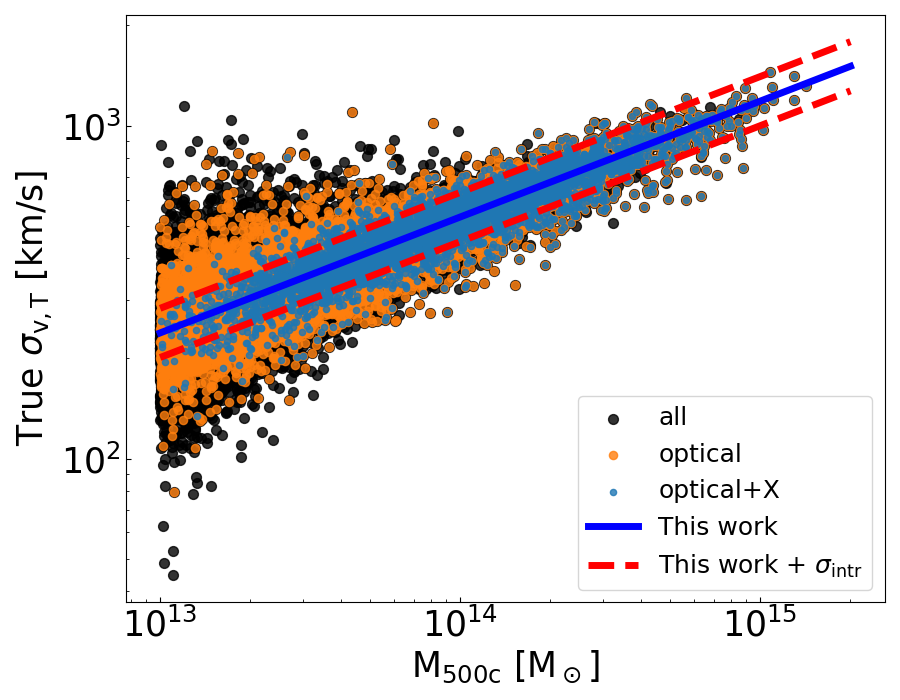}
    \includegraphics[width=0.8\columnwidth]{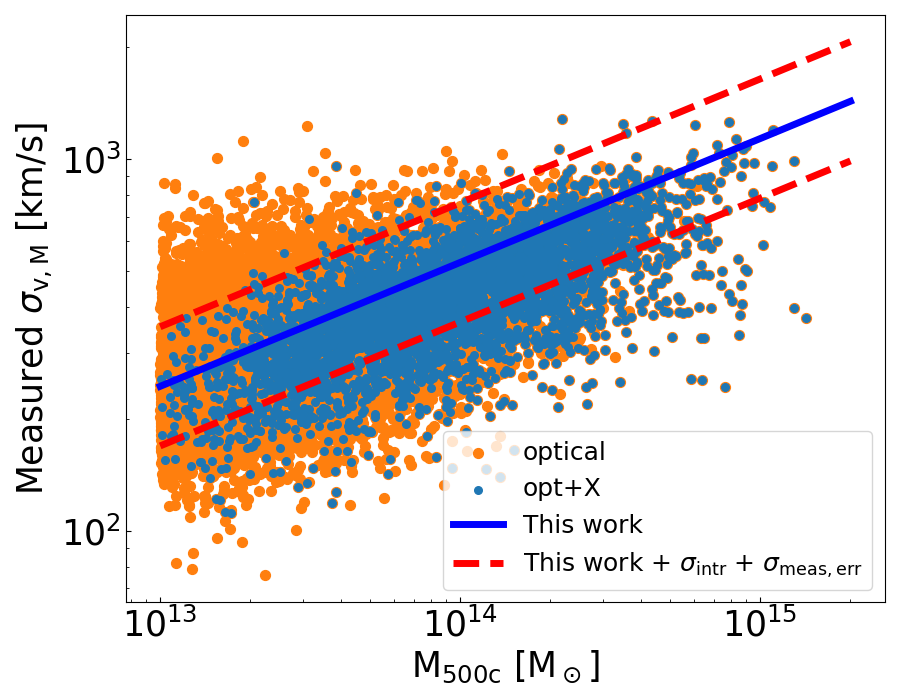}    
    \caption{Scaling relation between velocity dispersion and halo mass M$_{\rm 500c}$. The top and bottom panels are the same as in Fig. \ref{fig:veldisp_mass}, but the reference mass is M$_{\rm 500c}$ instead of M$_{\rm 200c}$.}
    \label{fig:veldisp_mass_500c}
\end{figure}

\begin{figure*}
    \centering
    \includegraphics[width=0.8\columnwidth]{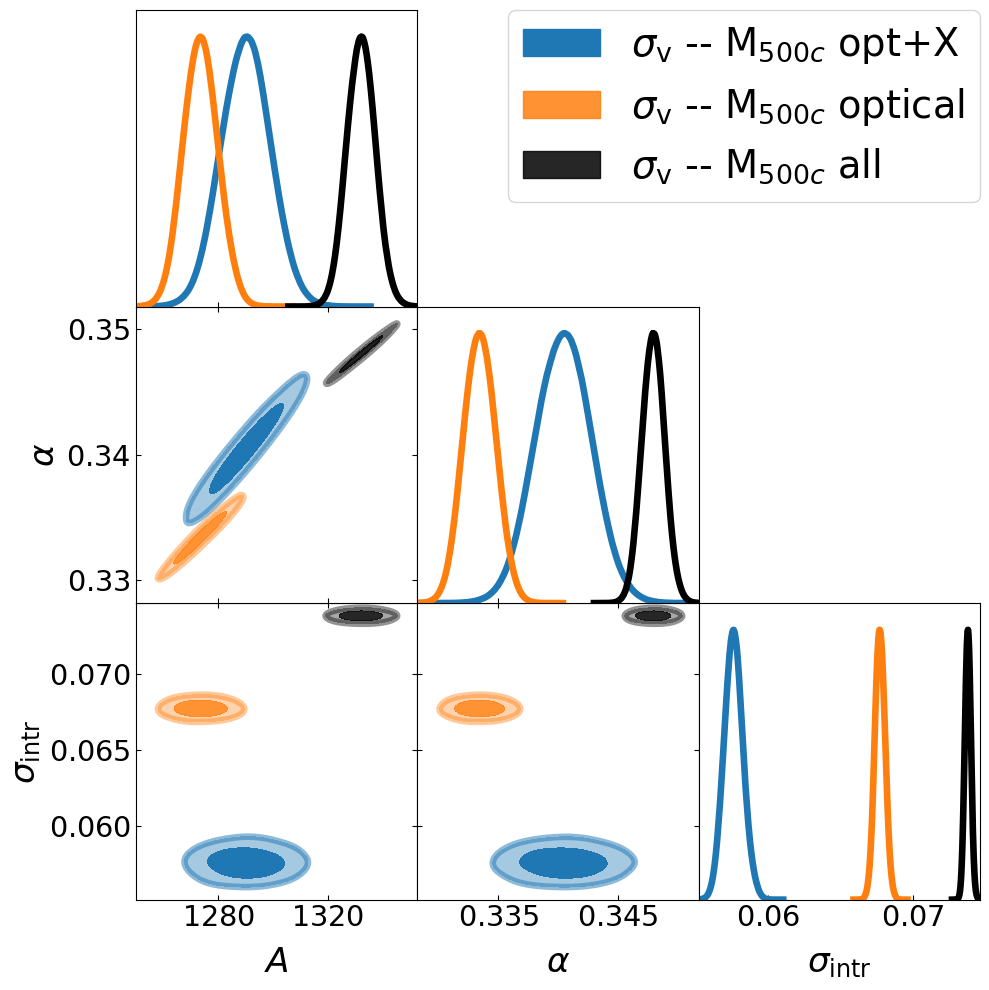}
    \includegraphics[width=0.8\columnwidth]{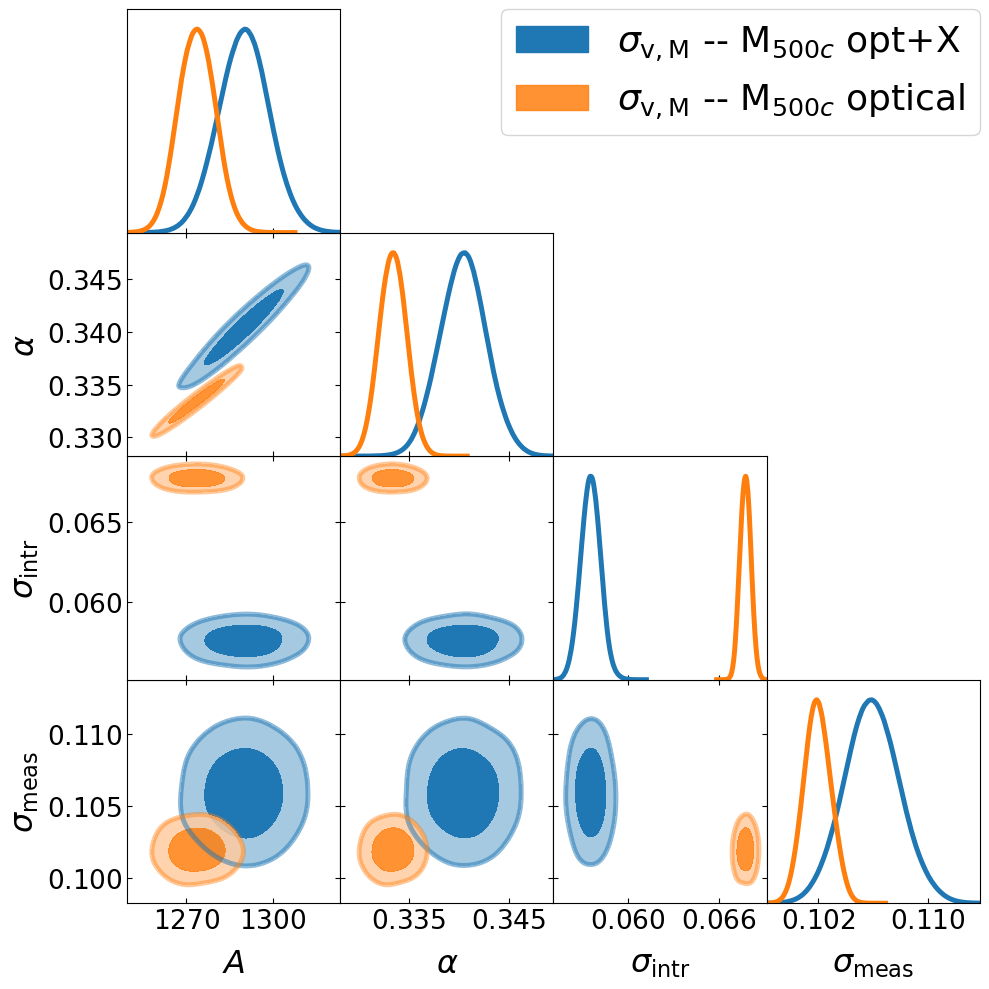}
    \caption{Same as Fig. \ref{fig:scalrel_posterior}, but referring to the scaling relation to M$_{\rm 500c}$.
    }
    \label{fig:scalrel_posterior_M500c}
\end{figure*}

\begin{table*}[]
    \centering
    \caption{Priors and posteriors of the scaling relation between halo mass M$_{\rm 500c}$ and velocity dispersion.}
    
    \begin{tabular}{c|c|c|c|c|c|c|c|}
    \hline
    \hline
       \textbf{Parameter}  & \textbf{Prior} & \textbf{$\sigma_{\rm v,T}$ All} & \textbf{$\sigma_{\rm v,T}$ Optical} & \textbf{$\sigma_{\rm v,T}$ Opt+X} & \textbf{$\sigma_{\rm v,M}$ Optical} & \textbf{$\sigma_{\rm v,M}$ Opt+X} \\
       \hline
       \rule{0pt}{2ex}    
        $A$ & $\mathcal{U}$(100, 2000) & 1332.2 $\pm$ 5.2 & 1273.6 $\pm$ 6.2 & 1290.1 $\pm$ 8.9 & 1273.6 $\pm$ 6.3 & 1290.0 $\pm$ 8.8 \\
        $\alpha$ & $\mathcal{U}$(0.1, 0.5) & 0.348 $\pm$ 0.001 & 0.333 $\pm$ 0.001 & 0.341 $\pm$ 0.002 & 0.333 $\pm$ 0.001 & 0.341 $\pm$ 0.002 \\
        $\sigma_{\rm intr}$ & $\mathcal{U}$(0.01, 1.0) & 0.073 $\pm$ 0.001 & 0.068 $\pm$ 0.001 & 0.058 $\pm$ 0.001 & 0.068 $\pm$ 0.001 & 0.058 $\pm$ 0.001 \\
        $\sigma_{\rm meas}$ & $\mathcal{U}$(0.01, 1.0) & - & - & - & 0.102 $\pm$ 0.001 & 0.106 $\pm$ 0.002 \\
        \hline
    \end{tabular}
    \tablefoot{The symbol $\mathcal{U}(M,N)$ denotes a uniform prior between the values M and N. Posterior values are reported from the third column onward, for each case labelled in the top row.}
    \label{tab:scaling_rel_pars_m500}
\end{table*}

Compared to other galaxy clusters samples, the strength of X-GAP resides in mapping the gas properties out to R$_{\rm 500c}$ with high signal to noise ratio for low mass haloes hosting galaxy groups. Therefore, estimating the halo mass at larger radii, such as M$_{\rm 200c}$ is challenging. 
As explained in the main body of the article, we focus first on M$_{\rm 200c}$ to compare our results with previous works, since most of the literature focuses on R$_{\rm 200c}$ for such studies. In addition, we use our mock to calibrate the scaling relation between velocity dispersion and M$_{\rm 500c}$. The formalism is the same as in Eq. \ref{eq:scal_rel}, we simply substitute M$_{\rm 200c}$ with M$_{\rm 500c}$.
We fit for $A$, $\alpha$, $\sigma_{\rm intr}$, and $\sigma_{\rm meas}$. We put uniform priors on the parameters as reported in Table \ref{tab:scaling_rel_pars_m500}.

The results for the two cases using M$_{\rm 200c}$ or M$_{\rm 500c}$ are qualitatively similar. For the M$_{\rm 500c}$ case, we find a higher normalisation compared to the M$_{\rm 200c}$ case equal to $A=1332.2 \pm 5.2$. This is expected, because the scaling relation is anchored at high mass of 10$^{\rm 15}$ M$_\odot$ and M$_{\rm 500c}$ is smaller than M$_{\rm 200c}$. The slope $\alpha = 0.348 \pm 0.001$ is slightly steeper, but compatible within the intrinsic scatter. The true intrinsic scatter $\sigma_{\rm intr} = 0.073 \pm 0.001$ is compatible within 1$\sigma$ with the M$_{\rm 200c}$ case. Similarly to the first case, the scatter is dominated by measurement uncertainty, with a $\sigma_{\rm meas} = 0.102 \pm 0.001$ and $\sigma_{\rm meas} = 0.106 \pm 0.002$ for the haloes detected in the optical and optical plus X-ray bands. Similarly to Fig. \ref{fig:veldisp_mass}, the full relation between velocity dispersion and M$_{\rm 500c}$ is reported in Fig. \ref{fig:veldisp_mass_500c}, with the top and bottom panels respectively showing the true and measured $\sigma_{\rm V}$ as a function of M$_{\rm 500c}$. Similarly, the full posterior distributions are shown in Fig. \ref{fig:scalrel_posterior_M500c}.

\section{Selection impact on scaling relations}
\label{appendix:sel_scalrel}

\begin{figure}
    \centering
    \includegraphics[width=0.8\columnwidth]{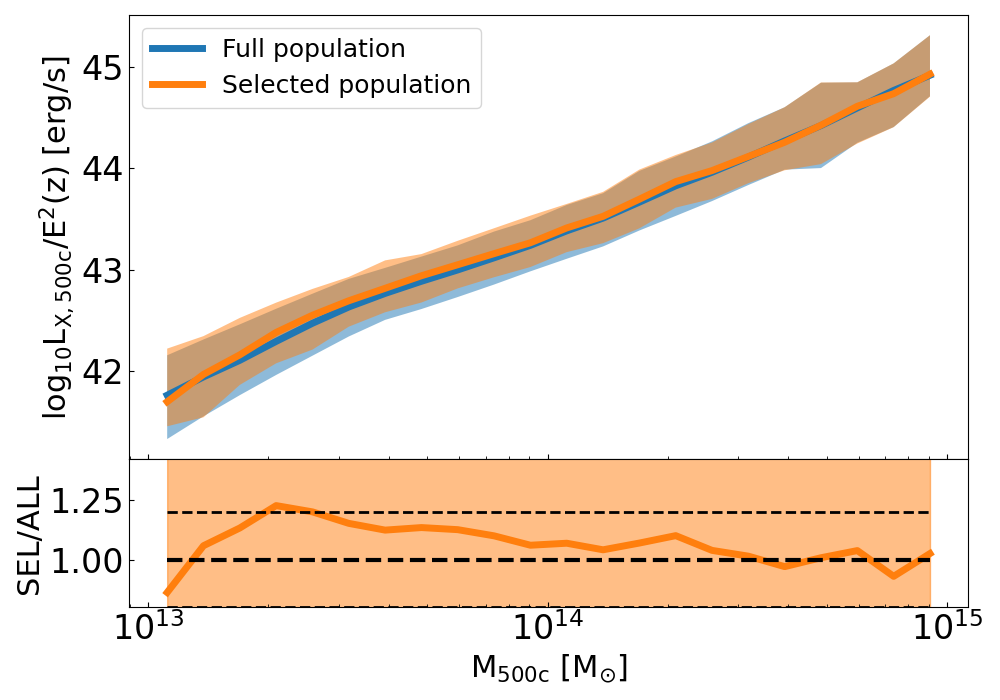}
    \includegraphics[width=0.8\columnwidth]{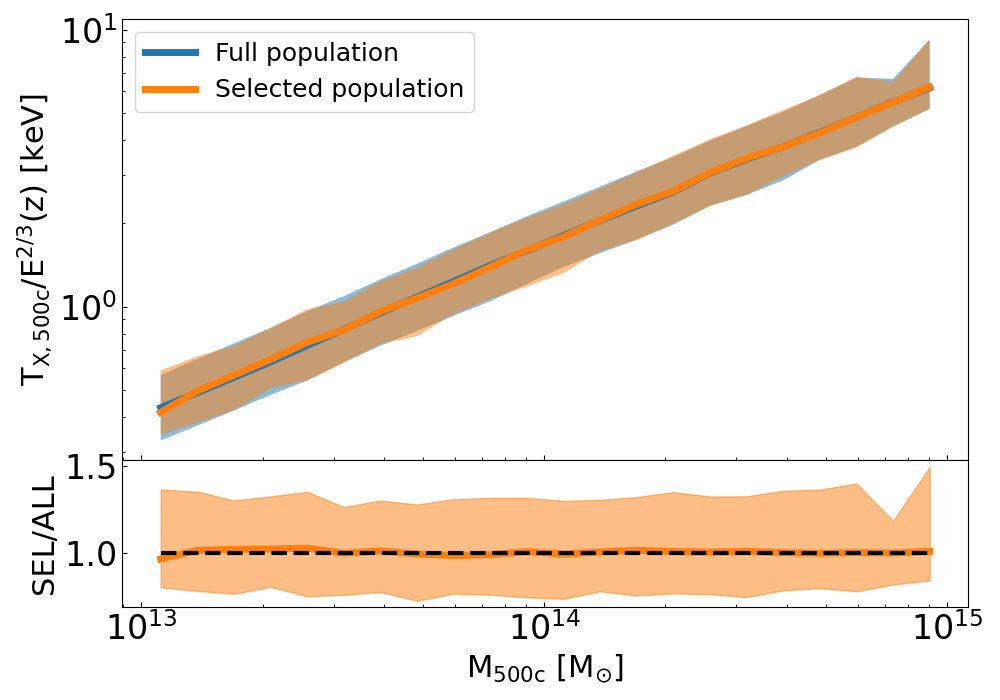}    
    \includegraphics[width=0.8\columnwidth]{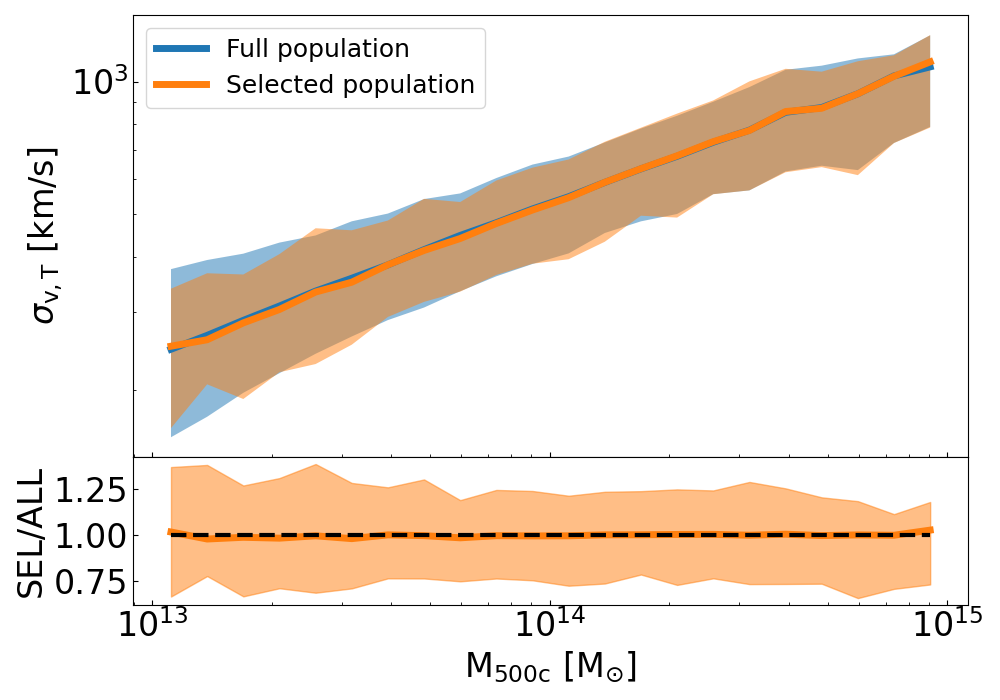}
    \caption{Scaling relation between observables and halo mass M$_{\rm 500c}$. The top panel shows the X-ray luminosity, the central one displays temperature, while the bottom one reports the velocity dispersion.}
    \label{fig:Xobs_mass}
\end{figure}

\begin{figure}
    \centering
    \includegraphics[width=0.8\columnwidth]{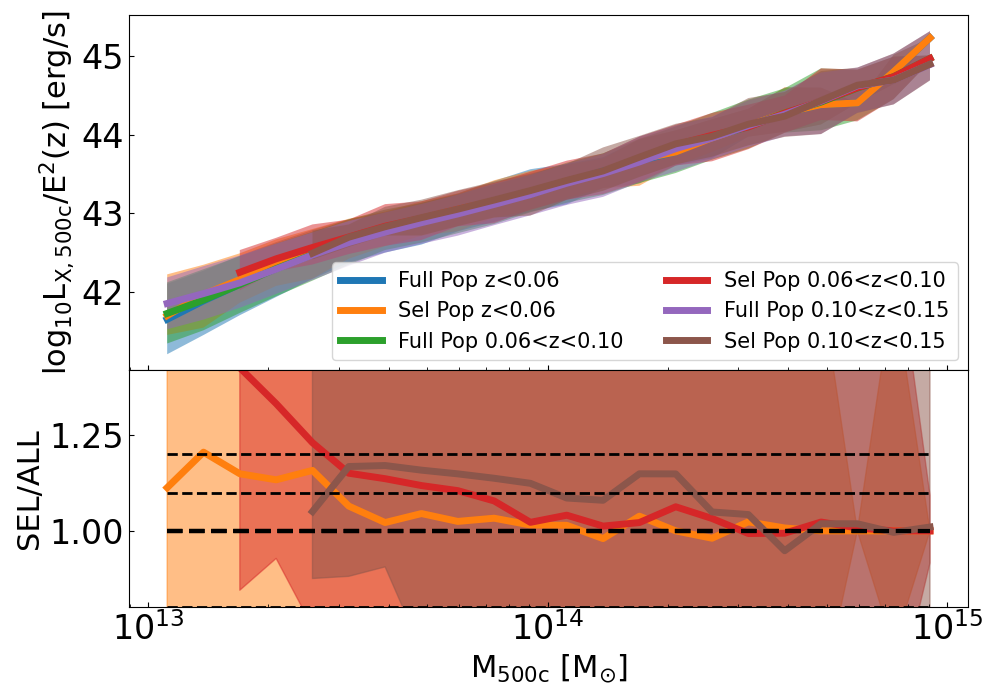} 
    \caption{Scaling relation between X-ray luminosity and halo mass, as in the top panel of Fig. \ref{fig:Xobs_mass}, but split in different redshift bins. The panel shows the comparison between full population in the light cone and the one selected from our detection scheme. Different colours denote various redshift intervals: below 0.06 (in blue and orange), then up to 0.1 (in green and red), and 0.15 (in purple and brown).}
    \label{fig:Lx_mass_zbins}
\end{figure}

In this section we present the impact of the selection on the scaling relations between halo mass and observables, i.e. temperature, luminosity, and velocity dispersion. In different mass bins, we compute the median observable, together with their 16th and 84th percentile points. We do it for the full population in the light cone and for the selected one after the optical and X-ray detection in our simulations. The result is collected in Fig. \ref{fig:Xobs_mass}. The top panels show the scaling relation rescaled by the self similar redshift evolution using E(z), the bottom panels display the ratio between the selected and the full population. We do not see an impact on the velocity dispersion to mass scaling relation, which confirms our previous findings in Sect. \ref{sec:probability_detection}, with the X-ray selection being the limiting factor in our setup, and Sect. \ref{sec:veldisp_mass}, with similar best fit parameters for the scaling relation before and after the selection. In addition, temperature does not impact the selection. The ratio between the selected and the full population is constant around 1. The same does not hold for luminosity. Moving towards lower masses, the selection is biased towards the brightest systems, with a 25$\%$ net effect 2$\times$10$^{13}$ M$_\odot$. This is expected, because luminosity is the intrinsic property related to the observable that mostly impact our detection probability, i.e. flux (see Sect. \ref{sec:probability_detection}).

We further investigate this trend by dividing our samples into multiple redshift bins. The result is shown in Fig. \ref{fig:Lx_mass_zbins}, where the redshift intervals are encoded in different colours: below 0.06 (in blue and orange), then up to 0.1 (in green and red), and 0.15 (in purple and brown). The bottom panel shows the ratio between the two populations in each bin, i.e. the orange (red, brown) line is divided by the reference blue (green, purple) one. 
We find that the low redshift population is the least biased. The ratio between the full and the selected population is close to one down to 4$\times$10$^{13}$ M$_\odot$, then it increases by about 20$\%$ down to about 1.5$\times$10$^{13}$ M$_\odot$. As redshift increases, the detection of faint systems at fixed mass is more challenging and the scaling relation is more biased. In the middle redshift bins, there is 10$\%$ bias at 6$\times$10$^{13}$ M$_\odot$, while in the higher redshift bin, the same bias is already reached at about 2$\times$10$^{14}$ M$_\odot$. This highlights the importance of accounting for the selection function in this type of analysis, which are naturally affected by selection effects in extragalactic surveys.

\section{Self similar emission measure profile}
\label{appendix:em_prof}

We detail the calculation of the self similar emission measure profiles in this appendix. We start from a cluster surface brightness profile, defined as count rate in different radial bins per unit surface, i.e. in units of cts/s/arcmin$^2$. We can obtain the APEC normalisation profile by dividing the surface brightness profile by the net count rate obtained from \texttt{xspec} by assuming a \texttt{Tbabs(APEC)} model with the proper values of $N_H$, temperature, redshift, metallicity and setting the normalisation to one, accounting for the response files of the instrument for the study \citep{Eckert2012A&A_outergas}.

The APEC normalisation is defined as

\begin{equation}
    \text{APEC NORM} = \frac{10^{-14}}{4\pi [d_A (1+z)]^2} \int n_e n_H dV,   [cm^{-5}] 
    \label{eq:apec_norm}
\end{equation}

with $d_A$ being the angular diameter distance at redshift z. 
The self similar emission measure integrated along the line of sight is

\begin{equation}
    EM_{\rm SS} = \frac{1}{\sqrt{kT/10\ \text{keV}} E(z)^3} \int n_e n_H dl.
    \label{eq:EM_SS}
\end{equation}

Combining equations \ref{eq:apec_norm} e \ref{eq:EM_SS} and splitting the volume integral between plane of the sky and line of sight we obtain

\begin{align}
    dV &= A \times dl \nonumber \\
    M &= \frac{10^{-14}}{4\pi [d_A (1+z)]^2} \nonumber \\
    N &= \sqrt{kT/10\ \text{keV}} E(z)^3 \nonumber \\
    \text{APEC NORM} &=  M \times N \times EM \times A.
    \label{eq:apec_EM}
\end{align}

Considering an area A of 1 arcmin$^2$ and the corresponding solid angle $\Omega$ with the proper conversion from arcmin to rad, we have
\begin{align}
    A &= \Omega \times d_A^2 \nonumber \\
    \Omega &= \Big(\frac{\pi}{180\times 60}\Big)^2.
    \label{eq:solid_angle}
\end{align}

Substituting equation \ref{eq:solid_angle} into \ref{eq:apec_EM} we get
\begin{equation}
    \text{APEC NORM} = \frac{10^{-14}\pi}{4(180\times 60)^2 (1+z)^2} \times N \times EM_{\rm SS}, 
\end{equation}

which we can invert to derive the final expression to infer the self similar emission measure profile from the APEC normalisation:
\begin{equation}
    EM_{\rm SS} = \frac{1}{\sqrt{kT/10\ \text{keV}} E(z)^3}  \frac{4(180\times 60)^2 (1+z)^2}{10^{-14}\pi} \times \text{APEC NORM}.
\end{equation}

\end{document}